\documentclass{ws-ijgmmp}
\usepackage{enumitem}
\usepackage{amsfonts}
\usepackage{amssymb}
\usepackage{graphicx}
\usepackage{array}
\usepackage{appendix}
\usepackage{mathrsfs}
\usepackage{hyperref}
\usepackage{amsmath}
\numberwithin{equation}{section}
\newtheorem{thm}{Theorem}[section]

\newtheorem{defn}{Definition}[section]

\begin{document}
\markboth{V. F. Bellino, G. Esposito}
{Fractional linear maps} 

\title{Fractional linear maps in general relativity and quantum mechanics}

\author{Vito Flavio Bellino}
\address{Dipartimento di Fisica ``Ettore Pancini'', Universit\`a degli Studi di Napoli ``Federico II'', 
Complesso Universitario di Monte
S. Angelo, Via Cintia Edificio 6, 80126 Napoli, Italy\\
\email{bell.vf95@gmail.com}}
\author{Giampiero Esposito}
\address{Dipartimento di Fisica ``Ettore Pancini'',
 Universit\`a degli Studi di Napoli ``Federico II'',
Complesso Universitario di Monte S. Angelo,
Via Cintia Edificio 6, 80126 Napoli, Italy\\  
Istituto Nazionale di Fisica Nucleare, Sezione di
Napoli, Complesso Universitario di Monte S. Angelo, 
Via Cintia Edificio 6, 80126 Napoli, Italy\\
\email{gesposit@na.infn.it}}

\maketitle

\begin{abstract}
This paper studies the nature of fractional linear transformations in
a general relativity context as well as in a quantum theoretical framework. 
Two features are found to deserve special attention: 
the first is the possibility of separating
the limit-point condition at infinity into loxodromic, hyperbolic, parabolic and
elliptic cases. This is useful in a context in which one wants to look for a
correspondence between essentially self-adjoint spherically symmetric Hamiltonians
of quantum physics and the theory of Bondi-Metzner-Sachs transformations in 
general relativity. The analogy therefore arising, suggests 
that further investigations might be performed for a theory in which the role of
fractional linear maps is viewed as a bridge between the quantum theory and
general relativity.
The second aspect to point out is the possibility of interpreting the limit-point
condition at both ends of the positive real line, for a second-order singular
differential operator, which occurs frequently in applied quantum mechanics, as
the limiting procedure arising from a very particular Kleinian group which is the
hyperbolic cyclic group. In this framework, this work finds that a
consistent system of equations can be derived and studied. Hence one is led to
consider the entire transcendental functions, from which it is possible to
construct a fundamental system of solutions of a second-order differential
equation with singular behavior at both ends of the positive real line, which 
in turn satisfy the limit-point conditions. Further developments in this
direction might also be obtained by constructing a fundamental system of solutions
and then deriving the differential equation whose solutions are the independent
system first obtained. This guarantees two important properties at the same time: 
the essential self-adjointness of a second-order differential operator and the
existence of a conserved quantity which is an automorphic function for the
cyclic group chosen. 
\end{abstract}

\section{Introduction}

Projective geometry was developed in the nineteenth century as a form of geometry
that describes graphical rather than metric properties 
\cite{Enriques,Spampinato}.
Nevertheless, over the years, it has been found to play a role in leading
to new directions both in differential geometry 
\cite{FubiniCech} and in pseudo-Riemannian geometry.
In the latter case, projective transformations play an important role in 
the asymptotic symmetry group of an asymptotically flat spacetime 
\cite{AE2018,EA2018} and hence in many asymptotic properties of classical and 
quantum field theories \cite{L1,L2,L3,L4,L5,L6,L7,L8,L9,L10,L11,L12,L13,L14,L15,L16,L17}.

In particular, the recent work in Ref. \cite{EA2018} has exploited the
analysis of fixed points of fractional linear transformations in order to
classify the Bondi-Metzner-Sachs transformations, and has even suggested
that a link exists between such transformations and their counterpart
in the theory of singular self-adjoint boundary-value problems. It has
been therefore our aim to understand whether such a correspondence is
truly conceivable, because it might imply that general relativity can
be seen as the bridge between classical and quantum physics.

For this purpose, Sect. 2 studies limit-point limit-circle theory and
its link with Bondi-Metzner-Sachs transformations. Section 3 is devoted
to the limit-point case at both ends of the positive real line, with the
associated hyperbolic cyclic groups. Eventually, our concluding remarks
are presented in Sect. 4.

\section{Limit Point, Limit Circle Theory and Bondi-Metzner-Sachs Transformations}

\subsection {Spherically Symmetric Hamiltonians}

We here introduce the limit-point, limit-circle 
theory for self-adjointness of Sturm-Liouville-like second-order 
differential operator on the real axis. The importance of this special 
class of operators in one particle quantum theory is well known, 
especially in the context of spherical symmetric Hamiltonians for bound states
\begin{equation}
\widehat{\cal{H}}=-{{{\rm{\hslash}}^{\rm{2}}}\over {2m}}\Delta 
+{\cal V}(r) , 
\label{(2.1)} 
\end{equation}
where the Euclidean $n$-dimensional Laplacian can be written in spherical coordinates as
\begin{equation}
-\Delta = - {{d^2}\over {dr^2}}-{{(n-1)}\over {r}}{{d}\over {dr}}-{\Delta}_S, 
\label{(2.2)}
\end{equation}
and the spherical Laplacian takes the form \cite{Chatterjee1990}
$$
-{\Delta}_S ={{{\hat{L}}^2}\over {{\rm{\hslash}}^2r^2}},
$$ 
in which ${\hat{L}}^2$ is the squared angular momentum operator of the particle. 
The Hilbert space to which the eigenfunctions of the operator in Eq. (2.1) 
should belong, is the closure \cite{ReedSimon1975} of the tensor space 
$$
{\cal{L}}^2\left({\Bbb{R}}^+,r^{\left(n-1\right)}dr\right)\otimes 
{\cal{L}}^2\left(S^{\left(n-1\right)},d\rm{\Omega}\right),
$$ 
in ${\cal{L}}^2\left({\Bbb{R}}^n,r^{\left(n-1\right)}dr \; d\rm{\Omega }\right)$, 
for which $S^{\left(n-1\right)}$ is the $\left(n-1\right)$-sphere embedded in 
${\Bbb{R}}^n$ and $d\rm{\Omega}$ is the surface element of such a 
sphere. The closure could be obtained by adjoining each limit of any 
sequence of functions of the space, to the space itself. The closure 
is thus a Hilbert space and coincides with 
${\cal{L}}^2 \left({\Bbb{R}}^n,r^{\left(n-1\right)}dr \; d\rm{\Omega}\right)$. 
Therefore, we can consider the eigenvalue problem for the operator (2.1)
$$
\widehat{\cal{H}}\varphi =E\varphi ,
$$ 
and from previous reasoning, we can factorize the eigenfunctions by the 
product of a purely angular function and a purely radial one
$$
\varphi \left(r;{\theta}_1,\dots ,{\theta}_{n-1}\right)
=\psi \left(r\right)\rm{\Theta }\left({\theta}_1,\dots ,{\theta}_{n-1}\right),
$$ 
but from Eq. (2.2), $\varphi$ is an eigenfunction of the operator (2.1) 
if and only if $\rm{\Theta}$ is an eigenfunction of ${\Delta}_S$, 
that is, if and only if one has \cite{Chatterjee1990}
$$
{{{\hat{L}}^2}\over {{\rm{\hslash}}^2r^2}}\rm{\Theta}
\left({\theta }_1,\dots ,{\theta}_{n-1}\right)={{l(l+n-2)}\over 
{{\rm{\hslash}}^2r^2}}\rm{\Theta}\left({\theta}_1,\dots 
,{\theta}_{n-1}\right),\; l=0,1,... .
$$ 
Thus, the eigenvalue equation reduces to
$$
\left[-\left({d^{2}\over dr^{2}}+{(n-1)\over r}{d \over dr}\right)
+{l(l+n-2)\over r^{2}}
+{2m \over \hbar^{2}} {\cal V}(r) \right]\psi(r)
={2m \over \hbar^{2}}E \psi(r),
$$ 
and one can see that the energy levels are strongly subjected to the 
angular momentum values and to the dimensionality of the Euclidean space 
under consideration. Therefore, it could be convenient to consider a family 
of two-integer-parameter operators by setting 
$$
{\kappa}_{n,l}=l\left(l+n-2\right),
$$ 
thus 
\begin{equation}
{\widehat{\cal{H}}}_r\left(n,l\right)=- \left[{d^{2}\over dr^{2}}+{(n-1)\over r}{d \over dr}\right]
+{\kappa_{n,l}\over r^{2}}
+{2m \over \hbar^{2}}{\cal V}(r). 
\label{(2.3)} 
\end{equation}
This is a family of operators acting on the 
${\cal{L}}^2\left({\Bbb{R}}^+,r^{\left(n-1\right)}dr\right)$ space. 
One can thus consider the following norm-preserving unitary map 
\cite{ReedSimon1975,EA2018}
\begin{equation}
\hat{U}:\;  \psi \in {\cal{L}}^2\left({\Bbb{R}}^+,r^{\left(n-1\right)}dr
\right)\longrightarrow \widetilde{\psi }=r^{{{\left(n-1\right)}\over {2}}}\psi 
\in {\cal{L}}^2\left({\Bbb{R}}^+,dr\right), 
\label{(2.4)} 
\end{equation}
which maps the operator (2.3) into
\begin{equation}
{\widehat{\widetilde{\cal{H}}}}_r\left(n,l\right)=\hat{U}{\widehat{\cal{H}}}_r
\left(n,l\right){\hat{U}}^{-1}=\left[-{d^{2}\over dr^{2}}+{(n-1)(n-3)\over 4r^{2}}
+{\kappa_{n,l}\over r^{2}}\right]+{2m \over \hbar^{2}} {\cal V}(r). 
\label{(2.5)} 
\end{equation}
By setting 
$$
{\lambda }_{n,l}=\left(l+{{1}\over {2}}\left(n-2\right)\right),
$$ 
and by using the equation for ${\kappa }_{n,l}$, we can write Eq. (2.5) as
\begin{equation}
\widehat{\widetilde{\cal{H}}}\left(n,l\right)=-{{d^2}\over {dr^2}}
+\left({{\left({\lambda }^2_{n,l}-{{1}\over {4}}\right)}\over {r^2}}
+{2m \over \hbar^{2}}{\cal V}(r) \right). 
\label{(2.6)} 
\end{equation}
The unitary operator (2.4) leaves the spectrum of 
${\widehat{\cal{H}}}_r\left(l,n\right)$ unaffected when the transformation 
(2.5) is applied. Hamiltonians of kind (2.6) have been studied in 
the literature for example in Refs. \cite{ReedSimon1975,deAlfaroRegge1965,Simon2015} 
and our attention is mainly focused on them because their form is suitable for the application of 
self-adjointness criterions first developed by H. Weyl in his early work \cite{Weyl1910}. 
In the subsequent developments we will review the limit-point, limit-circle 
theory and we hope it will be clear that the required self-adjointness 
for quantum mechanical Hamiltonians, does not only satisfy the empirical 
desire to conduct some sort of reasonable experiment, but also satisfies 
the curiosity of the theoretician which can investigate the matter of facts 
by looking with his mathematical lens, behind what is already known.

\subsection{Limit-Point, Limit-Circle Theory}

Throughout the present subsection we will investigate the spectral properties of 
the following differential operator on the real axis defined by
\begin{equation}
Lx=-\left(px'\right)'+qx, 
\label{(2.7)} 
\end{equation}
where the function $x$ is assumed to be a function of some $r$ variable 
on the real axis while $p^{-1},\ p'$ and $q$ are summable functions on any 
compact subinterval of interest \cite{BE2019}, 
and $p>0$ \cite{CoddingtonLevinson1955,Tit1962}. Note that the operator 
defined in (2.7) is analogous to the operator (2.6). We will call 
{\it singular points} of Eq. (2.7) each point which is a singular point for 
its coefficients or each point at infinity. For example, the operator (2.6) has 
two singular points: the point at infinity taken as the limit $r\to \infty $ and 
the point $r=0$ which is a singular point for
\begin{equation}
q\left(r\right)={{\left({\lambda }^2_{n,l}-{{1}\over {4}}\right)}\over 
{r^2}}+{2m \over \hbar^{2}} {\cal V}(r) .
\label{(2.8)}
\end{equation}
We note that in the case we are dealing with s-waves in three dimensions by picking up the 
operator ${\widehat{\widetilde{\cal{H}}}}_r\left(0,3\right)$, we have that 
${\lambda }^2_{3,0}={{1}\over {4}}$ and no singularity comes from the first 
term of the right-hand side of Eq. (2.8). Nevertheless, many physical 
potentials, for example the Coulomb potential, possess singular behaviour at 
$r=0$. Therefore, we will investigate operators of the type (2.7) by assuming 
such singular behaviour at both ends of the positive real axis. 

The limit-point, limit-circle theory treats singular self-adjoint problems of 
the second order whose differential equation is established in Eq. (2.7).  
For what follows, it is essential to consider the Green's formula which states 
that if $[r_1,r_2]$ is any interval in which the operator $L$ is defined and 
$f$ and $g$ are two functions for which $Lf$ and $Lg$ are meaningful, then
$$
\int^{r_2}_{r_1}{dr\left(\overline{g}Lf-f\overline{Lg}\right)}
=\left[fg\right]\left(r_2\right)-\left[fg\right]\left(r_1\right),
$$ 
where 
$$
\left[fg\right]\left(r\right)=p\left(r\right)\left(f\left(r\right)\overline{g}'
\left(r\right)-f'\left(r\right)\overline{g}\left(r\right)\right),
$$ 
and $\overline{f}$ is the complex conjugate of the function $f$.

\begin{defn}
Let $\tilde{r}$ be a singular point for Eq.(2.7)  
If for a particular complex number $l_0$ each solution of the equation
$$
Lx=l_0x,
$$ 
is square summable in some neighborhood of $\tilde{r}$, 
then $L$ is said to be in the ``limit-circle'' case at $\tilde{r}$. 
If this is not the case, then $L$ is said to be in the 
``limit-point'' case at $\tilde{r}$.
\end{defn}

The geometrical interpretation of this nomenclature will be clear soon. 
As already mentioned, we are mainly interested in only two singular points: 
the point at infinity and the point $r=0$ thus we will adapt each theorem 
and proof of Ref. \cite{CoddingtonLevinson1955} 
to these particular singular points.

\begin{thm}
Suppose that the only singular point in Eq. (2.7) 
is the point at infinity. If every solution of $Lx=l_0x$ is of 
class ${\cal{L}}^2\left(c,\infty \right)$ for some $c>0$ and some 
complex number $l_0$, then, for every arbitrary complex number $l$, 
every solution of $Lx=lx$ is of class ${\cal{L}}^2(c,\infty)$.
\end{thm}
\vskip 0.2cm
\noindent
{\it Proof.} Suppose $\varphi$ and $\psi$ are two linearly independent 
solutions of $Lx=l_0x$. Let $\chi $ be any solution of $Lx=lx$, or equivalently, of
$$
Lx=l_0x+\left(l-l_0\right)x.
$$ 
Upon multiplying $\varphi$ by a constant in order to achieve 
$\left[\varphi \overline{\psi}\right]\left(r\right)=1$ (note that if 
$f$ and $g$ are two solutions of $Lx=l_0x$ then $[\varphi \psi](r)$ is the 
Wronskian of the differential equation which is a constant for a fundamental 
system of solutions), the Lagrange variation of parameters formula yields
\begin{equation}
\chi \left(r\right)=c_1\varphi +c_2\psi +(l-l_0)\int^r_{\tilde{c}}
{dr'\left(\varphi \left(r\right)\psi \left(r'\right)-\varphi 
\left(r'\right)\psi \left(r\right)\right)\chi (r')},  
\label{(2.9)} 
\end{equation}
where $c_1,\ c_2$ and $\tilde{c}\ge c$ are three constants. If we set 
\begin{equation}
{\left\|\chi \right\|}^2_{\tilde{c}}=\int^r_{\tilde{c}}{dr'{|\chi |}^2},
\; r\ge \tilde{c},  
\label{(2.10)} 
\end{equation}
then there exists a constant $M$ such that 
${\left\|\varphi \right\|}_{\tilde{c}},{\left\|\psi \right\|}_{\tilde{c}}\ \le M$ 
for all $r>\tilde{c}$. The Schwarz inequality then gives
$$
\left|\int^r_{\tilde{c}}{dr'\left(\varphi \left(r\right)\psi \left(r'\right)
-\varphi \left(r'\right)\psi \left(r\right)\right)\chi (r')}\right|\le 
M\left(\left|\varphi \right|+\left|\psi \right|\right){\left\|\chi \right\|}_{\tilde{c}}.
$$ 
By using the Minkowski inequality jointly with the previous Schwarz inequality
$$
{\left(\int^r_{\tilde{c}}{dr'{(f+g)}^2}\right)}^{{{1}\over {2}}}\le 
{\left(\int^r_{\tilde{c}}{dr'f^2}\right)}^{{{1}\over {2}}}
+{\left(\int^r_{\tilde{c}}{dr'g^2}\right)}^{{{1}\over {2}}},
$$ 
into Eq. (2.9) we easily get
$$
{\left\|\chi \right\|}_{\tilde{c}}\le \left(\left|c_1\right|
+\left|c_2\right|\right)M+2M^2\left|l-l_0\right|\ {\left\|\chi \right\|}_{\tilde{c}},
$$ 
and if $\tilde{c}$ is chosen large enough so that 
$M^2\left|l-l_0\right|<{{1}\over {4}}$, then 
$$
{\left\|\chi \right\|}_{\tilde{c}}\le 2\left(\left|c_1\right|+\left|c_2\right|\right)M,
$$ 
and since the right-hand side of this inequality is independent of $r$, 
then $\chi \in {\cal{L}}^2(\tilde{c},\infty )$ for all $\tilde{c}\ge c$ and 
thus is ${\cal{L}}^2(c,\infty )${\it .}\hfill\hfill Q.E.D.

\begin{thm}
Suppose that the only singular point in Eq. (2.7)
is $r=0$. If every solution of $Lx=l_0x$ is of 
class ${\cal{L}}^2\left(0,c\right)$ for some $c>0$ and some 
complex number $l_0$, then, for every arbitrary complex number $l$, 
every solution of $Lx=lx$ is of class ${\cal{L}}^2(0,c)$.
\end{thm}
\vskip 0.2cm
\noindent
{\it Proof.} The proof is not different from that of the previous theorem. 
In this case Eq. (2.9) still holds but we are interested in the the inequality 
chain $0<r\le \tilde{c}\le c$ and this forces a modification for Eq. (2.10) 
which should be written as
$$
{\left\|\chi \right\|}^2_{\tilde{c}}=\int^{\tilde{c}}_r{dr'
{|\chi |}^2}, \; r \le \tilde{c}.
$$ 
Then there must exist a constant $M$ such that ${\left\|\varphi \right\|}_{\tilde{c}},
{\left\|\psi \right\|}_{\tilde{c}} \le M$ from which it follows the Schwarz 
and Minkowski inequality as stated above. Then, one can always chose a small 
enough $\tilde{c}$ such that $M^2\left|l-l_0\right|<{{1}\over {4}}$ and thus 
$\chi \in {\cal{L}}^2(0,\tilde{c})$ for every $\tilde{c}\le c$.\hfill\hfill Q,E.D.
 
The above theorems show that in the limit-point case, at most one linearly 
independent solution of $Lx=lx$ is of class ${\cal{L}}^2$ near the singular point, 
which we have chosen to be $r=0$ and infinity. Now we will show that in the 
limit-point case there is indeed one and only one square integrable function 
near the singular point for each $l$ such that the imaginary part ${\frak{I}}(l)\ne 0$. 
This proof will be carried out via a very powerful geometrical interpretation 
of the limit-point and limit-circle cases. 
\vskip 0.3cm
\noindent
{\it (a) Geometrical interpretation of the limit-point, limit-circle cases at infinity}

Suppose $Lx=lx$ to be defined in $[c,\infty [$ with $c>0$ and that the only singular 
point in this interval is the point at infinity. Let $\varphi $ and $\psi $ be 
two independent solutions satisfying 
$$
\varphi \left(c,l\right)={\sin \alpha}, \;  
\psi \left(c,l\right)={\cos \alpha},
$$ 
\begin{equation}
p\left(c\right)\varphi '\left(c,l\right)=-{\cos \alpha}, \;  
p\left(c\right)\psi '\left(c,l\right)= \sin \alpha , 
\label{(2.11)} 
\end{equation}
where $\alpha \in [0,\pi [$. Clearly, $\varphi $ and $\psi $ are linearly 
independent solutions. Note that for each $\alpha \in [0,\pi [$, conditions 
(2.11) can be always achieved by setting up a rather general Cauchy problem 
with initial point $c>0$. From general arguments about the existence of 
solutions for the equation $Lx=lx$, one can state that $\varphi ,\varphi ',\psi ,\psi '$ 
are entire functions of $l$ and continuous in the variables ($r,l$). 
Obviously, we have $\left[\varphi \overline{\psi }\right]\left(c\right)=1$ 
and thus $\left[\varphi \overline{\psi}\right]\left(r\right)=1$ for each $r$. 
These solutions are real for real $l$ and satisfy the following mixed boundary conditions in $c$:
$$
{\cos \alpha} \; \varphi \left(c,l\right)+{\sin \alpha} \; p
\left(c\right)\varphi '\left(c,l\right)=0,
$$ 
$$
\sin \alpha \; \psi \left(c,l\right)-{\cos 
\alpha} \; p(c)\psi '\left(c,l\right)=0.
$$ 
Every solution to $Lx=lx$ must be of the form
$$
\chi =\varphi +m\psi ,
$$ 
with some constant $m$ which depends upon $l$. Now consider the following 
boundary conditions at $b$ with $c<b<\infty $:
\begin{equation}
{\cos \beta} \; x(b)+{\sin \beta} \; p(b)
x'(b)=0,\; \; \;  \beta \in [0,\pi [. 
\label{(2.12}) 
\end{equation}
One can see that if $\chi$ must satisfy condition (2.12) then it must be
$$
m=-{\cot \beta \varphi \left(b,l\right)+p(b)
\varphi' \left(b,l\right) \over \cot \beta \psi 
\left(b,l\right)+p(b)\psi '(b,l)} ,
$$ 
which is a function of the triplet $(l,b,\beta )$. Since 
$\varphi ,\varphi ',\psi ,\psi '$ are entire and continuous functions of 
$(l,r)$, then it follows that $m$ is meromorphic in $l$ and real for real $l$. 
By setting $z=\cot \beta$, this function becomes
\begin{equation}
m=-{{Az+B}\over {Cz+D}}, 
\label{(2.13)} 
\end{equation}
where the coefficients $A,B,C,D$ are functions of the pair $(l,b)$ and one 
can easily see what these correspond to. Equation (2.13) is a fractional 
linear transformation when we freely let $z$ run on $\widehat{\Bbb{C}}$. We 
already know that such kind of transformations are responsible of a one-to-one 
mapping between circles of the complex plane. Therefore, the $z$ variable 
runs over the real line when we let $\beta$ vary on its range $[0,\pi [$ and 
the map (2.13) transforms such a line into a circle $C_b$ (note that the 
circle is strictly related to the coefficients appearing in Eq. (2.13) and 
thus to the upper boundary point $b$) on the $m$ complex plane. Thus, $\chi$ 
satisfies the condition (2.12) if and only if $m$ lies on the circle $C_b$.

The derivation of the equation for such a circle is not different from that 
obtained in Ref. \cite{Ford}, and it is
\begin{equation}
\left(\overline{A}+\overline{C}\overline{m}\right)\left(B+Dm\right)
-\left(A+Cm\right)\left(\overline{B}+\overline{D}\overline{m}\right)=0. 
\label{(2.14}) 
\end{equation}
One can show that the centre and the radius for $C_b$ must respectively be
$$
{\tilde{m}}_b={{A\overline{D}-B\overline{C}}\over {\overline{C}D-C\overline{D}}},
$$ 
$$
r_b={{\left|AD-BC\right|}\over {\left|\overline{C}D-C\overline{D}\right|}}.
$$ 
From the fact that 
$$
A=\varphi \left(b,l\right), \;  B=p(b) \varphi '\left(b,l\right),
$$ 
\begin{equation}
C=\psi \left(b,l\right), \;  D=p(b)\psi '\left(b,l\right), 
\label{(2.15)} 
\end{equation}
one can see that Eq. (2.14) can be written as
\begin{equation}
\left[\chi \chi \right](b)=0, 
\label{(2.16)} 
\end{equation}
while
$$
\left[\varphi \psi \right](b)=A\overline{D}-B\overline{C},
$$ 
$$
\left[\psi \psi \right]\left(b\right)=C\overline{D}-\overline{C}D,
$$ 
$$
\left[\varphi \overline{\psi}\right]\left(b\right)=AD-BC=1,
$$ 
thus 
\begin{equation}
{\tilde{m}}_b=-{{\left[\varphi \psi \right]\left(b\right)}\over 
{\left[\psi \psi \right](b)}},\ \ \ \ \ \ \ \ \ \ r_b={{1}
\over {\left|\left[\psi \psi \right](b)\right|}}. 
\label{(2.17)} 
\end{equation}
Since the coefficient of $m\overline{m}$ in Eq. (2.14) is $[\psi \psi ](b)$, 
it follows that the interior of $C_b$ is given by
\begin{equation}
{{[\chi \chi ](b)}\over {[\psi \psi ](b)}}<0. 
\label{(2.18)} 
\end{equation} 
Now, by using the following Green's formula:
$$
\int_{c}^{b}dr(\bar\psi L\psi-\psi \overline{L\psi})=(l-\bar l)\int_{c}^{b}dr\,
\psi\bar{\psi}=[\psi\psi](b)-[\psi\psi](c),
$$
and recalling that $[\psi\psi](c)=p(c)(\psi(c)\bar\psi'(c)-\psi'(c)\bar\psi(c))$, 
from Eqs. (2.11) one obtains
\begin{equation}
\left[\psi \psi \right](b)=2i{\frak{I}}(l)\int^b_c{dr'{|\psi |}^2}, 
\label{(2.19)} 
\end{equation}
as well as 
\begin{equation}
\left[\chi \chi \right](b)=2i{\frak{I}}\left(l\right)
\int^b_c{dr'{\left|\chi \right|}^2+\left[\chi \chi \right](c)}, 
\label{(2.20)} 
\end{equation}
and since $\left[\chi \chi \right](c)=-2i{\frak{I}}(m)$, Eq. (2.18) becomes 
\begin{equation}
\int^b_c{dr'{|\chi |}^2}<{{{\frak{I}}(m)}\over {{\frak{I}}(l)}},\ \ \ \ \ \ \ 
{\frak{I}}(l)\ne 0. 
\label{(2.21)} 
\end{equation} 
Hence, all interior points of $C_b$ are defined by the previous equation while all 
points on the circle $C_b$ satisfy the equality sign in place of the inequality 
sign into Eq. (2.21). The radius is thus
\begin{equation}
r_b={\left(2|{\mathfrak{I}}(l)|\int^b_c{dr'{|\psi |}^2}\right)}^{-1}. 
\label{(2.22)} 
\end{equation}
If one chooses some other upper end $\tilde{b}<b$, Eq. (2.14) defines 
another circle $C_{\tilde{b}}$ whose radius is larger than that of $C_b$. One 
can ask how $C_b$ and $C_{\tilde{b}}$ are related one to the other and it 
follows from Eq. (2.21) that
$$
\int^{\tilde{b}}_c{dr'{|\chi |}^2}<\int^b_c{dr'{|\chi |}^2}
\le {{{\mathfrak{I}}(m)}\over {{\mathfrak{I}}(l)}},
$$ 
thus all points of $C_b$ are contained in the interior of $C_{\tilde{b}}$. 
This means that while increasing the upper end $b$ taking the limit $b\to \infty $, 
this process lets the circles converge to a limit point $m_{\infty }$ or to a 
limit circle $C_{\infty }$. In the former case, the radius of the circle $C_b$ 
must converge to zero, i.e. $r_b\to 0$ and thus 
$$
\lim_{b \to \infty} \int^b_c dr'{|\psi|}^{2}=\infty ,
$$ 
and the function $\psi$ does not belong to ${\cal{L}}^2(c,\infty )$, i.e. not all 
solutions of the equation $Lx=lx$ are square summable in the neighbourhood of infinity 
and this coincides with the limit-point case previously defined. 
But from Eq. (2.21) one sees that
$$
\int^b_c{dr'{|\chi|}^2}<{{{\frak{I}}{(m}_{\infty})}\over {{\frak{I}}(l)}},
$$ 
where $m_{\infty }$ is the limit point. Therefore, by letting $b$ approach 
infinity in the previous equation, one deduces that $\chi \in {\cal{L}}^2(c,\infty )$, 
and from the fact that $\psi$ is not square summable we obtain that there is 
one and only one independent solution which is square summable near infinity, 
as we have already mentioned above. 

In the latter case, the radius $r_b$ approaches a limit $r_{\infty}>0$ and 
this implies that $\psi \in {\cal{L}}^2(c,\infty )$. If ${\hat{m}}_{\infty }$ is 
any point on the limit circle $C_{\infty }$, it obviously gives rise to the following equation:
$$
\int^b_c{dr'{|\chi |}^2}<{{{\frak{I}}({\hat{m}}_{\infty })}\over {{\frak{I}}(l)}},
$$ 
and by taking the limit $b \to \infty$, one deduces that besides $\psi $, also 
$\chi$ is square summable near infinity thus every solution is 
${\cal{L}}^2(c,\infty)$ and this coincides with the limit-circle case previously 
defined. In this case $m$ lies on $C_{\infty}$ if and only if
\begin{equation}
{\mathfrak{I}}\left(l\right)\int^{\infty }_c{dr'}{\left|\chi \right|}^2
={\mathfrak{I}}\left(m\right), 
\label{(2.23)} 
\end{equation}
and since $\left[\chi \chi \right]\left(c\right)=-2i{\frak{I}}(m)$, from Eqs. 
(2.20) and (2.21) we deduce that $m$ is on $C_{\infty }$ if and only if 
$\left[\chi \chi \right]\left(\infty \right)=0$.  We have thus proved the following theorem:

\begin{thm}
Let ${\mathfrak{I}}\left(l\right)\ne 0$ and $\varphi ,\psi$ 
be linearly independent solutions of $Lx=lx$, where the equation 
is defined on $[c,\infty [$ with $c>0$ and have its only singular 
point at infinity. Suppose that these solutions satisfy Eq. (2.11), then 
the solution $\chi=\varphi +m\psi$ satisfies the real boundary 
condition (2.12) if and only if $m$ lies on the 
circle $C_b$ in the complex plane whose equation is $\left[\chi \chi \right]
\left(b\right)=0$. As $b\to \infty$ either $C_b\to C_{\infty}$, 
a limit circle, or $C_b\to m_{\infty}$, a limit point. All solutions 
of $Lx=lx$ are ${\cal{L}}^2\left(c,\infty \right)$ in the former 
case, and if ${\frak{I}}\left(l\right)\ne 0$, there is exactly one linearly 
independent solution which is ${\cal{L}}^2\left(c,\infty \right)$ in the 
latter case. Moreover, in the limit-circle case, a point is on the limit 
circle $C_{\infty }\left(l\right)$ if and only if $\left[\chi 
\chi \right]\left(\infty \right)=0$.
\end{thm}

At this stage of the theory it is often convenient to state some criterion 
\cite{Weyl1910,ReedSimon1975}
for the establishment of the limit-point or limit-circle case in such a 
way that one can always obtain limiting properties by simply looking at the 
coefficients of the second order differential operator. Our aim is quite distinct 
here: we do not want to recover the maximal amount of information about limit-circle, 
limit-point properties for some special kind of operators, but we want to establish 
the fundamental fact that the requirement of self-adjointness for Eq. (2.7) is 
always accompanied by a pictorial geometrical interpretation which could suggest 
some investigation paths. This is the motivation for treating explicitly the 
geometrical interpretation of limit-point, limit-circle cases at $r=0$, which, as 
we will see, is carried out with a slight modification of some equations derived 
for the geometrical interpretation at infinity.

\vskip 0.3cm

{\it (b) Geometrical interpretation of the limit-point, limit-circle cases at the origin.}
\vskip 0.3cm

Suppose $Lx=lx$ to be defined in $]0,c]$ with $c>0$ and that the only singular 
point in this interval is at $r=0$. Let $\varphi$ and $\psi$ be two 
independent solutions of the equation. We generally want that such solutions 
coincide with that assumed for the case {\it (a)} from the fact that operators 
of type (2.6) are defined on the entire positive real line and thus they often 
possess singular behaviours at $r=0$ and infinity, as we have already mentioned. 
Therefore we require that Eqs. (2.11) should hold also for $\varphi $ and 
$\psi $ here introduced. We thus have $\left[\varphi \overline{\psi}\right]
\left(r\right)=1$ for all $r \in ]0,c]$ and 
$$
\cos \alpha \; \varphi \left(c,l\right)+\sin \alpha \; p(c)
\varphi'\left(c,l\right)=0,
$$ 
$$
\sin \alpha \; \psi \left(c,l\right)-{\cos \alpha} \; p(c)
\psi '\left(c,l\right)=0,
$$ 
as in the case (a). Every solution of the equation must be of the 
form $\chi =\varphi +m'\psi $. The boundary condition (2.11) must be modified 
by considering a point $a$ for which $0<a<c$. It is replaced by
\begin{equation}
{\cos \beta'} \; x(a)+{\sin \beta'} \; p(a)
x'(a)=0, \;  \beta' \in [0,\pi [, 
\label{(2.24)} 
\end{equation}
and if we require $\chi$ to satisfy Eq. (2.24), we must have
$$
m'=-{{{\rm{cot} \beta '\ }\varphi \left(a,l\right)+p(a)
\varphi '\left(a,l\right)}\over {{\rm{cot} \beta '}\psi 
\left(a,l\right)+p(a)\psi '\left(a,l\right)}},
$$ 
and it is a function of the triplet $(l,a,\beta ')$. Since 
$\varphi ,\varphi',\psi ,\psi'$ are entire and continuous functions 
of $(l,r)$, then it follows that $m'$ is also meromorphic in $l$ and real 
for real $l$. By setting $z'={\rm{cot} \beta '}$ we obtain the 
analogous equation of Eq. (2.13) which is
\begin{equation}
m'=-{{A'z'+B'}\over {C'z'+D'}}, 
\label{(2.25)} 
\end{equation}
and this last equation maps the real line into a circle $C'_a$ on the 
$m'$ plane as before. The equation for $C'_a$ is analogous to Eq. (2.14) 
by letting all quantities be primed. The center and the radius are
$$
{\tilde{m}'}_a={{A'\overline{D'}-B'\overline{C'}}\over 
{\overline{C'}D'-C'\overline{D'}}},
$$ 
$$
r_a={{\left|A'D'-B'C'\right|}\over {\left|\overline{C'}D'-C'\overline{D'}\right|}},
$$ 
where 
$$
A'=\varphi \left(a,l\right), \;  B'=p\left(a\right)\varphi '
\left(a,l\right),
$$ 
\begin{equation}
C'=\psi \left(a,l\right), \;  D'=p\left(a\right)\psi '
\left(a,l\right). 
\label{(2.26)} 
\end{equation}
from which
$$
\left[\varphi \psi \right](a)=A'\overline{D'}-B'\overline{C'},
$$ 
$$
\left[\psi \psi \right](a)=C'\overline{D'}-\overline{C'}D',
$$ 
$$
\left[\varphi \overline{\psi}\right](a)=A'D'-B'C'=1.
$$ 
The equation for ${C'}_a$ is equally given by the concise form 
$\left[\chi \chi \right](a)=0$ while the centre and the radius 
can be written as
$$
{\tilde{m}'}_a=-{{\left[\varphi \psi \right](a)}\over 
{\left[\psi \psi \right](a)}}, \;  r_a
={{1}\over {\left|\left[\psi \psi \right](a)\right|}}.
$$ 
The interior of ${C'}_a$ is obtained by a modification of Eq. (2.18) as
\begin{equation}
{{[\chi \chi ](a)}\over {[\psi \psi ](a)}}<0, 
\label{(2.27)} 
\end{equation}
and from the Green's formula we obtain a modification of Eq. (2.19) 
which in this case must be written as
\begin{equation}
\left[\psi \psi \right](a)=-2i{\frak{I}}\left(l\right)
\int^c_a{dr'{\left|\psi \right|}^2}, 
\label{(2.28)} 
\end{equation}
by making use of the Green's formula jointly with Eq. (2.11), as below:
$$
\int_{a}^{c}dr(\bar\psi L\psi-\psi\overline{L\psi})=(l-\bar l)\int_{c}^{b}dr\,
\psi\bar{\psi}=[\psi\psi](c)-[\psi\psi](a),
$$
while Eq. (2.20) is replaced by
\begin{equation}
\left[\chi \chi \right](a)=-2i{\frak{I}}\left(l\right)
\int^c_a{dr'{\left|\chi \right|}^2+\left[\chi \chi \right]\left(c\right)}, 
\label{(2.29)} 
\end{equation}
by also making use of the Green's formula
$$
\int_{a}^{c}dr(\bar\chi L\chi-\chi\overline{L\chi})
=(l-\bar l)\int_{c}^{b}dr\,\chi\bar{\chi}=[\chi\chi](c)-[\chi\chi](a).
$$
In this way, Eq. (2.27) must be written for all $l$ such that ${\frak{I}}(l)\ne 0$, as
\begin{equation}
\int^c_a{dr'{|\chi |}^2}<-{{{\frak{I}}\left(m'\right)}\over 
{{\frak{I}}\left(l\right)}},\; {\frak{I}}(l)\ne 0, 
\label{(2.30)} 
\end{equation}
and this last equation defines all the interior points of ${C'}_a$. 
The equality sign defines all points that lie on ${C'}_a$. One can see that 
Eq. (2.30) is very different from Eq. (2.21) because here 
${\frak{I}}\left(m'\right)$ is required to be opposite in sign to 
${\frak{I}}\left(l\right)$ while in Eq. (2.21) the same sign is required. 
Thus, if we fix the complex number $l$, then the functions $m(l,b,\beta )$ 
and $m'(l,a,\beta ')$ must lie on opposite complex half-planes and thus have 
opposite sign for their imaginary part. As before, the radius of ${C'}_a$ is
\begin{equation}
r_a={\left(2|{\frak{I}}(l)|\int^c_a{dr'{|\psi |}^2}\right)}^{-1}, 
\label{(2.31)} 
\end{equation}
which can approach a finite limit or tend to zero in the limiting procedure 
$a\to 0$. However, if $\tilde{a}$ is such that $0<a<\tilde{a}$, then it 
defines a circle ${C'}_{\tilde{a}}$ and from the fact that
$$
\int^c_{\tilde{a}}{dr'{|\chi |}^2}<\int^c_a{dr'{|\chi |}^2}<
-{{{\frak{I}}\left(m'\right)}\over {{\frak{I}}\left(l\right)}},
$$ 
we deduce that ${C'}_{\tilde{a}}$ contains ${C'}_a$. As it happened in 
the case (a) for the geometrical interpretation at infinity, we 
distinguish the limit-point case when the radius (2.31) approaches zero 
as $a\to 0$, from the limit-circle case when Eq. (2.31) approaches a finite 
value. In the limit-point case, $\psi$ is not square summable near the 
origin, while it happens that
$$
\int^c_0{dr'{|\chi |}^2}<-{{{\frak{I}}{(m'}_{\infty })}\over {{\frak{I}}(l)}},
$$ 
where ${m'}_{\infty }$ is the limit point. Thus there is one and only one 
solution which is square summable near the origin. In the limit-circle case, 
$\psi$ is square summable near the origin and if we choose a point 
${\hat{m}'}_{\infty }$ lying on the limit circle ${C'}_{\infty }$, then 
$$
\int^c_0{dr'{|\chi |}^2}<-{{{\frak{I}}{(\hat{m}'}_{\infty })}\over {{\frak{I}}(l)}},
$$ 
and we have that all solutions to the equation $Lx=lx$ are ${\cal{L}}^2(0,c)$.

We have thus proved the analogous of Theorem 2.3, i.e.

\begin{thm}
Let ${\mathfrak{I}}\left(l\right)\ne 0$ and  
$\varphi ,\psi$ be linearly independent solutions of $Lx=lx$, 
where the equation is defined in $]0,c]$ with $c>0$ and have 
its only singular point at $r=0$. Suppose that these solutions satisfy 
Eq. (2.11), then the solution $\chi =\varphi +m\psi$ satisfies 
the real boundary condition (2.24) if and only if $m$  
lies on the circle ${C'}_a$ in the complex plane whose equation 
is $\left[\chi \chi \right](a)=0$. As
$a\to 0$ either ${C'}_a\to {C'}_{\infty }$, a limit 
circle, or ${C'}_a\to {m'}_{\infty}$, a limit point. All 
solutions of $Lx=lx$ are ${\cal{L}}^2\left(0,c\right)$ in 
the former case, and if ${\frak{I}}\left(l\right)\ne 0$, there is 
exactly one linearly independent solution which is ${\cal{L}}^2\left(0,c\right)$ 
in the latter case. Moreover, in the limit-circle case, a point is 
on the limit circle ${C'}_{\infty }\left(l\right)$ if and only if
$\left[\chi \chi \right]\left(\infty \right)=0$.
\end{thm}

Now that we have provided the geometrical interpretation of the limit-point, 
limit-circle theory, we revert to another important question which can be 
answered by this theory, that is the self-adjointness of the operator (2.7) 
in the case it is singular at both ends of the real positive line.

\subsection {Singular Behavior at Both Ends of the Interval}

Here we consider the interval $(0,\infty )$ and suppose that the coefficients 
of the operator $L$ have a singular behaviour at $r=0$. Thus we are treating 
singular behaviour at both ends of the positive real line. We suppose that 
$p\left(r\right)>0$ on such a semi-infinite interval and that $p,p',q$ are 
real and continuous on ${\Bbb{R}}^+$ (these conditions can be relaxed somewhat). 
Let $c>0$ and let ${\varphi }_1,{\varphi }_2$ be two solutions to $Lx=lx$, 
real for real $l$, and satisfying the following conditions at $c$:
$$
{\varphi }_1\left(c,l\right)=1,\; {\varphi }_2\left(c,l\right)=0,
$$ 
$$
p\left(c\right){\varphi '}_1\left(c,l\right)=0,\; p
\left(c\right){\varphi '}_2\left(c,l\right)=1,
$$ 
then ${\varphi }_1$ and ${\varphi }_2$ form a fundamental system of 
solutions for the equation and they are also entire functions of $l$ for 
fixed $r$. Let $\delta =[a,b]\subset {\Bbb{R}}^+$ be a finite interval 
containing $c$ and consider the following self-adjoint problem:
\begin{equation}
\left \{
\begin{array}{rll}  Lx=lx , 
\\ 
& {\cos \beta'} \; x(a)+{\sin \beta'} \; p(a)x'(a)=0, 
\\ 
& {\cos \beta} \; x(b)+{\sin \beta} \; p(b)x'(b)=0,
\end{array} \right . 
\label{(2.32)} 
\end{equation}
with $\beta ,\beta '\in [0,\pi [$. Then there exists a countable sequence 
of eigenvalues $\left\{{\lambda }^{(\delta )}_n\right\}$, $n=1,2,\dots $, 
and a complete set of orthonormal eigenfunctions $\left\{h^{(\delta )}_n\right\}$ 
in ${\cal{L}}^2(\delta )$. If there is some degeneracy for any of the 
eigenvalues, we indicate all the eigenfunctions belonging to the eigenspace under 
consideration, by substituting the index $n$ with $n_m$, $m=1,2,\dots$ when it 
is necessary. When it is not specified, all summations over $n$ are meant 
to be summations over the complete set of eigenfunctions regardless of the 
order which can be established between eigenvectors belonging to the same 
eigenspace. In this way the Parseval equality can be written down as
\begin{equation}
\int_{\delta }{dr{|f(r)|}^2}=\sum^{\infty }_{n=1}{{\left|\int_{\delta}
{drf(r){\overline{h}}^{(\delta )}_n}\right|}^2}, 
\label{(2.33)} 
\end{equation}
while the Hilbert product in ${\cal{L}}^2(0,\infty )$ 
between $f_1$ and $f_2$ is given by
\begin{equation}
\int_{\delta }{drf_1(r){\overline{f}}_2(r)}=\sum^{\infty }_{n=1}
{\int_{\delta }{drf_1(r){\overline{h}}^{(\delta )}_n}\overline
{\int_{\delta }{drf_2(r){\overline{h}}^{(\delta )}_n}}}. 
\label{(2.34)} 
\end{equation}
But ${\varphi }_1$ and ${\varphi }_2$ form a fundamental system, thus
\begin{equation}
h^{\left(\delta \right)}_n\left(r\right)=t^{\left(\delta \right)}_{n,1}
{\varphi }_1\left(r,{\lambda }^{\left(\delta \right)}_n\right)
+t^{\left(\delta \right)}_{n,2}{\varphi }_2
\left(r,{\lambda }^{\left(\delta \right)}_n\right), 
\label{(2.35)} 
\end{equation}
where $t^{(\delta )}_{n,j}$ for every $n=1,2,\dots ,$ and $j=1,2$, are 
complex constants. By inserting Eq. (2.35) into Eq. (2.33) 
we can write this last equation as
$$
\int_{\delta }{dr{|f(r)|}^2}=\sum^{\infty }_{n=1}{\sum^2_{j,k=1}
{\int_{\delta }{drf(r){\overline{t}}^{\left(\delta \right)}_{n,j}
{\varphi }_j\left(r,{\lambda }^{\left(\delta \right)}_n\right)}
\overline{\int_{\delta }{drf(r){\overline{t}}^{\left(\delta 
\right)}_{n,k}{\varphi }_k\left(r,{\lambda }^{\left(\delta \right)}_n\right)}}}},
$$ 
(where we have used the fact that ${\varphi}_j$ are real functions) then we can set
\begin{equation}
g^{\left(\delta \right)}_j\left(\lambda \right)=\int_{\delta }
{drf\left(r\right)}{\varphi }_j\left(r,\lambda \right), 
\label{(2.36)} 
\end{equation} 
\begin{equation}
{\rho }^{(\delta )}_{jk}\left(\lambda \right)= \left \{ \begin{array}{rr} 
\ \ \ 0\; \; \ \ \ \ \ \ \ \ \ \ \ 
{\rm for}\ \lambda =0 \\
& \sum_m{{\overline{t}}^{\left(\delta \right)}_{n_m,j}
t^{\left(\delta \right)}_{n_m,k}}+{\rho }^{(\delta )}_{jk}
\left({\lambda }^{(\delta )}_{n-1}\right)\;  \rm{for} 
\lambda \in [{\lambda }^{\left(\delta \right)}_n,
{\lambda }^{\left(\delta \right)}_{n+1}[
\end{array} \right. 
\label{(2.37)} 
\end{equation}
where the summation over $m$ in Eq. (2.37) stands for a summation over 
all indices such that ${\lambda }^{(\delta )}_n={\lambda }^{(\delta )}_{n_m}$ 
when some degeneracy may occur. In terms of Eqs. (2.36) and (2.37), 
Eq. (2.33) can be written as
\begin{equation}
\int_{\delta }{dr{|f(r)|}^2}=\int^{\infty }_{-\infty }{d{\rho }^{\left
(\delta \right)}_{jk}(\lambda )\sum^2_{j,k=1}{{\overline{g}}^{\left
(\delta \right)}_j\left(\lambda \right)}g^{\left(\delta 
\right)}_k(\lambda )}. 
\label{(2.38)} 
\end{equation}
The matrix ${\rho }^{(\delta )}_{jk}\left(\lambda \right)$ is called 
{\it spectral matrix} associated to the self-adjoint problem (2.32) 
and it satisfies the following three requirements:
\vskip 0.3cm
\noindent 
(i) It is Hermitian, i.e. ${\rho }^{(\delta )}_{jk}\left(\lambda 
\right)={\overline{\rho }}^{(\delta )}_{kj}\left(\lambda \right)$.
\vskip 0.3cm
\noindent 
(ii) ${\rho }^{(\delta )}\left(\Delta \right)={\rho }^{(\delta )}
\left(\lambda \right)-{\rho }^{(\delta )}\left(\mu \right)$
is positive semidefinite if $\lambda >\mu$, where $\Delta =]\mu ,\lambda ]$.
\vskip 0.3cm
\noindent 
(iii) The total variation of ${\rho }^{(\delta )}_{jk}\left
(\lambda \right)$ is finite on every finite $\lambda$ interval.
\vskip 0.3cm
\noindent 
Any matrix satisfying (ii) is said to be {\it nondecreasing}.

By applying the Parseval equality (2.38) to any continuous 
function on ${\Bbb{R}}^+$ which vanishes outside some interval 
${\delta }_1\subset \delta $, one obtains the same Eq. (2.38) but instead 
of the function $g^{\left(\delta \right)}_j$ defined in Eq. (2.36) 
it is more convenient to use
\begin{equation}
g^{\left(\delta \right)}_j=\int^{\infty }_{-\infty }{dr \; f(r)}
{\varphi }_j\left(r,\lambda \right). 
\label{(2.39)} 
\end{equation}

We can now show that if Eq. (2.7) is in the limit-point case both at 
$r=0$ and infinity, there exists an unique matrix $\rho$ satisfying 
the properties (i), (ii) and (iii) such that ${\rho}^{(\delta)}\to \rho$ 
when the limit $\delta \to (0,\infty )$ is taken. Then, for every 
$f\in {\cal{L}}^2(0,\infty )$, Eq. (2.38) holds with $\rho$ in place 
of ${\rho}^{(\delta)}$ and Eq. (2.39) in place of Eq. (2.36). If 
one of the ends is in the limit-circle case, then the limiting spectral 
matrix still exists but the uniqueness is not guaranteed. 
\vskip 0.3cm
\centerline {\it The existence of a limiting spectral matrix.}
\vskip 100cm
The key for proving the existence of this limiting spectral matrix, resides 
in the possibility of showing that the integral 
$$
\int^{\infty}_{-\infty}{{{d{\rho }^{\left(\delta \right)}_{jk}
\left(\lambda \right)}\over {{|\lambda -l|}^2}}},
$$ 
is uniformly convergent when one takes the limit $\delta \to {\Bbb{R}}^{+}$. 
This is sufficient for proving the existence. In doing this, we must construct 
this particular type of integral and this will be our effort for the next few pages. 

Let ${\chi }_a={\varphi }_1+{m'}_a{\varphi }_2$ be a solution of 
the equation $Lx=lx$ which satisfies
$$
\cos \beta' \; x(a)+ \sin  \beta' \; p(a) x'(a)=0, 
$$ 
and ${\chi}_b={\varphi}_1+m_b{\varphi}_2$ another solution which satisfies
$$
\cos \beta \; x(b) + \sin \beta \; p(b)x'(b)=0,
$$ 
then $m_b$ and ${m'}_a$ lie on the circles $C_b$ and $C'_a$, respectively, of equations
$$
\left[{\chi }_b{\chi }_b\right]\left(b\right)=0,\; \;  
\left[{\chi }_a{\chi }_a\right]\left(a\right)=0.
$$ 
The Green's function for the problem (2.32) exists and it can be easily 
calculated, provided ${\frak{I}}\left(l\right)\ne 0$, and is given by
$$
G^{\left(\delta \right)}\left(r,\varrho ,l\right)= \left \{ \begin{array}{rr}
{{{\chi }_a(r,l){\chi }_b(\varrho ,l)}
\over {{m'}_a\left(l\right)-m_b(l)}} \; 0<r\le \varrho \\ 
& {{{\chi }_a(\varrho ,l){\chi }_b(r,l)}\over {{m'}_a\left
(l\right)-m_b(l)}} \; r>\varrho  
\end{array} \right .
$$ 
We now want to apply the completeness relation (2.34) to the following functions:
\begin{equation}
f_1\left(r\right)={{{\partial }^sG^{\left(\delta \right)}}
\over {\partial {\varrho }^s}}\left(r,c,l\right),\;  
f_2\left(r\right)={{{\partial }^pG^{\left(\delta \right)}}\over 
{\partial {\varrho }^p}}\left(r,c,l\right),\; \left(s,p=0,1\right). 
\label{(2.40)} 
\end{equation}
Take the above definition of the Green's function and calculate it at $\varrho =c$. We have
$$
G^{\left(\delta \right)}\left(r,c,l\right)= \left \{ \begin{array}{rr} 
{{{\chi }_a(r,l)}\over {{m'}_a\left(l\right)-m_b(l)}}\ \ \ \ \ 0<r\le c \\ 
& {{{\chi }_b(r,l)}\over {{m'}_a (l) -m_b(l)}}\ \ \ \ \ \ r>c
\end{array} \right .
$$
and one can also easily compute the first derivative of the Green function 
with respect to the $\varrho $ variable and calculate it for $\varrho =c$
$$
{{\partial G^{\left(\delta \right)}}\over {\partial \varrho }}
\left(r,c,l\right)= \left \{ \begin{array}{rr}  
{{m_b(l){\chi }_a(r,l)}\over {p(c)({m'}_a\left(l\right)-m_b(l))}}
\ \ \ \ \ 0<r\le c \\
& {{m_a(l){\chi }_b(r,l)}\over {p(c)({m'}_a\left(l\right)-m_b(l))}}
\ \ \ \ \ \ r>c
\end{array} \right .
$$
In order to derive the above functions, the previously stated conditions 
${\varphi }_1\left(c,l\right)=1$, $p\left(c\right){\varphi '}_1\left(c,l\right)=0$, 
${\varphi }_2\left(c,l\right)=0$ and $p\left(c\right){\varphi '}_2
\left(c,l\right)=1\ $have been implicitly used.

For example, we can evaluate one of the required integrals 
underlying the completeness relations we are looking for. Take 
$f_1=f_2=G^{\left(\delta \right)}\left(r,c,l\right)$ in Eq. (2.40) 
and apply to them the completeness relation (2.34)
$$
2i{\mathfrak{I}}\left(l\right)\int_{\delta }{dr{\left|G^{\left(\delta \right)}
\left(r,c,l\right)\right|}^2=2i{\mathfrak{I}}\left(l\right){\left|{m'}_a
\left(l\right)-m_b\left(l\right)\right|}^{-2}\left\{\int^c_a{dr{\left|
{\chi }_a\left(r,l\right)\right|}^2}\right.}
$$ 
$$
+ \left.\int^b_c{dr{\left|{\chi }_b
\left(r,l\right)\right|}^2}\right\}
$$ 
$$
={\left|{m'}_a\left(l\right)-m_b\left(l\right)\right|}^{-2}\left\{
\left[{\chi }_a{\chi }_a\right]\left(c\right)-[{\chi }_b{\chi }_b](c)\right\}
$$ 
$$
=2i{\frak{I}}\left({m'}_a\left(l\right)-m_b\left(l\right)\right){\left|{m'}_a
\left(l\right)-m_b\left(l\right)\right|}^{-2},
$$ 
where we have used the Green's formula jointly with the equations 
$\left[{\chi }_a{\chi }_a\right]$$\left(a\right)=0$ and $\left[{\chi }_b{\chi }_b\right]
\left(b\right)=0$ for the circles ${C'}_a$ and $C_b$, respectively. Thus
\begin{equation}
\int_{\delta }{dr{|G^{\left(\delta \right)}\left(r,c,l\right)|}^2}
={{\displaystyle{\frak{I}}\left({{{m'}_a\left(l\right)-m_b\left(l\right)}}
\right)^{-1}}\over {{\frak{I}}\left(l\right)}}. 
\label{(2.41)} 
\end{equation}
Similarly 
$$
\left({m'}_a\left(l\right)-m_b\left(l\right)\right)\left(l-{\lambda }^{\left(
\delta \right)}_n\right)\int_{\delta }{drG^{\left(\delta \right)}
\left(r,c,l\right)}{\overline{h}}^{\left(\delta \right)}_n\left(r\right)
$$ 
$$
=\left[{\chi }_bh^{\left(\delta \right)}_n\right]\left(b\right)
-\left[{\chi }_bh^{\left(\delta \right)}_n\right]\left(c\right)
+\left[{\chi }_ah^{\left(\delta \right)}_n\right]\left(c\right)
-\left[{\chi }_ah^{\left(\delta \right)}_n\right](a)
$$ 
$$
=\left({m'}_a\left(l\right)-m_b\left(l\right)\right)\left
[{\varphi }_2h^{\left(\delta \right)}_n\right]\left(c\right)
$$ 
$$
=\left({m'}_a\left(l\right)-m_b\left(l\right)\right){\overline{t}}^{(\delta )}_{n,1},
$$ 
where the Green's formula has been used for the passage from the first 
to the second line, while the relation $\left[{\chi }_bh^{\left(\delta 
\right)}_n\right]\left(b\right)=\left[{\chi }_ah^{\left(\delta \right)}_n
\right]\left(a\right)=0$ has been used in the passage from the second to 
the third line which follows from the fact that ${\chi}_b$ and 
$h^{\left(\delta \right)}_n$ satisfy the same boundary condition at $b$ and 
the same holds for ${\chi}_a$ and $h^{\left(\delta \right)}_n$ in $a$. 
The passage from the third to the fourth line follows from Eq. (2.35). 

One thus obtains
\begin{equation}
\int_{\delta }{drG^{\left(\delta \right)}\left(r,c,l\right)}{\overline{h}}^{\left
(\delta \right)}_n\left(r\right)={{{\overline{t}}^{(\delta )}_{n,1}}
\over {l-{\lambda }^{\left(\delta \right)}_n}}. 
\label{(2.42)} 
\end{equation}
From Eq. (2.37) one can show by using the standard theory of generalized functions, that
\begin{equation}
d{\rho }^{\left(\delta \right)}_{jk}\left(\lambda \right)
=\sum^{\infty }_{n=1}{d\lambda \ \delta \left(\lambda -{\lambda }^{(\delta )}_n
\right)\sum_m{{\overline{t}}^{\left(\delta \right)}_{n_m,j}t^{\left
(\delta \right)}_{n_m,k}}}, 
\label{(2.43)} 
\end{equation}
(where the sum over the $m$ index takes into account the degeneracy of 
the eigenvalue) hence, if we divide Eq. (2.43) by 
${|\lambda -l|}^2$ and integrate over the $\lambda $ variable in the 
range $\left(-\infty ,\infty \right)$ we get
$$
\int^{\infty }_{-\infty }{{{d{\rho}^{\left(\delta \right)}_{jk}
\left(\lambda \right)}\over {{|\lambda -l|}^2}}}=\sum^{\infty }_{n=1}
{\sum_m{{{{\overline{t}}^{\left(\delta \right)}_{n_m,j}t^{\left(
\delta \right)}_{n_m,k}}\over {{|{\lambda }^{\left(\delta \right)}_n-l|}^2}}}}.
$$ 
Now, recalling that
$$
\int_{\delta }{dr{|G^{\left(\delta \right)}\left(r,c,l\right)|}^2}
=\sum^{\infty }_{n=1}{\sum_m{\int_{\delta }{drG^{\left(\delta \right)}
\left(r,c,l\right)}{\overline{h}}^{\left(\delta \right)}_{n_m}
\left(r\right)\overline{\int_{\delta }{drG^{\left(\delta \right)}
\left(r,c,l\right)}{\overline{h}}^{\left(\delta \right)}_{n_m}\left(r\right)}}},
$$ 
and by using Eqs. (2.41) and (2.42) one can deduce that
$$
\int^{\infty }_{-\infty }{{{d{\rho }^{\left(\delta \right)}_{11}
\left(\lambda \right)}\over {{|\lambda -l|}^2}}}={{{\frak{I}}
\left(M^{(\delta )}_{11}\right)}\over {{\frak{I}}\left(l\right)}},
$$ 
with $M^{(\delta )}_{11}={\left({m'}_a\left(l\right)-m_b\left(l\right)
\right)}^{-1}$. At this stage, if one carries all calculations for 
completeness by using Eq. (2.34) between the other functions in Eqs. 
(2.40) by setting $s\neq p$, one finds that  
\begin{equation}
\int^{\infty }_{-\infty }{{{d{\rho }^{\left(\delta \right)}_{jk}
\left(\lambda \right)}\over {{|\lambda -l|}^2}}}={{{\frak{I}}
\left(M^{(\delta )}_{jk}\right)}\over {{\frak{I}}\left(l\right)}}, 
\label{(2.44)} 
\end{equation}
where 
$$
M^{(\delta )}_{11}={{1}\over {{m'}_a\left(l\right)-m_b\left(l\right)}},
\; M^{(\delta )}_{12}=M^{(\delta )}_{21}={{1}\over {2}}
{{{m'}_a\left(l\right)+m_b\left(l\right)}\over {{m'}_a\left(l\right)
-m_b\left(l\right)}},
$$ 
\begin{equation}
M^{(\delta )}_{22}={{{m'}_a(l)m_b(l)}\over {{m'}_a\left(l\right)
-m_b\left(l\right)}}, 
\label{(2.45)} 
\end{equation}

Now we have to show the existence of the limiting spectral matrix by applying 
some fundamental theorems such as the Helly selection theorem and one 
integration theorem, to Eq. (2.44). But first, let us recall that 
$m_b\left(l\right)$ and ${m'}_a\left(l\right)$ must lie on opposite half-planes 
from Eqs. (2.21) and (2.30). Suppose $l=i$. Then, the points ${m'}_a\left(i\right)$ 
must lie in ${C'}_1$ for $a<1$, whereas the points $m_b\left(i\right)$ must lie 
on $C_b$ which is in $C_1$ for $b>1$. Thus, there exists a positive constant 
$u$ such that $\left|{m'}_a\left(i\right)-m_b\left(i\right)\right|>u$ for 
$a<1$ and $b>1$. But since $m_b\left(l\right)$ and ${m'}_a\left(l\right)$ 
are uniformly bounded for $a<1$ and $b>1$, it follows from Eq. (2.44) that
$$
\int^{\infty}_{-\infty }{{{d{\rho}^{\left(\delta \right)}_{jj}
\left(\lambda \right)}\over {1+{\lambda }^2}}}={\frak{I}}\left(
M^{(\delta )}_{jj}\right)<K,\; j=1,2,
$$ 
for some constant $K$. By looking at the differential (2.43), one can 
easily see that the products of the type ${\overline{t}}^{\left(\delta 
\right)}_{n_m,j}t^{\left(\delta \right)}_{n_m,k}$ must be absolutely 
bounded from the law of cosines which yields 
$$
2\left|{\overline{t}}^{\left(\delta \right)}_{n_m,j}t^{\left(\delta 
\right)}_{n_m,k}\right|\le {\left|t^{\left(\delta \right)}_{n_m,j}
\right|}^2+{\left|t^{\left(\delta \right)}_{n_m,k}\right|}^2,
$$ 
and from which it follows:
\begin{equation}
\int^{\infty }_{-\infty }{{{\left|d{\rho }^{\left(\delta \right)}_{jk}
\left(\lambda \right)\right|}\over {{\lambda }^2+1}}}<K, 
\label{(2.46)} 
\end{equation}
that holds also for $j\ne k$. Now we introduce the following theorems 
which are the Helly selection theorem and a particular integration theorem, 
respectively \cite{CoddingtonLevinson1955}:
\begin{thm}
Let $\left\{h_n\right\}$, $n=1,2,\dots $, be 
a sequence of real nondecreasing functions on $\lambda \in \Bbb{R}$, and 
let $H$ be a continuous nonnegative function on the same interval. If
$$
\left|h_n(\lambda )\right|\le H\left(\lambda \right),\;  
n=1,2,\dots ,\; \lambda \in \Bbb{R},
$$ 
then there exists a subsequence $\left\{h_{n_k}\right\}$ and 
a nondecreasing function $h$ such that
$$
\left|h(\lambda )\right|\le H\left(\lambda \right),\; \lambda \in \Bbb{R},
$$ 
and
$$
\lim_{k\to \infty} h_{n_k}\left(\lambda \right)=h(\lambda ).
$$ 
\end{thm}
\begin{thm}
Suppose $\left\{h_n\right\}$ is a real, uniformly bounded, 
sequence of nondecreasing functions on a finite interval $\lambda \in [a,c]$, 
and assume
$$
\lim_{n \to \infty} h_n\left(\lambda \right)
=h\left(\lambda \right),\;  \lambda \in [a,c].
$$ 
If $f$ is any continuous function on $\lambda \in [a,c]$, then 
$$
\lim_{n \to \infty} \int^c_a{dh_n(\lambda )f(\lambda )}
=\int^c_a{dh(\lambda )f(\lambda )}.
$$ 
\end{thm}
\vskip 0.3cm
Consider now Eq. (2.46). Therefore, if we take any $\nu >0$ there must be
$$
\int^{\nu }_{-\nu }{\left|d{\rho }^{\left(\delta \right)}_{jk}
\left(\lambda \right)\right|}<K\left(1+{\nu }^2\right),
$$ 
and this, together with the condition ${\rho }^{\left(\delta \right)}_{jk}
\left(0\right)=0$ in Eq. (2.37) gives
$$
\left|{\rho }^{\left(\delta \right)}_{jk}\left(\lambda \right)\right|
\le K\left(1+{\lambda }^2\right).
$$ 
Now, if we apply Theorem 2.5 by choosing a sequence of intervals 
${\delta }_n=\left[a_n,b_n\right]$ such that ${\delta }_n\to {\Bbb{R}}^+$, 
there remains defined a sequence of real nondecreasing functions 
${\rho }^{\left({\delta }_n\right)}_{jk}\left(\lambda \right)$ 
(the reality and nondecreasing behaviour follows from property (ii) for 
spectral matrices and Eq. (2.37)) for which there exists a subsequence 
converging to a limit function ${\rho }_{jk}(\lambda)$ which is monotone, 
nondecreasing and satisfies 
$$
\left|{\rho }_{jk}(\lambda)\right|\le K\left(1+{\lambda}^2\right),
$$ 
that is a spectral matrix for which properties (i), (ii) and (iii) hold. 
It is also possible to show that the Parseval equality (2.38) still holds 
with ${\rho }_{jk}(\lambda)$ in place of ${\rho}^{\left({\delta}_n\right)}_{jk}
\left(\lambda \right)$ for every $f\in {\cal{L}}^2\left(0,\infty \right)$ 
by an application of the integration theorem 2.6. 

When the existence of the limiting spectral matrix is proved, then the 
following theorem also holds \cite{CoddingtonLevinson1955}:
\begin{thm}
Let $\rho$ be any limiting matrix of the set
$\left\{{\rho }^{(\delta )}\right\}$. If $f\in {\cal{L}}^2
\left(0,\infty \right)$, the vector $g=\left(g_1,g_2\right)$, where
$$
g_j\left(\lambda \right)=\int^{\infty }_{-\infty }{drf(r){\varphi }_j(r,\lambda )},
$$ 
and ${\varphi }_j$ with $j=1,2$ form a fundamental 
system of solutions for the equation $Lx=lx$ satisfying the conditions
$$
{\varphi }_1\left(c,l\right)=1,\; {\varphi }_2\left(c,l\right)=0,
$$ 
$$
p\left(c\right){\varphi '}_1\left(c,l\right)=0,\; p
\left(c\right){\varphi '}_2\left(c,l\right),
$$ 
for some $c>0$, then $g_j\left(\lambda \right)$
converges in ${\cal{L}}^2\left(\rho \right)$, i.e. in 
the Hilbert space of all square summable functions on the measure space whose 
measure is given by $\rho$, that is, there exists a $g\in 
{\cal{L}}^2\left(\rho \right)$ such that
$$
\left\|g-g_{cd}\right\|\longrightarrow 0\; {\rm for} \; c \; 
\longrightarrow 0\; {\rm and} \; d\longrightarrow \infty ,
$$ 
where
$$
g_{cd,j}\left(\lambda \right)=\int^d_c{drf\left(r\right){\varphi }_j
\left(r,\lambda \right)},\; c<d.
$$ 
In terms of this $g$, the Parseval equality
$$
\int^{\infty }_{-\infty }{dr{|f(r)|}^2}=\int^{\infty }_{-\infty }
{d{\rho }_{jk}(\lambda )\sum^2_{j,k=1}{{\overline{g}}_j
\left(\lambda \right)}g_k(\lambda )},
$$ 
and the expansion
$$
f\left(r\right)=\int^{\infty }_{-\infty }{d{\rho }_{jk}(\lambda )
\sum^2_{j,k=1}{{\varphi }_j\left(r,\lambda \right)g_k(\lambda )}},
$$ 
are valid, the latter integral converges in the
$\ {\cal{L}}^2\left(0,\infty \right)$ norm.
\end{thm}
\vskip 0.5cm
\centerline {\it Uniqueness of the limiting spectral matrix}
\vskip 0.3cm
The uniqueness for the spectral matrix relies on the existence of the 
following limit for every pair of continuity points $\lambda ,\mu 
\in {\Bbb{R}}^+$ for ${\rho }_{jk}$:
$$
{\rho}_{jk}\left(\lambda \right)-{\rho}_{jk}\left(\mu \right)
=\lim_{{\delta }_n\to {\Bbb{R}}^+} 
\left({\rho }^{\left({\delta }_n\right)}_{jk}
\left(\lambda \right)-{\rho }^{\left({\delta}_n\right)}_{jk}
\left(\mu \right)\right),\;  j,k=1,2,
$$ 
Now we will show that this limit exists if Eq. (2.7) is in limit-point 
case at both ends of the real positive line. This is possible if we show that 
\begin{equation}
{\rho }_{jk}\left(\lambda \right)-{\rho }_{jk}\left(\mu \right)
=\lim_{\epsilon \to 0^+} {{1}\over {\pi }}
\int^{\lambda }_{\mu }{d\nu \ {\frak{I}}\left(M_{jk}\left(\nu 
+i\epsilon \right)\right)}, 
\label{(2.47)} 
\end{equation}
because the limit-point case at both ends guarantees the uniqueness of the limits
$$
M^{(\delta )}_{jk}\longrightarrow M_{jk},\; {\rm {for}}
\ \ \delta \longrightarrow {\Bbb{R}}^+,\;  j,k=1,2,
$$ 
with $M^{(\delta )}_{jk}$ given in Eqs. (2.45), and thus the existence of the limit (2.47). 

In proving Eq. (2.47), let us consider Eq. (2.44). For any fixed 
$l$ with ${\frak{I}}(l)\ne 0$, there exists a constant $K$ such that
$$
\int^{\mu }_{-\mu }{{{d{\rho }^{\left(\delta \right)}_{jk}\left(
\lambda \right)}\over {{|\lambda -l|}^2}}}\le K,
\; j,k=1,2,
$$ 
for $b>1$ and $a<1$. Upon choosing a sequence ${\delta }_n\to {\Bbb{R}}^+$ 
as done before, it follows that the above equation is also true with 
${\rho }_{jk}$ in place of ${\rho }^{\left(\delta \right)}_{jk}$. Since 
the above equation holds for every $\mu >0$, then 
$$
\int^{\infty }_{-\infty }{{{d{\rho }_{jk}\left(\lambda \right)}\over 
{{|\lambda -l|}^2}}}<\infty ,\; j,k=1,2.
$$ 
From Eq. (2.46) there exists a constant $\tilde{K}$ such that, 
for $b>1$ and $a<1$, one has
$$
\int^{\infty }_{\mu }{{{d{\rho }^{\left(\delta \right)}_{jk}(\lambda )}
\over {{\lambda }^3}}<{{\tilde{K}}\over {\mu }}},\; j,k=1,2.
$$ 
This relation similarly holds if the integration is taken over $]-\infty ,-\mu [$. 

If ${\frak{I}}(l)\ne 0$ and ${\frak{I}}(l_0)\ne 0$ and the equation
\begin{equation}
\int^{\infty }_{-\infty }{d{\rho }^{\left(\delta \right)}_{jk}
\left(\lambda \right)\biggl({{1}\over {{\left|\lambda -l\right|}^2}}
-{{1}\over {{\left|\lambda -l_0\right|}^2}}\biggr)},\;  
j,k=1,2, 
\label{(2.48)} 
\end{equation}
is considered over the intervals $]-\infty ,-\mu [$, $]-\mu ,\mu [$ 
and $]\mu ,\infty [$, it follows that, if $\delta \to \infty $ through a 
chosen subsequence and if then $\mu \to \infty $, the integration Theorem
2.6 guarantees that Eq. (2.48) tends to
$$
\int^{\infty }_{-\infty }{d{\rho }_{jk}\left(\lambda \right)
\biggl({{1}\over {{\left|\lambda -l\right|}^2}}-{{1}\over {{\left|
\lambda -l_0\right|}^2}}\biggr)},\;  j,k=1,2.
$$ 
But if we make use of Eq. (2.44), we can write Eq. (2.48) as
$$
{{{\mathfrak{I}}\left(M^{\left(\delta \right)}_{jk}(l)\right)}\over 
{{\mathfrak{I}}\left(l\right)}}-{{{\frak{I}}\left(M^{\left(\delta 
\right)}_{jk}\left(l_0\right)\right)}\over {{\frak{I}}\left(l_0\right)}},
\; j,k=1,2,
$$ 
which tends to 
$$
{{{\mathfrak{I}}\left(M_{jk}(l)\right)}\over {{\frak{I}}\left(l\right)}}
-{{{\mathfrak{I}}\left(M_{jk}\left(l_0\right)\right)}\over 
{{\mathfrak{I}}\left(l_0\right)}},\;  j,k=1,2,
$$ 
where 
$$
M_{11}(l)={{1}\over {{m'}_{\infty }\left(l\right)-m_{\infty }
\left(l\right)}},\; M_{12}(l)=M_{21}(l)={{1}\over 
{2}}{{{m'}_{\infty }\left(l\right)+m_{\infty }\left(l\right)}\over 
{{m'}_{\infty }\left(l\right)-m_{\infty }\left(l\right)}},
$$ 
$$
M_{22}\left(l\right)={{{m'}_{\infty }\left(l\right)m_{\infty }
\left(l\right)}\over {{m'}_{\infty }\left(l\right)-m_{\infty }\left(l\right)}},
$$ 
and ${m'}_{\infty }\left(l\right)$, $m_{\infty }\left(l\right)$ are 
the limit points at the origin and at infinity, respectively. Therefore
\begin{equation}
{{{\mathfrak{I}}\left(M_{jk}(l)\right)}\over {{\mathfrak{I}}\left(l\right)}}
=\int^{\infty }_{-\infty }{{{d{\rho }_{jk}\left(\lambda \right)}\over 
{{\left|\lambda -l\right|}^2}}}+c_{jk},\; j,k=1,2, 
\label{(2.49)} 
\end{equation}
where $c_{jk}$ are four constants independent of $l$, provided 
${\mathfrak{I}}(l)\ne 0$. Now, letting ${\frak{R}}\left(l\right)=0$ and 
${\mathfrak{I}}(l)\to \infty $, it readily follows that $c_{jk}=0$. Now, let 
$\lambda ,\mu$ be points of continuity for ${\rho }_{jk}$. Then, from Eq. 
(2.49) it follows
$$
\lim_{\epsilon \to 0^+} \int^{\lambda}_{\mu} d\nu \; 
{\mathfrak{I}}\left(M_{jk}\left(\nu +i\epsilon \right)\right)
=\lim_{\epsilon \to 0^+} \int^{\lambda}_{\mu}
\int^{\infty}_{-\infty} d\nu
{\epsilon d\rho_{jk}(\sigma) \over 
(\sigma-\nu)^{2}+\epsilon^{2}} 
$$
$$
=\lim_{\epsilon \to 0^+} \int^{\infty}_{-\infty}
d{\rho }_{jk}(\sigma )\left[{{\rm{tan}}^{-1} \left({{\lambda -\sigma }
\over {\epsilon }}\right)\ }-{{\rm{tan}}^{-1} \left({{\mu -\sigma }
\over {\epsilon }}\right)\ }\right]
$$ 
$$=\pi \left({\rho }_{jk}\left(\lambda \right)-{\rho }_{jk}(\mu )
\right),\; j,k=1,2.
$$ 

This proves Eq. (2.47) and hence the uniqueness of the limiting spectral 
matrix in the limit-point case at both ends of the real positive line. 
Thus, we have proved the following fundamental theorem:
\begin{thm}
Let $L$ be in the limit point case 
at $r=0$ and $r=\infty$. There exists a nondecreasing Hermitian 
matrix $\rho =\left({\rho }_{jk}\right)$ whose elements are of 
bounded variation on every finite $\lambda$ interval, and which 
is essentially unique in the sense that 
$$
{\rho }_{jk}\left(\lambda \right)-{\rho }_{jk}\left(\mu \right)
={\mathop{\rm{lim}}_{{\delta }_n\to {\Bbb{R}}^+} \left({\rho }^{\left(
{\delta }_n\right)}_{jk}\left(\lambda \right)-{\rho }^{\left(
{\delta }_n\right)}_{jk}\left(\mu \right)\right)\ },\; j,k=1,2,
$$ 
at points of continuity $\lambda ,\mu$ of ${\rho }_{jk}$. Furthermore,
\begin{equation}
{\rho }_{jk}\left(\lambda \right)-{\rho }_{jk}\left(\mu \right)
={{1}\over {\pi }}\int^{\lambda }_{\mu }{d\nu \ {\frak{I}}\left(M_{jk}
\left(\nu +i\epsilon \right)\right)},\; j,k=1,2, 
\label{(2.50)} 
\end{equation}
where 
$$
M_{11}\left(l\right)={{1}\over {{m'}_{\infty }\left(l\right)
-m_{\infty }\left(l\right)}},
$$ 
$$
M_{12}\left(l\right)=M_{21}\left(l\right)={{1}\over {2}}{{{m'}_{\infty }
\left(l\right)+m_{\infty }\left(l\right)}\over {{m'}_{\infty }
\left(l\right)-m_{\infty }\left(l\right)}},
$$ 
\begin{equation}
M_{22}\left(l\right)={{{m'}_{\infty}\left(l\right)m_{\infty}
\left(l\right)}\over {{m'}_{\infty}\left(l\right)-m_{\infty}
\left(l\right)}}. 
\label{(2.51)} 
\end{equation}
\end{thm}
\vskip 0.3cm
The spectrum associated with a problem for which $\rho$ is uniquely determined, 
is the set of nonconstancy points of $\rho$, that is, the set of all nonconstancy 
points of all elements ${\rho }_{jk}$. Since $\rho$ is Hermitian and 
nondecreasing, it follows that
$$
{\left|{\rho}_{jk}(\Delta )\right|}^2\le {\rho }_{jj}\left(\Delta 
\right){\rho}_{kk}\left(\Delta \right),
$$ 
where 
$$
{\rho}_{jk}\left(\Delta \right)={\rho }_{jk}\left(\lambda \right)
-{\rho}_{jk}\left(\mu \right),\; \Delta =]\mu ,\lambda ].
$$ 
Hence the set of nonconstancy points of all elements of the limiting 
spectral matrix is the same of all nonconstancy points for its diagonal 
elements. Clearly the spectrum is a closed set. The {\it point spectrum} 
is the set of all discontinuity points of $\rho$, and the {\it continuous spectrum} 
is the set of continuity points of $\rho$. Points in the spectrum are 
called eigenvalues and the solutions to the eigenvalue problem 
for such points are called eigenfuntions.

It is essential to remark that each physical problem whose Hamiltonian is 
given by Eq. (2.6) and for which $r=0$ and $r=\infty $ are singular 
points (thus we are taking aside the case ${\lambda }_{3,0}={{1}\over {2}}$ 
into Eq. (2.6) with a nonsingular potential $\cal V$) must possess one and 
only one limiting spectral matrix in such a way that every experiment about 
spherical symmetric quantum particles, would be always reproducible. This is 
the core of the predictability of quantum mechanics: the energy levels are always 
theoretically established within a margin of error that can be estimated only 
through experimental data.  

Taking aside the case ${\lambda}_{3,0}={{1}\over {2}}$, quantum 
mechanical Hamiltonians of type (2.6) must be in limit-point case at both 
$r=0$ and $r=\infty$ and this leads us to the possibility to give a 
geometrical interpretation for such problems. We recall that this kind of 
uniqueness for self-adjoint problems we are treating, is often called in the 
literature {\it essential self-adjointness} \cite{ReedSimon1975,Simon2015},
i.e., the closure of the operator is self-adjoint, which implies in turn
that the self-adjoint extension exists and is unique. From the fact that 
the establishment of an essentially self-adjoint problem, within the theory 
we have developed, is carried out by a limiting procedure of self-adjoint problems
$$
\delta =\left[a,b\right]\longrightarrow {\Bbb{R}}^+,\;  a
\longrightarrow 0,\;  b\longrightarrow \infty ,
$$ 
we can always choose a sequence of intervals $\left\{{\delta }_n=\left[
a_n,b_n\right]\right\}$ which converges to the real positive line in the 
limit $n\to \infty$ and for which there remain defined two families of 
circles $\left\{{C'}_{a_n}\right\}$ and $\left\{C_{b_n}\right\}$, where 
${C'}_{a_n}$ lies on the ${m'}_{a_n}(l)$ plane while $C_{b_n}$ lies on the 
$m_{b_n}(l)$ plane, and which lie on opposite half-planes for each fixed $n$. 
The equations for these circles are $\left[{\chi}_{a_n}{\chi}_{a_n}\right](a_n)=0$ 
for ${C'}_{a_n}$ and $\left[{\chi }_{b_n}{\chi }_{b_n}\right](b_n)=0$ 
for $C_{b_n}$. As we already know, the limit-point cases at both ends 
(case which must occur for spherically symmetric quantum mechanical Hamiltonians) 
are characterized by the existence of limit points 
$$
{C'}_{a_n}\longrightarrow {m'}_{\infty },\; 
C_{b_n}\longrightarrow m_{\infty},
$$ 
thus, for each essentially self-adjoint quantum mechanical problem for 
Hamiltonians (2.6) with singular behavior at both ends of the positive 
real line, there remain defined two families of circles lying on opposite 
half-planes and such that each family is formed by circles enclosed one into the 
other whose radii approach zero.
\vskip 0.3cm
\centerline {\it Singular behavior at infinity only}
\vskip 0.3cm
As the last argument of the limit-point, limit-circle theory we will give 
some fundamental results for the case in which Eq. (2.7) has singular behavior 
only at infinity. This is the case of three-dimensional $s$-waves, i.e. 
${\lambda }_{3,0}={{1}\over {2}}$ in Eq. (2.6) with nonsingular potential 
$\cal V$. The proof for the existence and uniqueness of spectral functions, 
retraces the method we have established in the last two sections, 
thus we will omit the explicit proof.

Let us consider the following self-adjoint problem:
\begin{equation}
\left \{ \begin{array}{rl} Lx=lx \\ 
& \cos \alpha \; x(c) + \sin \alpha \; p(c) x'(c)=0 \\ 
& \cos \beta \; x(b) + \sin \beta \; p(b)x'(b)=0, 
\end{array} \right. 
\label{(2.52)} 
\end{equation}
with $\alpha,\beta \in [0,\pi [$ and $0<c<b$. The problem here is 
identical to Eq. (2.32) but we will fix throughout the exposition, 
the value of $c<b$. Hence, there exists a countable sequence of eigenvalues 
$\left\{{\lambda }^{(b)}_n\right\}$, $n=1,2,\dots$, and a complete set of 
orthonormal eigenfunctions $\left\{h^{(b)}_n\right\}$ in ${\cal{L}}^2(c,b)$. 
Let $\varphi $ and $\psi $ be two independent solutions to $Lx=lx$ satisfying 
conditions (2.11), thus $\psi$ satisfies the first boundary condition of 
Eq. (2.52) and no solution independent of $\psi$ can satisfy this condition. 
Therefore, each eigenfunction must be of the form
$$
h^{(b)}_n=t^{\left(b\right)}_n\psi \left(r,{\lambda }^{(b)}_n\right),
\; n=1,2,\dots ,
$$ 
with $t^{\left(b\right)}_n$ complex constants independent of $r$. 
If $f$ is any continuous function on $(c,b)$, then the Parseval 
equality is written as
$$
\int^b_c{dr{\left|f(r)\right|}^2}=\sum^{\infty }_{n=1}{{\left|t^{(b)}_n
\right|}^2}{\left|\int^b_c{drf(r)\psi \left(r,{\lambda }^{(b)}_n\right)}\right|}^2.
$$ 
Let 
$$
g\left(\lambda \right)=\int^{\infty }_0{drf(r)\psi (r,\lambda)},
$$ 
and let ${\rho }^{(b)}$ be a monotone nondecreasing step function of 
$\lambda $ having a jump of ${\left|t^{(b)}_n\right|}^2$ at each eigenvalue 
${\lambda }^{(b)}_n$
\begin{equation}
{\rho}^{\left(b\right)}(\lambda )= \left \{ \begin{array}{rr} 0 \; 
{\rm for} \; \lambda =0 \\ 
& {\ \ \left|t^{(b)}_n\right|}^2+{\rho}^{\left(b\right)}\left(
{\lambda}^{(b)}_{n-1}\right) \; {\rm for} \; \lambda \in 
[{\lambda }^{\left(b\right)}_n,{\lambda }^{\left(b\right)}_{n+1}[
\end{array} \right. 
\label{(2.53)} 
\end{equation}
which is called {\it spectral function} for the problem (2.52).
Then the Parseval equality should be written as
\begin{equation}
\int^{\infty}_c{dr{\left|f(r)\right|}^2}=\int^{\infty}_{-\infty}
{{d{\rho}^{\left(b\right)}(\lambda)\left|g(\lambda)\right|}^2}. 
\label{(2.54)} 
\end{equation}
At this stage, the fundamental idea behind the generalization of Eq. 
(2.54) to the case of the entire positive real line, is to show the 
existence of a nondecreasing function $\rho$ which is the limit 
${\rho}^{(b)}\to \rho$ when $b\to \infty $, and such that Eq. (2.54) 
holds when we replace ${\rho }^{(b)}$ with $\rho$ in it. The following 
theorem, of which we omit the proof, holds:
\begin{thm}
Let $L$ be in the limit-point case at 
infinity. Then there exists a monotone nondecreasing function $\rho(\lambda)$ 
on $\Bbb{R}$ such that it is unique in the sense of
\begin{equation}
\rho \left(\lambda \right)-\rho \left(\mu \right)={\mathop{
\rm{lim}}_{b\to \infty } \left({\rho }^{\left(b\right)}\left(
\lambda \right)-{\rho }^{\left(b\right)}\left(\mu \right)\right)\ }, 
\label{(2.55)} 
\end{equation}
at the points of continuity $\lambda ,\mu$ of $\rho$. Furthermore
\begin{equation}
\rho \left(\lambda \right)-\rho \left(\mu \right)={\mathop{\rm{lim}}_{\epsilon 
\to 0^+} {{1}\over {\pi }}\ }\int^{\lambda }_{\mu }{d\nu \ 
{\frak{I}}\left(m_{\infty }\left(\nu +i\epsilon \right)\right)}, 
\label{(2.56)} 
\end{equation}
where $m_{\infty }\left(l\right)$ is the limit point at infinity for $l$ fixed.
\end{thm}

The proof is similar to that given for singularities at both ends of the 
positive real line. The uniqueness of the spectral function is established 
in Eqs. (2.55) and (2.56).

As it happens for singular behaviors at both ends of the positive real line, 
when we apply these arguments to that class of Hamiltonians among the family 
(2.6) which have singular behavior only at infinity, then we expect that 
these should be in limit-point case. The geometrical interpretation is as 
follows: the limiting procedure $b\to \infty$ gives rise to a family of 
circles $\left\{C_b\right\}$ which are contained one into the other and which 
lie in one of the two complex half-planes of opposite imaginary part, 
and their radii approach zero as $b \to \infty$.

\subsection {Connection Between Limit-Point, Limit-Circle Theory and BMS Transformations}

In this section we are characterizing the linear transformations (2.13), 
showing which are the basic requirements to be made for their coefficients in 
order to establish the limit-point cases at infinity. Of course, we must 
require $r_b\to 0$ as $b\to 0$ where 
\begin{equation}
r_b={{1}\over {\left|\overline{C}D-C\overline{D}\right|}}, 
\label{(2.57)} 
\end{equation}
and the coefficients $C,D$ are given in Eqs. (2.15).

We can write Eq. (2.13) in the form
\begin{equation}
m_b={{\alpha z+\beta }\over {\gamma z+\delta }}, \; 
\left \{ \begin{array}{rrrr}  \alpha =-A\tau  \\ 
& \beta =-B\tau  \\ 
& \gamma =\tau C \\ 
& \delta =\tau D 
\end{array} \right.
\; \;  \tau =\pm i, 
\label{(2.58)} 
\end{equation}
and in this way we ensure that $AD-BC=1$, thus we can always deal with 
the fractional linear transformations in terms of $\alpha ,\beta ,\gamma ,  
\delta$. The trace of this transformation is $j=\alpha +\delta$ and from the 
classification given in Ref. \cite{Ford}, we know that if $j$ is real 
and $\left|j\right|<2$ the transformation is elliptic, if $j$ is real and 
$\left|j\right|=2$ then it is parabolic while if $j$ is real and $\left|j\right|>2$, 
it is hyperbolic. In the case $j^2\notin [0,\infty [$, or equivalently $j$ 
is not real, the transformation is loxodromic. From Eq. (2.57), we observe that
\begin{equation}
r^{-1}_b=\left|\overline{C}D-C\overline{D}\right|=\left|\overline{\gamma }
\delta -\gamma \overline{\delta }\right|=2\left|{\frak{I}}\left({\gamma }
\overline\delta \right)\right|, 
\label{(2.59)} 
\end{equation}
and in the limit-point case at infinity, it must be $\left|{\frak{I}}\left({\gamma}
\overline\delta \right)\right|\to \infty$ as $b\to \infty$. 
This implies two possibilities: 
\vskip 0.3cm
\noindent 
(1) The modulus of the $\delta$ variable must tend to infinity;
\vskip 0.3cm
\noindent 
(2) The modulus of the $\gamma$ variable must tend to infinity.
\vskip 0.3cm
In the case (1) we observe that the trace $j=\alpha +\delta $ 
of Eq. (2.58) must diverge, thus, the transformations (2.58) must 
reduce to hyperbolic or loxodromic as $b$ increases. 
For what follows, it is helpful to set up the following nomenclature:
$$
\alpha (b)={\alpha}_1(b)+i{\alpha}_2(b),\; 
\beta (b)={\beta}_1(b)+i{\beta}_2(b),
$$ 
\begin{equation}
\gamma (b)={\gamma}_1(b)+i{\gamma}_2(b),\; \delta (b)
={\delta}_1(b)+i{\delta}_2(b), 
\label{(2.60)} 
\end{equation}
although the $b$ dependence will be explicitly omitted hereafter.
\vskip 0.3cm
\leftline {(1) {\it The $\delta$ variable approaching infinity}}

We can make use of the equation $\alpha \delta -\beta \gamma =1$ in Eq. (2.58), 
and solve it in terms of the $\delta $ variable
\begin{equation}
\delta ={{1+\beta \gamma}\over {\alpha}}, 
\label{(2.61)} 
\end{equation}
from which 
$$
{\delta}_1+i{\delta}_2={{[1+\left({\beta}_1+i{\beta}_2\right)\left(
{\gamma}_1+i{\gamma}_2\right)]}\over {({\alpha}_1+i{\alpha}_2)}}
={{[1+({\beta}_1+i{\beta}_2)({\gamma}_1+i{\gamma}_2)](
{\alpha}_1-i{\alpha}_2)}\over {{|\alpha |}^2}}
$$ 
$$
={{{\alpha}_1-i{\alpha}_2+[{\beta}_1{\gamma}_1+i{\beta}_1{\gamma}_2
+i{\beta}_2{\gamma}_1-{\beta}_2{\gamma}_2]
({\alpha}_1-i{\alpha}_2)}\over {{|\alpha |}^2}}
$$ 
and thus
$$
{\delta}_1={|\alpha |}^{-2}\left[\left(1+{\beta}_1{\gamma}_1
-{\beta}_2{\gamma}_2\right){\alpha}_1+\left({\beta}_1{\gamma}_2
+{\beta}_2{\gamma}_1\right){\alpha}_2\right],
$$ 
\begin{equation}
{\delta}_2={|\alpha |}^{-2}\left[\left({\beta}_1{\gamma}_2
+{\beta}_2{\gamma}_1\right){\alpha}_1+\left(-1+{\beta}_2{\gamma}_2
-{\beta}_1{\gamma}_1\right){\alpha}_2\right], 
\label{(2.62)} 
\end{equation}
which can be put in the following matrix form:
\begin{equation}
\left( 
\begin{matrix} {\delta }_{1} \\ {\delta}_{2} \end{matrix} \right)
={{1}\over {{|\alpha |}^2}} \left( \begin{matrix} 
1+{\beta}_1{\gamma}_1
-{\beta}_2{\gamma}_2 & {\beta}_1{\gamma}_2+{\beta}_2 {\gamma}_1 \\ 
{\beta}_1 {\gamma}_2+{\beta}_2 {\gamma}_1 & -1-{\beta}_1 {\gamma}_1
+{\beta}_2 {\gamma}_2
\end{matrix} 
\right) 
\left(
\begin{matrix} {\alpha}_{1} \\ {\alpha}_{2}
\end{matrix} 
\right) . 
\label{(2.63)} 
\end{equation}
There are two mutually exclusive cases arising from the requirement 
${\delta}_2\to \infty $, which can be treated: the hyperbolic case and the 
loxodromic case (hereafter we will write H for hyperbolic and L for loxodromic)
\vskip 0.3cm
\noindent{\it Subcase} (1.H)
\vskip 0.3cm
\noindent 
The hyperbolic case is obtained by evaluating 
$$
j^2={(\alpha +\delta )}^2={({\alpha }_1+{\delta }_1)}^2-{\left({\alpha }_2
+{\delta }_2\right)}^2+2i\left({\alpha }_1+{\delta }_1\right)
\left({\alpha }_2+{\delta }_2\right),
$$ 
and then requiring 
$$
j^2>4.
$$
Thus we must set
\begin{equation}
{\alpha}_{2}=-{\delta}_{2}, 
\label{(2.64)} 
\end{equation}
to obtain $j^2\in \Bbb{R}$ while the condition $j^2>4$ is automatically ensured 
in the limit $b\to \infty$ when $\delta \to \infty$ as in the case we are 
treating. In this case, Eq. (2.63) takes the form  
$$
\left(
\begin{matrix} {\delta}_{1} \\ {\delta}_{2} \end{matrix} 
\right)
={{1}\over {{\alpha}^2_1+{\delta}^2_2}} \left( 
\begin{matrix} 1+{\beta}_1{\gamma}_1
-{\beta}_2{\gamma}_2 & {\beta}_1{\gamma}_2+{\beta}_2{\gamma}_1 \\ 
{\beta}_1{\gamma}_2+{\beta}_2{\gamma}_1 & -1-{\beta}_1{\gamma}_1
+{\beta}_2{\gamma}_2 \end{matrix} 
\right)
\left(
\begin{matrix} {\alpha}_{1} \\ -{\delta}_{2} \end{matrix} 
\right).
$$ 
We are mainly interested in the limit value of all variables at infinity, 
therefore we can set 
$$
{\mathop{\rm{lim}}_{b\to \infty } {\alpha }_k={\widehat{\alpha }}_k\ },
\;  {\mathop{\rm{lim}}_{b\to \infty } {\beta }_k\ }={\widehat{\beta }}_k,
$$ 
\begin{equation}
{\mathop{\rm{lim}}_{b\to \infty } {\gamma }_k\ }={\widehat{\gamma }}_k,
\; {\mathop{\rm{lim}}_{b\to \infty } {\delta }_k\ }
={\widehat{\delta }}_k, 
\label{(2.65)} 
\end{equation}
for $k=1,2$. Of course, in this case we must have ${\widehat{\delta}}_{2}=\infty$, 
but if some estimates of the order of infinity of ${\widehat{\delta }}_2$ are 
needed, as well as the limits in Eq. (2.65), it is convenient to write 
Eq. (2.63) in the limit point case as
\begin{equation}
\left(
\begin{matrix} {\widehat{\delta}}_{1} \\ {\widehat{\delta}}_{2}
\end{matrix} \right)
={{1}\over {{\widehat{\alpha }}^2_1+{\widehat{\delta }}^2_2}}
\left( 
\begin{matrix} 1+{\widehat{\beta}}_{1} {\widehat{\gamma }}_{1}
-{\widehat{\beta}}_{2} {\widehat{\gamma}}_{2} 
& {\widehat{\beta}}_{1} {\widehat{\gamma}}_{2}
+{\widehat{\beta}}_{2} {\widehat{\gamma}}_{1} \\ 
{\widehat{\beta}}_{1} {\widehat{\gamma}}_{2}
+{\widehat{\beta}}_{2} {\widehat{\gamma}}_{1} & 
-1-{\widehat{\beta}}_{1}{\widehat{\gamma}}_{1}
+{\widehat{\beta}}_{2}{\widehat{\gamma}}_{2} 
\end{matrix} \right)
\left(
\begin{matrix} {\widehat{\alpha}}_{1} \\ -{\widehat{\delta}}_{2}
\end{matrix} \right). 
\label{(2.66)} 
\end{equation}
\vskip 0.3cm
\noindent {\it Subcase}(1.L)
\vskip 0.3cm
\noindent
The loxodromic case is obtained by the property
$$
j^2={(\alpha +\delta)}^2={({\alpha}_1+{\delta}_1)}^2-{\left({\alpha}_2
+{\delta}_2\right)}^2+2i\left({\alpha}_1+{\delta}_1\right)
\left({\alpha}_2+{\delta}_2\right)\in {\Bbb{C}-\Bbb{R}}^{+},
$$ 
thus we must require
\begin{equation}
\left({\alpha}_1+{\delta}_1\right)\left({\alpha}_2+{\delta}_2 \right)\ne 0, 
\label{((2.67)} 
\end{equation}
and this automatically ensures $j^2\in {\Bbb{C}}-{\Bbb{R}}^{+}$. In this case Eq. 
(2.63) does not require any modification. The limiting equation can be written as
\begin{equation}
\left(
\begin{matrix} {\widehat{\delta}}_{1} \\ {\widehat{\delta}}_{2}
\end{matrix} \right)
={{1}\over {{|\widehat{\alpha }|}^2}}
\left( 
\begin{matrix} 1+{\widehat{\beta}}_{1}
{\widehat{\gamma}}_1-{\widehat{\beta}}_{2} {\widehat{\gamma}}_{2} & 
{\widehat{\beta}}_{1}{\widehat{\gamma}}_{2}
+{\widehat{\beta}}_{2} {\widehat{\gamma}}_{1} \\ 
{\widehat{\beta}}_{1}{\widehat{\gamma}}_{2}
+{\widehat{\beta}}_{2}{\widehat{\gamma}}_{1} 
& -1-{\widehat{\beta}}_{1} {\widehat{\gamma}}_{1}
+{\widehat{\beta}}_{2} {\widehat{\gamma}}_{2}
\end{matrix} \right)
\left(
\begin{matrix} {\widehat{\alpha}}_{1} \\ 
{\widehat{\alpha}}_{2}
\end{matrix} \right). 
\label{(2.68)} 
\end{equation}
\vskip 0.3cm
\noindent
(2) {\it The $\gamma$ variable approaching infinity}

In this case it is useful to solve the equation $\alpha \delta -\beta \gamma =1$ 
in terms of the $\gamma$ variable. Simple calculations, which are similar 
to those which led us to Eq. (2.62), show that
$$
{\gamma }_1={|\beta |}^{-2}\left[\left(-1+{\alpha }_1{\delta }_1
-{\alpha }_2{\delta }_2\right){\beta }_1+\left({\alpha }_1{\delta }_2
+{\alpha }_2{\delta }_1\right){\beta }_2\right],
$$ 
\begin{equation}
{\gamma }_2={|\beta |}^{-2}\left[\left({\alpha }_1{\delta }_2
+{\alpha }_2{\delta }_1\right){\beta }_1+\left(1+{\alpha }_2{\delta }_2
-{\alpha }_1{\delta }_1\right){\beta }_2\right], 
\label{(2.69)} 
\end{equation}
thus we can write Eq. (2.68) in matrix form as
\begin{equation}
\left(\begin{matrix} {{\gamma }_1} \cr {{\gamma }_2}
\end{matrix} \right)
={{1}\over {{|\beta |}^2}}\left( 
\begin{matrix} -1+{\alpha }_1{\delta }_1
-{\alpha }_2{\delta }_2 & {\alpha }_1{\delta }_2+{\alpha }_2{\delta }_1 \cr 
{\alpha }_1{\delta }_2+{\alpha }_2{\delta }_1 & 1-{\alpha }_1{\delta }_1
+{\alpha }_2{\delta }_2 \end{matrix} 
\right)\left(\begin{matrix} {{\beta}_1} \\ 
{{\beta}_2}\end{matrix} \right). 
\label{(2.70)} 
\end{equation}
In this case we can distinguish four subcases: the hyperbolic, the loxodromic, 
the parabolic and the elliptic cases (hereafter, we write H for hyperbolic, 
L for loxodromic, P for parabolic and E for elliptic).
\vskip 0.3cm
\noindent {\it Subcase} (2.H)
 
The hyperbolic case is obtained by evaluating 
$$
j^2={(\alpha +\delta )}^2={({\alpha }_1+{\delta }_1)}^2-{\left({\alpha }_2
+{\delta }_2\right)}^2+2i\left({\alpha }_1+{\delta }_1\right)
\left({\alpha }_2+{\delta }_2\right),
$$ 
and then requiring 
$$
j^2>4.
$$
Thus we must set
$$
{\alpha }_2=-{\delta }_2, 
$$ 
jointly with the condition
$$
({\alpha }_1+{\delta }_1)^2>4.
$$
In this case, Eq. (2.70) takes the form
$$
\left(
\begin{matrix} {\gamma}_{1} \\ {\gamma}_{2}
\end{matrix} \right)
={{1}\over {{|\beta |}^2}}
\left( 
\begin{matrix} -1+{\alpha}_{1}{\delta}_{1}
+{\delta}^2_2 & {\alpha}_1 {\delta}_2-{\delta}_2 {\delta }_1 \\ 
{\alpha}_1 {\delta}_2-{\delta}_2 {\delta}_1 & 
1-{\alpha }_1{\delta }_1-{\delta}^2_2
\end{matrix} \right)
\left(
\begin{matrix} {\beta}_{1} \\ {\beta}_{2}
\end{matrix} \right) ,  
$$
which in the limit-point case can be written as
$$
\left(
\begin{matrix} {\hat \gamma}_{1} \\ {\hat \gamma}_{2}
\end{matrix} \right)
={{1}\over {{|\hat\beta |}^2}}
\left( 
\begin{matrix} -1+{\hat \alpha}_1 {\hat \delta}_1
+{\hat \delta}^2_2 & {\hat \alpha}_1 {\hat \delta}_2
-{\hat \delta}_2 {\hat \delta}_1 \\ 
{\hat \alpha}_1 {\hat \delta}_2 -{\hat \delta}_2 {\hat \delta}_1 & 
1-{\hat \alpha}_1 {\hat \delta}_1-{\hat \delta}^2_2
\end{matrix} \right)
\left(
\begin{matrix} {\hat \beta}_{1} \\ {\hat \beta}_{2}
\end{matrix} \right).  
$$
\vskip 0.3cm
\noindent
{\it Subcase }(2.L)
 
The loxodromic case is obtained by the property
$$
j^2={(\alpha +\delta )}^2={({\alpha }_1+{\delta }_1)}^2
-{\left({\alpha }_2+{\delta }_2\right)}^2+2i\left({\alpha }_1+{\delta }_1\right)
\left({\alpha }_2+{\delta }_2\right)\in {\Bbb{C}-\Bbb{R}}^+,
$$ 
thus we must require
$$
\left({\alpha }_1+{\delta }_1\right)\left({\alpha }_2+{\delta }_2\right)\ne 0.
$$
In this case, Eq. (2.70) does not need any modification. 
In the limit-point case, we must set
$$
\left(
\begin{matrix} {\hat\gamma}_{1} \\ {\hat\gamma}_{2}
\end{matrix} \right)
={{1}\over {{|\hat\beta |}^2}}
\left( 
\begin{matrix} -1+{\hat \alpha}_1 {\hat \delta }_1
-{\hat \alpha}_2 {\hat \delta}_2 & {\hat \alpha}_1 {\hat \delta}_2
+{\hat \alpha}_2 {\hat \delta}_1 \\ 
{\hat \alpha}_1 {\hat \delta}_2 + {\hat \alpha}_2 {\hat \delta}_1 & 
1-{\hat \alpha}_1 {\hat \delta}_1 + {\hat \alpha}_2 {\hat \delta}_2 
\end{matrix} \right)
\left(
\begin{matrix} {\hat \beta}_{1} \\ {\hat \beta}_{2}
\end{matrix} \right). 
$$
\vskip 0.3cm
\noindent {\it Subcase }(2.P)

In the parabolic case, we start with the equation
$$
j^2={(\alpha +\delta )}^2={({\alpha }_1+{\delta }_1)}^2
-{\left({\alpha }_2+{\delta }_2\right)}^2+2i\left({\alpha }_1
+{\delta }_1\right)\left({\alpha }_2+{\delta }_2\right)=4,
$$ 
and this can be fulfilled if and only if 
$$
{\alpha }_2=-{\delta }_2,
$$ 
\begin{equation}
{({\alpha }_1+{\delta }_1)}^2=4 \;  \Longrightarrow \;  
{\alpha }_1=\pm 2-{\delta }_1, 
\label{(2.71)} 
\end{equation}
and hence Eq. (2.69) reduces to
$$
\left( \begin{matrix} 
\gamma_{1} \\ \gamma_{2}
\end{matrix}\right)
={1 \over |\beta|^{2}}
\left( \begin{matrix}
-1 \pm 2 \delta_{1}-\delta_{1}^{2}+\delta_{2}^{2} &
\pm 2 \delta_{2}-2 \delta_{2} \delta_{1} \\
\pm 2 \delta_{2} - 2 \delta_{2} \delta_{1} &
1 \mp 2 \delta_{1}+\delta_{1}^{2}-\delta_{2}^{2}
\end{matrix} \right)
\left( \begin{matrix}
\beta_{1} \\ \beta_{2} 
\end{matrix}\right)
$$
which in the limit-point case, by using Eq. (2.67), 
can be written as
\begin{equation}
\left( \begin{matrix}
{\widehat \gamma}_{1} \\ {\widehat \gamma}_{2}
\end{matrix} \right)
={1 \over |{\widehat \beta}|^{2}}
\left( \begin{matrix}
-1 \pm 2 {\widehat \delta}_{1}
-{\widehat \delta}_{1}^{2}+{\widehat \delta}_{2}^{2} 
& \pm 2 {\widehat \delta}_{2}-2{\widehat \delta}_{2}
{\widehat \delta}_{1} \\
\pm 2 {\widehat \delta}_{2}-2{\widehat \delta}_{2}
{\widehat \delta}_{1} & 
1 \mp 2 {\widehat \delta}_{1}
+{\widehat \delta}_{1}^{2}-{\widehat \delta}_{2}^{2}
\end{matrix} \right)
\left( \begin{matrix}
{\widehat \beta}_{1} \\ {\widehat \beta}_{2}
\end{matrix}\right) ,
\label{(2.72)}
\end{equation}
in which some analysis of the behaviors of 
${\widehat \delta}_{k}$ and ${\widehat \beta}_{k}$
for $k=1,2$ would be necessary for further developments.
\vskip 0.3cm
\noindent
{\it Subcase} (2.E)
\vskip 0.3cm
In the elliptic case, the squared trace
$$
j^{2}=(\alpha+\delta)^{2}=(\alpha_{1}+\delta_{1})^{2}
-(\alpha_{2}+\delta_{2})^{2}
+2i (\alpha_{1}+\delta_{1})(\alpha_{2}+\delta_{2}),
$$
must lie in $[0,4[$ and hence
$$
\alpha_{2}=-\delta_{2},
$$
\begin{equation}
\alpha_{1}+\delta_{1} \in ]-2,2[ ,
\label{(2.73)}
\end{equation}
and Eq. (2.70) reduces to
$$
\left( \begin{matrix} 
\gamma_{1} \\ \gamma_{2}
\end{matrix}\right)
={1 \over |\beta|^{2}}
\left( \begin{matrix}
-1 +\alpha_{1} \delta_{1}+\delta_{2}^{2} 
& \alpha_{1} \delta_{2} - \delta_{2} \delta_{1} \\
\alpha_{1} \delta_{2}-\delta_{2} \delta_{1} 
& 1-\alpha_{1}\delta_{1}-\delta_{2}^{2} 
\end{matrix} \right)
\left( \begin{matrix}
\beta_{1} \\ \beta_{2} 
\end{matrix} \right) ,
$$
which in the limit-point case can be written as
\begin{equation}
\left( \begin{matrix}
{\widehat \gamma}_{1} \\ {\widehat \gamma}_{2}
\end{matrix} \right)
={1 \over |{\widehat \beta}|^{2}}
\left( \begin{matrix}
-1+ {\widehat \alpha}_{1} {\widehat \delta}_{1}
+{\widehat \delta}_{2}^{2} 
& {\widehat \alpha}_{1} {\widehat \delta}_{2}
-{\widehat \delta}_{2} {\widehat \delta}_{1} \\
{\widehat \alpha}_{1}{\widehat \delta}_{2}
-{\widehat \delta}_{2} {\widehat \delta}_{1}
& 1 - {\widehat \alpha}_{1} {\widehat \delta}_{1}
-{\widehat \delta}_{2}^{2}
\end{matrix} \right) 
\left( \begin{matrix}
{\widehat \beta}_{1} \\ {\widehat \beta}_{2}
\end{matrix} \right) ,
\label{(2.74)}
\end{equation} 
and also in this case, the behaviors of $\alpha_{k},\beta_{k}$ and
$\delta_{k}$, for $k=1,2$ are necessary for further developments.

The evaluation of the limits (2.66), (2.68), (2.72) and (2.74) 
might be very involved because the limit-point, limit-circle theory does 
not impose any a priori restriction on the behavior of $\alpha$, 
$\beta$, $\gamma$ and $\delta$ defined in Eq. (2.58) at infinity. 
This gives us only some advice on the square integrability near infinity 
for the functions $\varphi$ and $\psi$ (thus on $\alpha$ and $\gamma$, 
respectively). The lack of square integrability near infinity for the 
function $\psi$, results from Eq. (2.22) which, in the limit-point case 
we are treating, coincides with the requirement $r_b\to 0$ for large $b$. 
The lack of square integrability near infinity for the function $\varphi$, 
can be appreciated by using the method developed in 
Ref. \cite{Tit1962}, where it is shown, by making use of
$$
\int^b_0 dr'|\varphi +m \psi|^{2} \le 
{|m| \over {\mathfrak I}(l)}
$$ 
that it must be
$$
{{1}\over {2}}{|m|}^2\int^b_c{dr'{|\psi |}^2}-\int^b_0{dr'{\left|\varphi 
\right|}^2}\le \int^b_0{dr'}{|\varphi +m\psi |}^2\le {{\left|m\right|}
\over {{\mathfrak{I}}\left(l\right)}},
$$ 
which is a second degree algebraic inequality for the variable $|m|^2$, 
thus we must also have
$$
{|m|}^2 \int^b_c {dr'{|\psi |}^2} -2{{\left|m\right|}\over 
{{\mathfrak{I}}\left(l\right)}}
-2\int^b_0{dr'{\left|\varphi \right|}^2} \leq 0.
$$ 
The admissible roots lie within the interval defined by the associated 
algebraic equation but we must of course rule out the negative root, 
finding therefore
\begin{equation}
\left|m\right|\le {{1}\over {{\mathfrak{I}}\left(l\right)\int^b_0{dr'{
\left|\psi \right|}^2}}}+{\left\{{{2\int^b_0{dr'{\left|\varphi 
\right|}^2}}\over {\int^b_0{dr'{\left|\psi \right|}^2}}}+{{1}\over 
{{{\mathfrak{I}}^{2}\left(l\right)}{\left(\int^b_0{dr'{\left|\psi \right|}^2}
\right)}^2}}\right\}}^{{{1}\over {2}}}. 
\label{(2.75)} 
\end{equation}
From the fact that ${\mathfrak{I}}(m)\ne 0$ for every $b>0$, then $|m|$ is 
always positive also in the limit $b\to \infty$ and this can be reached 
if the ${\cal{L}}^2$ squared norm near infinity of $\varphi $, has the 
same order of infinity of that of $\psi $. Therefore, $\varphi$ is not 
square integrable near infinity, as well as $\psi$.

We can provide an interpretation of cases (1.H), (1.L), (2.P) and (2.E) by 
looking at the type of BMS transformations which we can call ``purely hyperbolic'', 
``purely loxodromic'', ``purely parabolic'' or ``purely elliptic''. Recall 
that a BMS transformation is given by the 
pair of transformations \cite{EA2018}
\begin{equation}
\zeta \to {{a\zeta +b}\over {c\zeta +d}}, 
\label{(2.76)} 
\end{equation}
\begin{equation}
u\to {{\left(1+{\left|\zeta \right|}^2\right)[u+\alpha (\zeta ,\overline{\zeta})]}
\over {{|a\zeta +b|}^2+{|c\zeta +d|}^2}}, 
\label{(2.77)} 
\end{equation}
for the conformal infinity of an asymptotically flat space time. The nontrivial 
point here, is that such maps can be equally well described if we choose 
a particular form for the fractional linear transformation in them. It is 
well known from Ref. \cite{Ford}, that each fractional 
linear transformation with one fixed point only, i.e. a parabolic transformation, 
can be always reduced to a pure translation of the form 
\begin{equation}
\zeta '=\zeta +b, 
\label{(2.78)} 
\end{equation}
while all the other transformations with two fixed points, can be always written in the form 
\begin{equation}
\zeta '=K\zeta , 
\label{(2.79)} 
\end{equation}
with $K=Ae^{i\theta }$. The ``purely elliptic'' case corresponds to 
setting $A=1$, the ``pure hyperbolic'' case corresponds to setting 
$\theta =2k\pi$ with $k\in \Bbb{Z}$, while the ``pure loxodromic'' 
case corresponds to setting $A\ne 1$ and $\theta \ne 2k\pi $ with 
$k\notin \Bbb{Z}$. We also recall that the procedure which enables us 
to write Eqs. (2.78) and (2.79), is a process which admits the 
possibility to conjugate the fixed points of the transformation, to 
particular points which are often chosen as the point $\zeta =\infty$ 
and $\zeta =0$, when a 2-fixed-point transformation is considered, while 
in the case of a transformation with one fixed point only, the fixed point 
is conjugated to the point at infinity. In this way, despite the position 
of the fixed points for the superrotation in Eq. (2.76), we can always 
reduce it to one among Eq. (2.78) or (2.79). Once this step is done, 
we can distinguish four cases as below \cite{EA2018}:
\vskip 0.3cm
\noindent
(H) {\it Hyperbolic BMS}
$$
\zeta \to A\zeta ,\;  A\in {{\Bbb{R}}^+}-\{1\},
$$ 
$$
u\to F_H\cdot \left[u+\alpha \right],\;  F_H
={{\left(1+{\left|\zeta \right|}^2\right)}\over 
{\left[1+A^2{\left|\zeta \right|}^2\right]}}.
$$ 
\vskip 0.3cm
\noindent
(L) {\it Loxodromic BMS}
$$
\zeta \to Ae^{i\theta }\zeta ,\;  
A\in {{\Bbb{R}}^+}-\{1\},\;  \theta \ne 2k\pi ,\ \ \ k\in \Bbb{Z},
$$ 
$$
u\to F_L\cdot \left[u+\alpha \right],\;  F_L
={{\left(1+{\left|\zeta \right|}^2\right)}\over 
{\left[1+A^2{\left|\zeta \right|}^2\right]}}.
$$ 
\vskip 0.3cm
\noindent
(P) {\it Parabolic BMS}
$$
\zeta \to \zeta +\beta ,
$$ 
$$
u\to F_P\cdot \left[u+\alpha \right],\;  F_P
={{\left(1+{\left|\zeta \right|}^2\right)}\over {\left[1
+{\left|\zeta +\beta \right|}^2\right]}}.
$$ 
\vskip 0.3cm
\noindent
(E) {\it Elliptic BMS}
$$
\zeta \to e^{i\theta }\zeta ,\;  \theta \ne 
2k\pi ,\ \ \ k\in \Bbb{Z},
$$ 
$$
u\to F_E\cdot \left[u+\alpha \right],\; 
F_E={{\left(1+{\left|\zeta \right|}^2\right)}\over 
{\left(1+{\left|\zeta \right|}^2\right)}}=1.
$$ 

This characterization of BMS transformations, leads to a correspondence 
between half of the BMS transformations and singular second order 
self-adjoint problems in quantum mechanics. If we can solve a limit-point, 
limit-circle problem for a given Hamiltonian in quantum mechanics, and 
thus we can calculate the functions $\varphi$ and $\psi$ in the 
limit-point case, then Eq. (2.59) forces the parameters $\alpha ,\beta ,
\gamma$ and $\delta$ occurring in Eq. (2.58) to fall back in one of 
the cases (1.H), (1.L), (2.H), (2.L), (2.P) and (2.E) for $b\to \infty$ 
and thus to give rise to ``purely hyperbolic'', ``purely loxodromic'', 
``purely parabolic'' or ``purely elliptic'' BMS transformations, whose 
functional form is expressed in the transformations (H), (L), (P) and (E) 
above. We can thus suggest that the limit-point case at infinity admits
a profound interpretation in terms of symmetries of the space-time itself 
and thus that a self-adjoint problem in quantum mechanics is strictly related 
to a particular class of BMS transformations for an asymptotically flat 
space-time. Suppose first to have solved a limit-point case at infinity in such 
a way that the two independent solutions $\varphi$ and $\psi$ are known. 
From our previous analysis, it is clear that the limit-point requirement 
must force these two independent solutions to fall back in one of the cases 
(1.H), (1.L), (2.H), (2.L), (2.P) and (2.E). This means that there must exist 
a lower bound $M$ such that for each $b>M$, all the fractional linear 
transformations in Eq. (2.58) are of one special kind and cannot be of 
some other kind. This ensures us that all such transformations, for $b>M$, 
can give rise to one and only one type of BMS transformations between (H), (L), 
(P) and (E). Further developments in this direction, can be accomplished only 
if the particular limit-point theory is solved and thus if a concrete case is 
chosen as an example of application for the theory here treated. 

Other questions which can arise from solving a concrete limit-point, 
limit-circle problem, can regard the possibility of constructing precise 
discrete subgroups of $PSL(2,\Bbb{C})$ in the limit $b\to \infty$ or 
equivalently, if it is possible to construct a bounded sequence $\{b_n\}$ on 
the real positive line, chosen in such a way that Eq. (2.58) falls back 
in one of the cases discussed above for each $b_n$ and, at the same time, 
which forces the BMS transformations such arising, to form a Kleinian group. 

\section {Limit-Point Case at Both Ends of the Positive Real Line and Hyperbolic
Cyclic Groups}

\subsection {The Arrangement of the Isometric Circles in a Hyperbolic Cyclic Group}

In this section we try to obtain a relation between the limit-point, 
limit-circle theory and the theory of hyperbolic cyclic groups of fractional 
linear transformations. As we have already established in the previous section, 
each second-order self-adjoint problem which is singular at both ends of the 
positive real line, is accompanied by two families of circles which reduce to 
a pair of points on the complex plane when the interval in which the problem 
is first studied is stretched out by covering the whole positive real line. 
We have also mentioned the importance of singular self-adjoint problems in quantum 
mechanics thus, by solving one of such boundary problems in the framework of the 
limit-point, limit-circle theory one can also obtain some insight about other 
aspects of physics, which cannot be immediately viewed but nevertheless can 
arise in a very elegant way as it happened in the connection between limit-point, 
limit-circle theory and the BMS transformations discussed in the last subsection 
of the previous section. In this framework we can expect to interpret our 
previous results and all of that which will come next, as an example of how 
some areas of quantum theory can give rise to a non-trivial connection with 
some areas of the theory of gravitation.

The basic idea that we will follow hereafter, relies on an evident similitude 
between the arrangement of the isometric circles of a hyperbolic cyclic group 
(that we denote by $I_n$), and the arrangement of all 
circles $C_b$ in the limit-point, limit-circle theory. Actually, a hyperbolic 
cyclic group gives rise to two families of isometric circles, families which 
we can indicate as $\left\{I_n\right\}$ and $\left\{I_{-n}\right\}$ referring 
to $I_n$ as the isometric circle of the hyperbolic transformation $f^n$ while 
referring to $I_{-n}$ as the isometric circle of the inverse $f^{-n}$ (note  
that $f^n\circ f^{-n}=\rm{l}$), which satisfies the following requirements:
\vskip 0.3cm
\noindent 
{\it (i)} Each member of the family $\left\{I_n\right\}$ is exterior 
to each member of the family $\left\{I_{-n}\right\}$.
\vskip 0.3cm
\noindent 
{\it (ii)} For $m>n$, $I_m$ is contained in the interior of $I_n$ and 
$I_{-m}$ is contained in the interior of $I_{-n}$.
\vskip 0.3cm
\noindent 
{\it (iii)} In the limit $n\to \infty $, both $I_n\to {\xi}_1\in 
\Bbb{C}$ and $I_{-n}\to {\xi}_2\in \Bbb{C}$, where ${\xi}_1$ and ${\xi}_2$ 
are the fixed points of the generating hyperbolic transformation $f$.
\vskip 0.3cm

We also know that each isometric circle of a hyperbolic cyclic group, 
contains at most one fixed point of the generating transformation. We will 
show actually, that all members of a disjoint family of circles 
contain the same fixed point.
All properties {\it (i), (ii)} and {\it (iii)} above, are very similar to that 
established for the arrangement of the circles $C_b$ and ${C'}_a$ for a 
second-order self-adjoint problem on the positive real line which may occur 
if and only if the second-order differential operator of the theory is in limit 
point case at both ends of the interval. In that case, we can establish three 
properties that should hold in order to get the limit-point case at both ends.
\vskip 0.3cm
\noindent 
{\it (i')} Each member of the family $\left\{C_b\right\}$ is 
exterior to each member of the family $\left\{{C'}_a\right\}$.
\vskip 0.3cm
\noindent 
{\it (ii')} For $b>\tilde{b}$ and $a<\tilde{a}$, $C_b$ is contained 
in the interior of $C_{\tilde{b}}$ and ${C'}_a$ is contained in the 
interior of ${C'}_{\tilde{a}}$.
\vskip 0.3cm
\noindent 
{\it (iii')} In the limit $b\to \infty $ and $a\to 0$, both 
$C_b\to m_{\infty}\in \Bbb{C}$ and ${C'}_a\to {m'}_{\infty}\in \Bbb{C}$, 
where $m_{\infty}$ and ${m'}_{\infty}$ are the limit-points at 
$r=\infty $ and $r=0$, respectively.
\vskip 0.3cm
From previous discussions, we know that all circles belonging to 
the same disjoint family, contain one and only one of the two limit points. 
We can thus construct somehow, a monotonic increasing sequence $\left\{b_n\right\}$ 
and a monotonic decreasing sequence $\left\{a_n\right\}$, in such a way that 
${\delta}_n=[a_n,b_n]\to {\Bbb{R}}^+$, which is the goal of the limit-point 
theory at both ends of the positive real line. In this way, properties {\it (i'), 
(ii') }and {\it (iii') }can be reformulated as follows.
\vskip 0.3cm
\noindent 
{\it (i'')} Each member of the family $\left\{C_{b_n}\right\}$ is 
exterior to each member of the family $\left\{{C'}_{a_n}\right\}$.
\vskip 0.3cm
\noindent 
{\it (ii'') } For $m>n$, $C_{b_m}$ is contained in the interior of 
$C_{b_n}$ and ${C'}_{a_m}$ is contained in the interior of ${C'}_{a_n}$.
\vskip 0.3cm
\noindent 
{\it (iii'')} In the limit $n\to \infty $, both $C_{b_n}\to m_{\infty }
\in \Bbb{C}$ and $C_{a_n}\to {m'}_{\infty }\in \Bbb{C}$, where $m_{\infty }$ 
and ${m'}_{\infty }$ are the limit-points at $r=\infty $ and $r=0$, respectively.
\vskip 0.3cm
Now, one can see from the similar meanings of {\it i)} and {\it i''), ii)} 
and {\it ii'')}, {\it iii)} and {\it iii'')}, that a complete identification 
can be obtained if we impose the following restrictions:
$$
I_n=C_{b_n},\;  I_{-n}={C'}_{a_n},\;  \forall n\in \Bbb{N},
$$
\begin{equation} 
{\xi}_1=m_{\infty},\;  {\xi }_2={m'}_{\infty}. 
\label{(3.1)} 
\end{equation}
Despite the similarity first mentioned, these last conditions are not 
trivial for the different nature of circles $C_{b_n}$ and $I_n$, thus we 
will see that a consistent set of restrictions must be imposed is such a way 
that our desired relations (3.1) could be satisfied. The point here is that 
we are guided by the analogy between relations {\it i), ii), iii) } 
and {\it (i''), (ii''), (iii'')}, but nothing ensures that each limit-point 
condition at both ends of the positive real line could return us a hyperbolic 
cyclic group, hence some restriction may occur for the fundamental system of 
solutions ${\varphi}_1$ and ${\varphi}_2$ to the equation $Lx=lx$ as well 
as for the existence of two monotonic sequences of points $\left\{a_n\right\}$ 
and $\left\{b_n\right\}$ (which will be clear in the following).

\vskip0.3cm

In the remainder of this section, we will try to recover some further notions 
about hyperbolic cyclic groups. First, we must prove properties {\it (i), (ii) }
and {\it (iii)}. We will be concerned with the following type 
of fractional linear transformation:
\begin{equation}
f\left(z\right)={{az+b}\over {cz+d}}, \;  c\ne 0. 
\label{(3.2)} 
\end{equation}
As we already know, the requirement $c\ne 0$ corresponds to demanding that 
the map (3.2) has two finite fixed points. In this case, the isometric 
circle of the map (3.2) is given by
\begin{equation}
I:\ \ \ \left|cz+d\right|=1, \;  c\ne 0, 
\label{(3.3)} 
\end{equation}
and it is the locus of points of the complex plane whose arcs are unaltered 
in length when Eq. (3.2) is applied. Following the arguments contained 
in Ref. \cite{Ford}, the case $c\ne 0$ is appropriate for a powerful 
conjugation of fixed points of Eq. (3.2). Call its fixed points 
${\xi}_1$ and ${\xi}_2$ and use the four-point ratio
\begin{equation}
{{(z'-{z'}_1)}\over {(z'-{z'}_2)}}{{({z'}_2-{z'}_3)}\over {({z'}_1-{z'}_3)}}
={{(z-z_1)}\over {(z-z_2)}}{{(z_2-z_3)}\over {(z_1-z_3)}}, 
\label{(3.4)} 
\end{equation}
by setting 
$$
{z'}_1=z_1={\xi }_1,\; {z'}_2=z_2={\xi }_2,\; {z'}_3={{a}\over {c}},
$$ 
and solving it with respect to the $z'=f(z)$ variable
\begin{equation}
z'={{{\xi }_1-{\xi }_2K\left({{z-{\xi }_1}\over {z-{\xi }_2}}\right)}
\over {1-K\left({{z-{\xi }_1}\over {z-{\xi }_2}}\right)}}, 
\label{(3.5)} 
\end{equation}
where $K$ is the multiplier 
$$
K={{\left({\xi }_1-{{a}\over {c}}\right)}\over {\left({\xi }_2-{{a}\over {c}}\right)}}.
$$ 
We can obtain the same result in Eq. (2.5) by using the conjugation 
process which involves the following variables:
\begin{equation}
Z=g\left(z\right)={({z-{\xi}_1})\over ({z-{\xi}_2})},\; 
Z'=g\left(z'\right)={({z'-{\xi}_1})\over ({z'-{\xi}_2})}, 
\label{(3.6)} 
\end{equation}
which conjugate the fixed points to $z=0$ and $z=\infty $. Equation (3.4) reduces to
\begin{equation}
g\left(z'\right)=Kg\left(z\right), 
\label{(3.7)} 
\end{equation}
and by defining
\begin{equation}
K\left(z\right)=Kz, 
\label{(3.8)} 
\end{equation}
Eq. (3.4) reduces to
$$
g\left(z'\right)=K\left(g\left(z\right)\right)=K\circ g\left(z\right),
$$ 
from which 
$$
z'=g^{-1}\circ K\circ g\left(z\right),
$$ 
but $g^{-1}$ can be written as
$$
z=g^{-1}\left(Z\right)={{({\xi }_2Z-{\xi }_1})\over ({Z-1})},
$$ 
thus 
$$
z'=g^{-1}\left(KZ\right)=g^{-1}\left(K{{z-{\xi }_1}\over {z-{\xi }_2}}\right)
={{{\xi }_2K\left({{z-{\xi }_1}\over {z-{\xi }_2}}\right)-{\xi }_1}
\over {K\left({{z-{\xi }_1}\over {z-{\xi }_2}}\right)-1}},
$$ 
from which we recover Eq. (3.5). We recall the fact that a hyperbolic 
transformation is characterized by a real and positive value of $K$ not 
equal to one, i.e. $K\in {\Bbb{R}}^+-\left\{1\right\}$; for $K\in ]0,1[$ 
we get a contraction about the fixed point ${\xi }_1$ and a dilation 
about ${\xi}_2$, while if $K\in ]1,\infty [$ we get a dilation about 
${\xi}_1$ and a contraction about ${\xi}_2$ thus, in the former case 
${\xi}_1$ is an attractive point and ${\xi}_2$ is a repulsive point while 
in the latter the reverse holds. We also know that for a hyperbolic 
transformation each circle through the fixed points is mapped into another such 
circle, the interior of a circle through the fixed points is mapped into itself, 
any circle orthogonal to any circle through the fixed points is mapped into 
another such circle and that the fixed points are inverse one to the other 
with respect to each circle orthogonal to any circle through the fixed points. 
Therefore, if a fractional linear map is a hyperbolic transformation, 
its trace $j=a+c$ must satisfy 
$$
\left|j\right|>2.
$$ 
We are interested in cyclic groups hence, given Eq. (3.2), we can construct 
the variables (3.6). Now, by defining the following sequence of transformations 
for the $Z$ plane defined in Eq. (3.6)
\begin{equation}
Z^{''}=K^2Z,\; Z^{'''}=K^3Z,\; \dots \ ,\;  
Z^{\left(n\right)}=K^nZ, 
\label{(3.9)} 
\end{equation}
we are interested in finding which are the corresponding transformations, 
whose general mapping is given in Eq. (3.2), for the $z$ plane. 
Obviously, from previous reasoning, we have
\begin{equation}
f=g^{-1}\circ K\circ g, 
\label{(3.10)} 
\end{equation}
where the $K$ transformation is given in Eq. (3.8). One can see that the following maps
$$
f^2=g^{-1}\circ K\circ g\circ g^{-1}\circ K\circ g=g^{-1}\circ K^2\circ g,
$$ 
$$
f^3=g^{-1}\circ K\circ g\circ g^{-1}\circ K\circ g\circ g^{-1}\circ K\circ g=g^{-1}\circ K^3\circ g,
$$ 
$$
\dots \ \ \dots \ \ \dots \ \ \dots \ \ \dots \ \ \dots \ \ \dots \ \ \dots \ \ \dots \ ,
$$ 
$$
f^n=\underbrace{g^{-1}\circ K\circ g\circ \dots \circ g^{-1}\circ K\circ g}_{n}
=g^{-1}\circ K^n\circ g,
$$ 
are of the form given in Eq. (3.10) and thus suffice for defining 
transformations of type (3.7), which are explicitly given in Eq. (3.9). 
We have thus recovered the fact that the $n$-th power of some linear 
transformation $f$ (with two fixed points), has a multiplier which is the 
{\it n}th power of the multiplier for the map $f$. We can thus write the 
following expression from Eq. (3.5):
\begin{equation}
f^n\left(z\right)=z^{(n)}={{{\xi }_1-{\xi }_2K^n\left({{z-{\xi }_1}
\over {z-{\xi }_2}}\right)}\over {1-K^n\left({{z-{\xi }_1}
\over {z-{\xi }_2}}\right)}}, 
\label{(3.11)} 
\end{equation}
and hence, the net effect of several applications of the same transformation 
becomes merely a substitution of the multiplier for the original transformation. 
We also stress the fact that the multiplier is the only object which defines 
a particular transformation (once his fixed points are chosen) and the way in which 
it varies, while we are applying several copies of the same map, is independent 
of the representation used whatever is given in Eqs. (3.9) or in Eq. (3.11). 

We can also discuss some generalization of Eq. (3.10) taking into 
account the possibility of conjugating any triplet $z_1$, $z_2$ and 
$z_3$ to any other triplet ${z'}_1$, ${z'}_2$ and ${z'}_3$ despite of 
the particular choice which leads us to the multiplier $K$, by using Eq. 
(3.4). Notice that Eq. (3.4) has always a finite value even if we choose 
some of the points $z_1$, $z_2$, $z_3$ or ${z'}_1$, ${z'}_2$, ${z'}_3$ 
to be not finite. But if we set 
$$
\sigma ={{{z'}_2-{z'}_3}\over {{z'}_1-{z'}_3}},\; \rho 
={{z_2-z_3}\over {z_1-z_3}},\; \gamma ={{\rho}\over {\sigma}},
$$ 
then Eq. (3.4) can be written as
$$
\left({{z'-{z'}_1}\over {z'-{z'}_2}}\right)=\gamma \left({{z-z_1}\over {z-z_2}}\right),
$$ 
and by using the notation
$$
Z=h\left(z\right)={{z-z_1}\over {z-z_2}},\; Z'
=h\left(z'\right)={{z'-{z'}_1}\over {z'-{z'}_2}}
$$ 
jointly with the following definition:
$$
\gamma \left(Z\right)=\gamma Z,
$$ 
we can write the map $f\left(z\right)=z'$ as
\begin{equation}
f=h^{-1}\circ \gamma \circ h. 
\label{(3.12)} 
\end{equation}
Of course, by solving the equation $Z'=\gamma Z$ in the variable $z'$, 
the map $f$ takes the form
$$
f\left(z\right)=z'={{{z'}_1-{z'}_2\gamma \left({{z-z_1}\over 
{z-z_2}}\right)}\over {1-\gamma \left({{z-z_1}\over {z-z_2}}\right)}},
$$ 
and the same arguments given above, hold for subsequent applications 
of the same map (3.12), thus we can state that subsequent applications 
of the same map $f$ lead to a substitution of the multiplier 
$\gamma $ by powers of it and hence, the maps
$$
Z^{''}={\gamma }^2Z,\; \dots ,\; Z^{\left(n\right)}={\gamma }^nZ,
$$ 
correspond to the maps
$$
f^2=g^{-1}\circ K^2\circ g,\; \dots ,\; f^n=g^{-1}\circ K^n\circ g,
$$ 
on the $z$ plane, where 
\begin{equation}
f^n\left(z\right)=z^{(n)}={{{z'}_1-{z'}_2{\gamma }^n\left({{z-z_1}\over 
{z-z_2}}\right)}\over {1-{\gamma }^n\left({{z-z_1}
\over {z-z_2}}\right)}}. 
\label{(3.13)} 
\end{equation}
Now we want to derive the expression for the isometric circle for 
the transformation (3.11) which will be the isometric circle of the 
$n$-th power of any transformation whose multiplier $K$ is already 
known. First, let us note that the determinant of Eq. (3.5) is not 
equal to one, as well for Eq. (3.11), thus we must divide and multiply this latter by 
$$
K^{{{n}\over {2}}}{({\xi }_1-{\xi }_2)},
$$ 
hence Eq. (3.11) becomes
\begin{eqnarray}
z^{(n)}&=& {{{\xi }_1-{\xi }_2K^n\left({{z-{\xi}_1}\over {z-{\xi}_2}}\right)}
\over {1-K^n\left({{z-{\xi}_1}\over {z-{\xi}_2}}\right)}}
={{\left(K^n{\xi}_2
-{\xi }_1\right)z+\xi_{1}\xi_{2}(1-K^n)}\over {\left(K^n-1\right)z+({\xi}_2-K^n{\xi}_1)}}
\nonumber \\
&=& {{\ \left[{{\left(K^n{\xi}_2-{\xi}_1\right)}\over {K^{{{n}\over 
{2}}}({\xi}_1-{\xi}_2)}}\right]z+\left[{{\xi_{1}\xi_{2}(1-K^n)}\over {K^{{{n}
\over {2}}}({\xi}_1-{\xi}_2)}}\right]\ }\over {\ \left[{{\left(K^n-1\right)}
\over {K^{{{n}\over {2}}}({\xi}_1-{\xi}_2)}}\right]z
+\left[{{({\xi}_2
-K^n{\xi}_1)}\over {K^{{{n}\over {2}}}({\xi}_1-{\xi}_2)}}\right]\ }}, 
\label{(3.14)} 
\end{eqnarray}
and the last expression has determinant equal to one. The general case in Eq. 
(3.14) can also be worked out. In this case we can write Eq. (3.14) as
$$
z^{(n)}={{{z'}_1-{z'}_2{\gamma}^n\left({{z-z_1}\over {z-z_2}}\right)}
\over {1-{\gamma}^n\left({{z-z_1}\over {z-z_2}}\right)}}={{{z'}_1z-{z'}_1z_2
-{z'}_2{\gamma}^nz+{z'}_2z_1{\gamma}^n}\over {z-z_2-{\gamma}^nz+{\gamma }^nz_1}}
$$ 
$$
= {{\left({z'}_2{\gamma}^n-{z'}_1\right)z+\left({z'}_1z_2
-{z'}_2z_1{\gamma}^n\right)}\over {\left({\gamma}^n-1\right)z
+\left(z_2-{\gamma}^nz_1\right)}},
$$ 
whose determinant is
$$
t=\left({z'}_2{\gamma}^n-{z'}_1\right)\left(z_2-{\gamma}^n z_1\right)
-\left({\gamma}^n-1\right)\left({z'}_1 z_2-{z'}_2 z_1{\gamma}^n\right),
$$ 
and we have just to multiply and divide by $\sqrt{t}$ to get a fractional 
linear transformation with determinant equal to one
\begin{equation}
z^{(n)}={{\ {{\left({z'}_2{\gamma }^n-{z'}_1\right)}\over 
{\sqrt{t}}}z+{{\left({z'}_1z_2-{z'}_2z_1{\gamma}^n\right)}\over 
{\sqrt{t}}}\ }\over {{{\left({\gamma}^n-1\right)}\over 
{\sqrt{t}}}z+{{\left(z_2-{\gamma}^nz_1\right)}\over {\sqrt{t}}}}}, 
\label{(3.15)} 
\end{equation}
which follows at once from the relation
$$
{{a}\over {\sqrt{t}}}{{d}\over {\sqrt{t}}}-{{b}\over {\sqrt{t}}}{{c}
\over {\sqrt{t}}}={{t}\over {t}}=1.
$$ 
Taking into account Eq. (3.3), the isometric circle of the map (3.14) must be written as
\begin{equation}
I_n: \; \left|z+{{({\xi }_2-K^n{\xi }_1)}\over {\left(K^n-1\right)}}\right|
=\left|{{{\xi }_1-{\xi }_2}\over {K^{{{n}\over {2}}}
-K^{-{{n}\over {2}}}}}\right|, 
\label{(3.16)} 
\end{equation}
while the inverse transformation of Eq. (3.14) is obviously written as
\begin{equation}
f^{-n}\left(z\right)=z^{(-n)}={{\ -\left[{{({\xi }_2-K^n{\xi }_1)}
\over {K^{{{n}\over {2}}}({\xi }_1-{\xi }_2)}}\right]z
+\left[{{\xi_{1}\xi_{2} (1-K^n)}
\over {K^{{{n}\over {2}}}({\xi }_1-{\xi }_2)}}\right]\ \ \ }
\over {\ \ \left[{{\left(K^n-1\right)}\over {K^{{{n}
\over {2}}}({\xi }_1-{\xi }_2)}}\right]z-\left[{{\left(K^n{\xi }_2
-{\xi }_1\right)}\over {K^{{{n}\over {2}}}({\xi }_1-{\xi }_2)}}\right]}}, 
\label{(3.17)} 
\end{equation}
whose isometric circle is written as
\begin{equation}
I_{-n}: \; \left|z+{{({\xi }_1-K^n{\xi}_2)}\over {
\left(K^n-1\right)}}\right|=\left|{{{\xi}_1-{\xi}_2}\over 
{K^{{{n}\over {2}}}-K^{-{{n}\over {2}}}}}\right|, 
\label{(3.18)} 
\end{equation}
and $I_n$ and $I_{-n}$ have the same radius but different centre. 
In the case of a hyperbolic or loxodromic cyclic group with $K=Ae^{i\theta}$, 
with $A\in {\Bbb{R}}^+-\{1\}$ we can observe that the radii of circles 
$I_n$ and $I_{-n}$ approach zero as $n\to \infty$. From the theory  
developed in Ref. \cite{Ford}, we know that the limit points 
of a Kleinian group are all the points in the neighborhoods of 
which an infinity of arcs of isometric circles fall. Take for example 
a hyperbolic or loxodromic cyclic group, then we have
$$
A>1, \; I_n\longrightarrow 
{\xi }_1, \;  I_{-n}\longrightarrow {\xi }_2, 
$$ 
$$
A<1, \;  I_n
\longrightarrow {\xi }_2, \;  I_{-n}
\longrightarrow {\xi }_1,
$$ 
thus, the isometric circles wrap up at the fixed points showing that 
these are the limit points of the cyclic group. This proves 
the relation {\it (iii)} we have stated so far for hyperbolic cyclic groups.

In the following we will be mainly interested in purely hyperbolic cyclic 
groups, i.e. cyclic groups generated by a hyperbolic transformation. 
The reason is that we can easily show property {\it (i)} for hyperbolic 
cyclic groups by assuming $K=A\in {\Bbb{R}}^+-\{1\}$. This will 
be explicitly done later but for the moment we want to establish the 
fundamental fact that if one of the fixed points, say ${\xi}_1$, is in the 
interior of any $I_m$ for fixed $m$, then each member of the family 
$\left\{I_n\right\}$ contains in its interior the same fixed point. 

From the standard theory developed Ref. \cite{Ford}, we know that the 
isometric circle of a hyperbolic transformation contains one and only one 
fixed point while the isometric circle of its inverse contains the other 
and that such circles are exterior one to the other. Hence, it is not 
useless to establish that if ${\xi}_1$ (or ${\xi}_2$) is interior to 
any $I_m$, then it is also contained in the interior of $I_n$ for each 
$n$ and thus each family $\left\{I_n\right\}$ and $\left\{I_{-n}\right\}$ 
contains one and only one fixed point. This will pave the way for the proof 
of properties {\it (i)} and {\it (ii)} stated at the beginning of this section 
for the hyperbolic cyclic groups. 

All the interior points of of the isometric circle $I_m$ satisfy
$$
\left|z+{{({\xi}_2-K^m{\xi}_1)}\over {\left(K^m-1\right)}}
\right|<\left|{{{\xi }_1-{\xi }_2}\over {K^{{{m}\over {2}}}
-K^{-{{m}\over {2}}}}}\right|,
$$ 
and our aim is to obtain all conditions which admit the possibility 
that ${\xi}_2$ (or ${\xi}_1$) lies in the interior of $I_m$ for some $m$, thus
$$
\left|{\xi }_2+{{({\xi}_2-K^m{\xi}_1)}\over {\left(K^m-1\right)}}
\right|<\left|{{{\xi}_1-{\xi}_2}\over {K^{{{m}\over {2}}}-K^{-{{m}\over {2}}}}}\right|
$$ 
should hold and the previous equation can be put in the form
\begin{equation}
\left|q_mr_m\right|<\left|r_m\right|, 
\label{(3.19)} 
\end{equation}
where 
$$
q_m=K^{{{m}\over {2}}}, \;  r_m={{({\xi}_1-{\xi}_2)}\over 
{\left(K^{{{m}\over {2}}}-K^{-{{m}\over {2}}}\right)}}.
$$ 
Equation (3.19) is fulfilled if and only if $\left|q_m\right|<1$ 
and this can be accomplished if the modulus of $K$ is less than 1, thus
$$
\left|q_m\right|<1\ \Longrightarrow \left|K\right|
=\left|Ae^{i\theta }\right|=A<1,
$$ 
but this case is independent of the index $m$, i.e. if $A<1$, then 
$\left|q_m\right|<1$ for every $m\in \Bbb{N}$. We have thus proved 
that all members of the family $\left\{I_n\right\}$ contain the fixed 
point ${\xi}_2$ (and hence every member of the family $\left\{I_{-n}\right\}$ 
contains the other fixed point) when $A<1$. Same reasoning can be applied 
if we try to fulfil the request that ${\xi}_1$ lies in the interior 
of $I_m$ for some $m$. In this case, 
\begin{equation}
\left|r_m\right|<\left|q_mr_m\right| 
\label{(3.20)} 
\end{equation}
should hold instead of Eq. (3.19) and hence $A>1$ is also required. 
The fulfilment of Eq. (3.20) is independent of the $m$ index once 
any $A>1$ is chosen. Therefore, each member of the family 
$\left\{I_n\right\}$ contains in its interior the fixed point 
${\xi }_1$ (and hence every member of the family $\left\{I_{-n}\right\}$ 
contains the other fixed point) when $A>1$.

We will now prove the fundamental fact that $I_m$ is always in the 
interior of $I_n$ for $n<m$. This statement, follows at once by proving 
that $I_n$ and $I_{-m}$ are always exterior for each $n$ and $m$ and by 
applying Theorem 5.18 of Ref. \cite{Ford}, which we report here below:
\vskip 0.3cm
\noindent
Let $\ I_h,\ I_g,\ {I'}_g,\ {I'}_{gh}$ be the isometric 
circles of the transformations $h,\ g,g^{-1},g\circ h$, respectively. 
If $I_h$ and $I'_{gh}$ are exterior to one another, 
then $I_{gf}$ is contained in $I_g$.

We are thus concerned with the transformations $h=f^l$, $g=f^n$ and 
$g\circ h=f^m$ where $m>n$ and $\left(m-n\right)=l$. We should set up all 
conditions which ensure that $I_l$ is exterior to $I_{-m}$ in such a way 
that we can establish the fact that $I_m$ is in the interior of 
$I_n$ for $m>n$. We will soon see that this condition is independent 
of the choices made for $l,\ m,n$ and thus $I_m$ is always in the interior 
of $I_n$ for every $m>n$.
 
Take into account Eqs. (3.16) and (3.18) for $I_l$ and $I_{-m}$. 
The requirement that these two circles be exterior one to the other, follows 
from requiring that the distance between their centres exceeds 
the sum of their radii, i.e.,
\begin{equation}
\left|{{({\xi}_2-K^l{\xi}_1)}\over {\left(K^l-1\right)}}
-{{({\xi}_1-K^m{\xi}_2)}\over {\left(K^m-1\right)}}\right|>
\left|{{{\xi}_1-{\xi}_2}\over {K^{{{l}\over {2}}}-K^{-{{l}\over 
{2}}}}}\right|+\left|{{{\xi}_1-{\xi}_2}\over {K^{{{m}\over {2}}}
-K^{-{{m}\over {2}}}}}\right|. 
\label{(3.21)} 
\end{equation}
We can deal with such inequality, by setting
\begin{equation}
{\alpha }_s=K^{{{s}\over {2}}}-K^{-{{s}\over {2}}}, \;  {\gamma}_s
={\xi}_2 K^{-{{s}\over {2}}}-{\xi}_1 K^{{{s}\over {2}}}, \;  
{\gamma}_{-s}={\xi}_2K^{{{s}\over {2}}}-{\xi}_1 K^{-{{s}\over {2}}}, 
\label{(3.22)} 
\end{equation}
from which Eq. (3.21) reads as
\begin{equation}
\left|{{{\gamma}_l {\alpha}_m+{\gamma}_{-m} {\alpha}_l}\over 
{{\alpha}_l {\alpha}_m}}\right|<\left|{\xi}_1-{\xi}_2\right|
\left({{1}\over {|{\alpha}_l|}}+{{1}\over {|{\alpha}_m|}}\right), 
\label{(3.23)} 
\end{equation}
but some algebraic calculations make it possible to simplify the numerator 
of the left-hand side of Eq. (3.23)
$$
{\gamma}_l {\alpha}_m+{\gamma}_{-m} {\alpha}_l=\left({\xi}_2-{\xi}_1\right)
\left(K^{{{m+l}\over {2}}}-K^{-{{m+l}\over {2}}}\right),
$$ 
and also let us write Eq. (3.23) in the elegant form
$$
\left|{\alpha}_l\right|+\left|{\alpha}_m\right|<\left|K^{{{m+l}\over 
{2}}}-K^{-{{m+l}\over {2}}}\right|\ \Longrightarrow 
$$ 
\begin{equation}
\left|K^{{{l}\over {2}}}-K^{-{{l}\over {2}}}\right|+\left|K^{{{m}\over 
{2}}}-K^{-{{m}\over {2}}}\right|<\left|K^{{{m+l}\over {2}}}
-K^{-{{m+l}\over {2}}}\right|. 
\label{(3.24)} 
\end{equation}
We can take the move from Eq. (3.24) only if we suppose some 
functional form for the multiplier $K$. We can obtain some interesting 
identities in the case of a loxodromic generating transformation with 
$K=Ae^{i\theta }$, but this is not the easiest one which we can deal with, 
and thus we postpone the discussion about the general loxodromic case 
for the fulfilment of Eq. (3.24) and take into account the easier case 
of a hyperbolic generating transformation with $K=A\in {\Bbb{R}}^+-\{1\}$. 
Therefore, we must distinguish two cases: $A>1$ and $A<1$.
\vskip 0.3cm
\centerline{{\it Case}$\ A>1$}
\vskip 0.3cm
\noindent 
We have
$$
\left|A^{{{l}\over {2}}}-A^{-{{l}\over {2}}}\right|+\left|A^{{{m}\over 
{2}}}-A^{-{{m}\over {2}}}\right|<\left|A^{{{m+l}\over {2}}}
-A^{-{{m+l}\over {2}}}\right|,
$$ 
but $A^k-A^{-k}>0$ for $k>0$, hence 
\begin{equation}
A^{{{l}\over {2}}}-A^{-{{l}\over {2}}}+A^{{{m}\over {2}}}-A^{-{{m}\over {2}}}
<A^{{{m+l}\over {2}}}-A^{-{{m+l}\over {2}}}. 
\label{(3.25)} 
\end{equation}
But one has also the majorization
\begin{equation}
A^{{{l}\over {2}}}+A^{{{m}\over {2}}}<A^{{{m+l}\over {2}}}, 
\label{(3.26)} 
\end{equation}
from the fact that the product of quantities bigger than 1, exceeds 
always their sum. We have also that
$$
-A^{{{l}\over {2}}}-A^{{{m}\over {2}}}<-A^{{{m+l}\over {2}}},
$$ 
which follows from
\begin{equation}
{\left({{1}\over {A}}\right)}^{{{m}\over {2}}}+{\left({{1}\over 
{A}}\right)}^{{{l}\over {2}}}>{\left({{1}\over 
{A}}\right)}^{{{m+l}\over {2}}}. 
\label{(3.27)} 
\end{equation}
By noticing that the sum of quantities less then 1, exceeds always 
their product, Eqs. (3.26) and (3.27) completely prove the fulfilment 
of the inequality (3.25). Note that the fulfilment of (3.25) is 
independent of the choice of $l$ and $m$. This shows that in the present 
case, the circle $I_l$ is always exterior to the circle $I_{-m}$ and 
therefore, from Ref. \cite{Ford}, $I_m$ must lie in the interior of $I_n$ 
with $m=l+n>n$. From the arbitrariness of $l$, we can say that if $K=A>1$, 
then $I_{n+1}$ is contained in the interior of $I_n$ for each 
$n\in \Bbb{N}$. In this case, all properties {\it (i) }and {\it (ii)} for 
the isometric circles of a hyperbolic cyclic group stated at the beginning 
of the present section, are satisfied (which was our main interest for 
further developments). Of course, the case $A<1$ can equally be treated 
by establishing the fundamental fact that $I_l$ is exterior to $I_{-m}$ 
for each $l$ and $m$, and that $I_{n+1}$ is contained in the interior of 
$I_n$ for each $n\in \Bbb{N}$.
\vskip 0.3cm
\centerline{{\it Case}$\ A<1${\it .}}

Inequality (3.25) is replaced by
\begin{equation}
A^{{{l}\over {2}}}-A^{-{{l}\over {2}}}+A^{{{m}\over {2}}}
-A^{-{{m}\over {2}}}>A^{{{m+l}\over {2}}}-A^{-{{m+l}\over {2}}}, 
\label{(3.28)} 
\end{equation}
while conditions (3.26) and (3.27) should be replaced by
$$
A^{{{l}\over {2}}}+A^{{{m}\over {2}}}>A^{{{m+l}\over {2}}},
$$ 
$$
{\left({{1}\over {A}}\right)}^{{{m}\over {2}}}+{\left({{1}\over 
{A}}\right)}^{{{l}\over {2}}}<{\left({{1}\over {A}}\right)}^{{{m+l}\over {2}}},
$$ 
which lead to the fulfilment of Eq. (3.28). Thus, properties {\it (i)} and 
{\it (ii)} are satisfied also in case of a hyperbolic cyclic group with $A<1$.

We have just proved properties {\it (i)}, {\it (ii)} and {\it (iii)} for a 
generic hyperbolic cyclic group. This circumstance will make it possible for 
us to obtain a consistent set of equations which can be derived by imposing 
the validity of Eqs. (3.1).

\subsection {Some Remarks on Loxodromic Cyclic Groups}

We revert to Eq. (3.24) and discuss the possibility to fulfil such equation 
in the case of a loxodromic cyclic group for which $K=Ae^{i\theta }$. When 
$\theta =0$, such an inequality is always satisfied and thus the question 
arises about which circumstances might lead to its fulfilment when $\theta \ne 0$. 
Note that in the limits $l\to \infty$ or $m\to \infty$, Eq. (3.24) is 
trivially satisfied letting $I_{-m}$ be exterior to $I_l$. However, nothing 
ensures that for low values of $l$ and $m$ this could occur, but some development 
might be obtained as well although the complexity which arises from direct 
calculations suggests a very strong correlation between the values of $l,m$ and 
$\theta$, and this complexity can be faced only via numerical 
computation whenever needed. 

Let us note the presence of the following recurrent function in Eq. 
(3.24), which we can write as
\begin{equation}
F\left(\lambda \right)=\left|K^{\lambda }-K^{-\lambda }\right|, 
\label{(3.29)} 
\end{equation}
from which Eq. (3.24) reads as
\begin{equation}
F\left({{l}\over {2}}\right)+F\left({{m}\over {2}}\right)
<F\left({{l+m}\over {2}}\right). 
\label{(3.30)} 
\end{equation}
We want to study the function (3.29) and obtain useful relations 
which might be used for further developments. Let us note that
$$
{\rm arg} \left(K^{\lambda}\right)=\lambda \theta .
$$ 
We will be mainly interested in the smallest angle between the directions 
of $K^{\lambda }$ and $K^{-\lambda }$ which we will call ${\theta}_i$.
\vskip 0.3cm
\noindent
(1) Suppose that
\begin{equation}
0<\lambda \theta <{{\pi }\over {2}}. 
\label{(3.31)} 
\end{equation}
The interpretation for the value of the function $F(\lambda )$ is given in Fig. 1
\begin{figure}[h]
\begin{center}
\includegraphics[width=6cm, height=6cm]{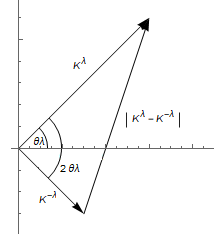}
\caption{Case 1}
\end{center}
\end{figure}
and we are thus interested in the value of the angle 
${\theta }_i=2\lambda \theta$.
\vskip 0.3cm
\noindent
(2) Suppose now that
\begin{equation}
\lambda \theta ={{\pi}\over {2}}. 
\label{(3.32)} 
\end{equation}
Thus the geometrical meaning of $F(\lambda )$ can be given 
as in Fig. 2.
\begin{figure}[h]
\begin{center}
\includegraphics[scale=0.7]{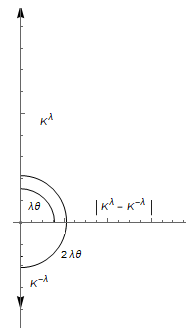}
\caption{Case 2}
\end{center}
\end{figure}
and the angle in which we are interested is ${\theta}_i=\pi $, 
while the function $F(\lambda)$ reduces to
$$
F\left(\lambda \right)=A^{\lambda }+A^{-\lambda }.
$$ 
\vskip 0.3cm
\noindent
(3) Suppose that
\begin{equation}
{{\pi }\over {2}}<\lambda \theta <\pi , 
\label{(3.33)} 
\end{equation}
hence the geometrical interpretation is the one shown in Fig. 3.
\begin{figure}[h]
\begin{center}
\includegraphics[scale=0.7]{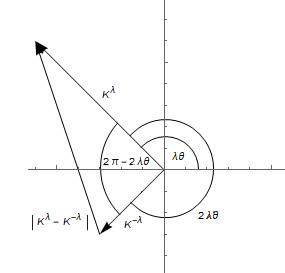}
\caption{Case 3}
\end{center}
\end{figure}
where ${\theta}_i=2\pi -2\lambda \theta$.
\vskip 0.3cm
\noindent
(4) In the case
\begin{equation}
\lambda \theta =\pi , 
\label{(3.34)} 
\end{equation}
this leads to the picture in Fig. 4
\begin{figure}[h]
\begin{center}
\includegraphics[scale=0.7]{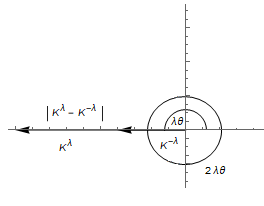}
\caption{Case 4}
\end{center}
\end{figure}
where ${\theta }_i=0$ and 
$$
F\left(\lambda \right)=\left|A^{\lambda}-A^{-\lambda}\right|.
$$ 
\vskip 0.3cm
\noindent
(5) The case 
\begin{equation}
\pi <\lambda \theta <{{3\pi }\over {2}}, 
\label{(3.35)} 
\end{equation}
is shown in Fig. 5
\begin{figure}[h]
\begin{center}
\includegraphics[scale=0.7]{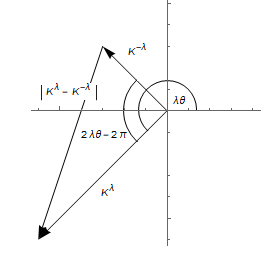}
\caption{Case 5}
\end{center}
\end{figure}
and the angle ${\theta}_i=2\left[\pi -\left(2\pi -\lambda \theta 
\right)\right]=2\lambda \theta -2\pi$ is obtained.
\vskip 0.3cm
\noindent
(6) Suppose next that
\begin{equation}
\lambda \theta ={{3\pi}\over {2}}, 
\label{(3.36)} 
\end{equation}
One therefore obtains the picture in Fig. 6
\begin{figure}[h]
\begin{center}
\includegraphics[scale=0.7]{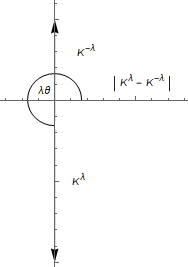}
\caption{Case 6}
\end{center}
\end{figure}
Then ${\theta}_i=\pi $ and 
$$
F\left(\lambda \right)=A^{\lambda }+A^{-\lambda}.
$$ 
\vskip 0.3cm
\noindent
(7) The case 
\begin{equation}
{{3\pi}\over {2}}<\lambda \theta <2\pi , 
\label{(3.37)} 
\end{equation}
is represented in Fig. 7
\begin{figure}[h]
\begin{center}
\includegraphics[scale=0.7]{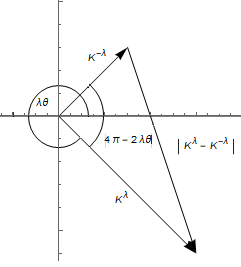}
\caption{Case 7}
\end{center}
\end{figure}
and the angle ${\theta}_i=2\left(2\pi -\lambda \theta \right)
=4\pi -2\lambda \theta $, is obtained as well.
\vskip 0.3cm
\noindent
(8) The last case is 
\begin{equation}
\lambda \theta =2\pi , 
\label{(3.38)} 
\end{equation}
in FIG. 8
\begin{figure}[h]
\begin{center}
\includegraphics[scale=0.7]{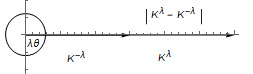}
\caption{Case 8}
\end{center}
\end{figure}
where ${\theta}_i=0$ and 
$$
F\left(\lambda \right)=\left|A^{\lambda}-A^{-\lambda}\right|.
$$ 

All the above cases are very useful for evaluating the function $F(\lambda)$. 
From the theorem of cosines, we can set
\begin{equation}
{F(\lambda)}^2=A^{2\lambda}+A^{-2\lambda}
-2 \cos {\theta}_{i}, 
\label{(3.39)} 
\end{equation}
and one can obtain $F(\lambda)$ once the angle ${\theta}_i$ is known. 
From the above discussion it is evident that ${\theta}_i$ is a function of 
the $\lambda$ variable and one can recover the fact that such a function 
is a continuous, periodic and bounded function of $\lambda$ whose 
functional expression can be easily obtained. 
Let us sum up all the cases obtained so far:
$$
(1) \; \lambda \theta \in \left]0,{{\pi}\over {2}}\right[
\; \Longrightarrow \; {\theta}_i=2\lambda \theta .
$$ 
$$
(2) \; \lambda \theta ={{\pi }\over {2}}\; \Longrightarrow \; {\theta}_i=\pi .
$$ 
$$
(3) \; \lambda \theta \in \left]{{\pi}\over {2}},\pi \right[
\; \Longrightarrow \; {\theta}_i=2\pi -2\lambda \theta .
$$ 
$$
(4) \; \lambda \theta =\pi \; \Longrightarrow \; {\theta}_i=0.
$$ 
$$
(5) \; \lambda \theta \in \left]\pi ,{{3\pi}\over {2}}\right[
\; \Longrightarrow \; {\theta}_i=2\lambda \theta -2\pi .
$$ 
$$
(6) \; \lambda \theta ={{3\pi}\over {2}}\; \Longrightarrow \; {\theta}_i=\pi .
$$ 
$$
(7) \; \lambda \theta \in \left]{{3\pi}\over {2}},2\pi \right[
\; \Longrightarrow \; {\theta}_i=4\pi -2\lambda \theta .
$$ 
$$
(8) \; \lambda \theta =2\pi \; \Longrightarrow \; {\theta}_i=0.
$$ 

We notice that for $\lambda \theta >2\pi $, several cases may occur 
but all of them return us a value of the ${\theta}_i$ function which falls 
back into one of the cases from (1) to (8). This enables us to state 
the periodicity (with period $2\pi$ in the $\lambda \theta$ variable) 
of the ${\theta}_i(\lambda \theta)$ function. A remark is also needed 
for cases (2), (4), (6) and (8): since Eq. (3.39) is continuous in the 
${\theta}_i$, it is also continuous at the point 
$$
{\theta}_i=0,\pi ,
$$ 
thus we can incorporate the case (2) into case (1), the case (4) into case 3), 
the case (6) into case (5) and the case (8) into case (7) by simply looking 
at the functional form of ${\theta}_i$ given in cases (1), (3), (5) and (7). 
Therefore, there are just four cases left:
\vskip0.2cm
$$
(1')\; \lambda \theta \in \left]0,{{\pi }\over {2}}\right]
\; \Longrightarrow \; {\theta}_i=2\lambda \theta .
$$ 
$$
(2')\; \lambda \theta \in \left]{{\pi }\over {2}},\pi \right]
\; \Longrightarrow \; {\theta}_i=2\pi -2\lambda \theta .
$$ 
$$
(3')\; \lambda \theta \in \left]\pi ,{{3\pi}\over {2}}\right]
\; \Longrightarrow \; {\theta}_i=2\lambda \theta -2\pi .
$$ 
$$
(4')\; \lambda \theta \in \left]{{3\pi}\over {2}},2\pi \right]
\; \Longrightarrow \; {\theta}_i=4\pi -2\lambda \theta .
$$ 
The functional form of ${\theta}_i(\lambda \theta)$ is 
(here $k \in \Bbb{Z}$)
\begin{equation}
{\theta}_i\left(\lambda \theta \right)=\left \{ 
\begin{array}{rr} 
2\lambda \theta 
-2k\pi , \; \lambda \theta \in \left]k\pi ,\left(2k+1\right){{\pi}
\over {2}}\right] \\
& -2\lambda \theta +2k\pi , \; \lambda \theta \in 
\left]\left(2k+1\right){{\pi}\over {2}},\left(k+1\right)\pi \right]
\end{array} 
\right . 
\label{(3.40)} 
\end{equation}
and it is represented in Fig. 9
\begin{figure}[h]
\begin{center}
\includegraphics[scale=0.8]{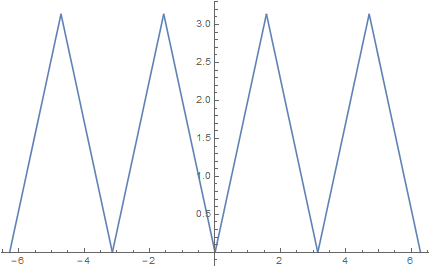}
\caption{Plot of the function describing the internal angle}
\end{center}
\end{figure} 

The inequality (3.30), which is
$$
F\left({{l}\over {2}}\right)+F\left({{m}\over {2}}\right)
<F\left({{l+m}\over {2}}\right),
$$ 
might be proved with the help of Eqs. (3.39) and (3.40). By recalling that 
such an inequality is satisfied in the limit $l\to \infty$ (and thus 
$m\to \infty$ from the relation $m=n+l$), one can ask whether there exists 
a minimum value of $l$ such that for any fixed $n$, the inequality is 
disproved. Notice that there is a strong correlation between the values of 
$l,m$ and $\theta$, thus in the perspective of further developments, it 
would be convenient to fix a definite value of $\theta$ and try to obtain a 
lower bound, for the $l$ variable and for a fixed $n$, below which the inequality 
is not satisfied. Take for example the configuration of isometric circles of 
Ref. \cite{Ford}. In that case, the isometric circles intersect each other for low 
values of $n$, thus it should be convenient to start with low values 
of the indices $n,l,m$.

As far as we can see, further developments are not immediate but nevertheless 
they might be obtained by taking the move from the results of the present subsection.  

\subsection {Correspondence Between Hyperbolic Cyclic Groups and Limit-Point Case at Both Ends}

Our aim is now to establish the identification stated in Eq. (3.1) between 
the circles arising from the limit-point theory at both ends of the positive real 
line and the isometric circles of a hyperbolic cyclic group with multiplier 
$K=A\in {\Bbb{R}}^+-\{1\}$. Suppose that the operator $L$ 
$$
Lx(r)=-\left(p\left(r\right)x\left(r\right)\right)'+q(r)x(r),
$$ 
defined in ${\Bbb{R}}^+$, is in the limit-point case at both $r=0$ and $r=\infty$. 
If ${\varphi}_1$ and ${\varphi}_2$ are two linearly independent solutions of the equation
\begin{equation}
Lx=lx, \; l\in \Bbb{C}, 
\label{(3.41)} 
\end{equation}
satisfying the conditions
$$
{\varphi}_1\left(c,l\right)=1, \; {\varphi}_2\left(c,l\right)=0,
$$ 
$$
p\left(c\right){\varphi '}_1\left(c,l\right)=0, \; p
\left(c\right){\varphi '}_2\left(c,l\right)=1,
$$ 
then we can set up a self-adjoint problem on the interval $\delta =[a,b]
\subset {\Bbb{R}}^+$, with $c\in \delta $, given in Eq. (2.32) and thus there 
exists a countable set of eigenfunctions and eigenvalues of $L$ from which the 
completeness and unitarity relation (also called Parseval equality) follows for 
every $f\in {\cal{L}}^2(\delta )$. Following the arguments of Sect. 2, 
we know that if we choose two solutions in the form
$$
{\chi}_a\left(r\right)={\varphi}_1+{m'}_a{\varphi}_2, \;
{\chi}_b\left(r\right)={\varphi}_1+m_b{\varphi}_2, 
$$ 
of Eq. (3.41), which also satisfy the following conditions:
$$
{\cos \beta'} \; {\chi}_a\left(a\right)+{\sin \beta'} \; p(a)
{\chi'}_a (a)=0, \; \beta '\in [0,\pi [,
$$ 
$$
{\cos \beta} \; {\chi}_b (b) +{\sin \beta} \; p(b)
{\chi'}_b (b) =0,\;  \beta \in [0,\pi [,
$$ 
then, there remain defined two circles ${C'}_a$ and $C_b$, in the complex 
plane of the ${m'}_a$ and $m_b$ variables, whose equations are 
$$
\left[{\chi}_a{\chi}_a\right](a)=0, \; 
\left[{\chi}_b{\chi}_b\right](b)=0,
$$ 
or equivalently, whose centres and radii are
$$
{C'}_a \; : \;  {\tilde{m}'}_a=\left({{[{\varphi}_1 {\varphi}_2](a)}
\over {2i{\frak{I}}(l)\int^c_a{dr'{|{\varphi}_2|}^2}}}\right),
\; r_a={\left(2i{\frak{I}}(l)\int^c_a{dr'{|{\varphi}_2|}^2}\right)}^{-1},
$$ 
\begin{equation}
C_b \; : \;  {\tilde{m}}_b=-\left({{\left[{\varphi}_1{\varphi}_2\right](b)}
\over {2i{\mathfrak{I}}\left(l\right)\int^b_c{dr'{\left|{\varphi}_2\right|}^2}}}
\right),\ \ \ \ \ \ r_b={\left(2i{\mathfrak{I}}\left(l\right)\int^b_c
{dr'{\left|{\varphi}_2\right|}^2}\right)}^{-1}. 
\label{(3.42)} 
\end{equation}

The limit point case is characterized by the following limiting 
values for centres and radii:
$$
{\mathop{\rm{lim}}_{a\to 0} r_a=0}, \;  
{\mathop{\rm{lim}}_{a\to 0} {\tilde{m}'}_a={m'}_{\infty}\in \Bbb{C}},
$$ 
\begin{equation}
{\mathop{\rm{lim}}_{b\to \infty} r_b=0}, \; 
{\mathop{\rm{lim}}_{b\to \infty} {\tilde{m}}_b=m_{\infty}
\in \Bbb{C}}, 
\label{(3.43)} 
\end{equation}
and ${m'}_{\infty}$ and $m_{\infty}$ are the values of the limit points at 
$r=0$ and $r=\infty$, respectively. 

We recall that the limit point condition at both ends of the interval 
guarantees the uniqueness of the spectral matrix of the problem, therefore 
the essential self-adjointness of the operator $L$ on the whole positive real 
line. The existence of a complete set of eigenfunctions and eigenvalues belonging 
to the point spectrum and the continuum spectrum is ensured, hence every 
$f\in {\cal{L}}^2({\Bbb{R}}^+)$ has a unique spectral decomposition in terms 
of the eigenfunctions of $L$ over ${\cal{L}}^2({\Bbb{R}}^+)$.

We want to obtain more useful relations by imposing the restriction 
contained in Eq. (3.1) in such a way that the circles ${C'}_a$ and $C_b$ 
so arising can be interpreted as the isometric circles of a hyperbolic 
cyclic group. In doing this, the discreteness of such a Kleinian group should 
let us reinterpret the limiting procedure of Eq. (3.43): we define a monotonic 
decreasing sequence of points $\left\{a_n\right\}$ such that $a_n\to 0$ and 
a monotonic increasing sequence of points $\left\{b_n\right\}$ such that $b_n\to \infty$ 
$$
a_1>a_2>\dots >a_n>...>0,
$$ 
\begin{equation}
b_1<b_2<\dots <b_n<\dots <\infty , 
\label{(3.44)} 
\end{equation}
and thus, in each interval ${\delta }_n=[a_n,b_n]\subset {\Bbb{R}}^+$, Eq. 
(2.32) defines a self-adjoint problem for the operator $L$. The limit-point 
condition at both ends guarantees the essential self-adjointness of $L$ when the 
limit ${\delta }_n\to {\Bbb{R}}^+$ is taken and hence, by using Eq. (3.42), the 
following circles remain defined:
$$
{C'}_{a_n} \; : \; \left|z+{{[{\varphi}_1{\varphi}_2](a_n)}\over 
{[{\varphi}_2 {\varphi}_2](a_n)}}\right|={{1}\over {\left|
{[\varphi}_2{\varphi}_2](a_n)\right|}},
$$ 
\begin{equation}
C_{b_n} \; : \;  \left|z+{{[{\varphi}_1{\varphi}_2](b_n)}\over 
{[{\varphi}_2{\varphi}_2](b_n)}}\right|={{1}\over {\left|{[\varphi}_2
{\varphi}_2](b_n)\right|}}, 
\label{(3.45)} 
\end{equation}
in which the concise form has been used for
\begin{equation}
\left[{\varphi}_2 {\varphi}_2 \right](a)=-2i{\mathfrak{I}}
\left(l\right)\int^c_a{dr'{\left|{\varphi}_2\right|}^2}, \; 
\left[{\varphi}_2 {\varphi}_2\right](b)=2i{\mathfrak{I}}
\left(l\right)\int^b_c{dr'{\left|{\varphi}_2\right|}^2}, 
\label{(3.46)} 
\end{equation}
derived in Sect. 2. The limiting relations in Eq. (3.43) 
must be replaced by the following:
$$
{\mathop{\rm{lim}}_{n\to \infty } r_{a_n}=0}, \;  
{\mathop{\rm{lim}}_{n\to \infty } {\tilde{m}'}_{a_n}={m'}_{\infty}
\in \Bbb{C}},
$$ 
$$
{\mathop{\rm{lim}}_{n\to \infty } r_{b_n}=0\ }, \;  
{\mathop{\rm{lim}}_{n\to \infty } {\tilde{m}}_{b_n}=m_{\infty }\in \Bbb{C}}.
$$ 
By making use of Eqs. (3.45), (3.16) and (3.18), conditions (3.1) 
lead us to the system of equations
\begin{equation}
\left \{ \begin{array}{rrrr} 
\int^{b_n}_c{dr'{\left|{\varphi}_2\right|}^2}
\triangleq \int^c_{a_n}{dr'}{\left|{\varphi}_2\right|}^2 
\\
& {\eta}_n={{({\xi}_1-K^n{\xi}_2)}\over {\left(K^n-1\right)}}
\triangleq {{[{\varphi}_1 {\varphi}_2](a_n)}\over 
{[{\varphi}_2 {\varphi}_2](a_n)}}=-{\tilde{m}'}_{a_n} \\
& {{\mu }_n={{({\xi}_2-K^n{\xi}_1)}\over 
{\left(K^n-1\right)}}\triangleq {{[{\varphi }_1{\varphi }_2](b_n)}\over 
{[{\varphi }_2{\varphi }_2](b_n)}}=-{\tilde{m}}_{b_n}} \\ 
& {\left|{{{\xi}_1-{\xi}_2}\over {K^{{{n}\over {2}}}-K^{-{{n}\over 
{2}}}}}\right|\triangleq {{1}\over {2{\frak{I}}\left(l\right)
\int^{b_n}_c{dr'{\left|{\varphi}_2\right|}^2}}}}
\end{array}
\right. 
\label{(3.47)} 
\end{equation}
The first equation corresponds to requiring that the radii of the 
isometric circles $I_n$ and $I_{-n}$ coincide (and thus $r_{a_n}=r_{b_n}$). 
Notice that $r^{-1}_b$ is a monotonic increasing function of the $b$ variable 
and it takes all values between $0$ and $\infty$. The same holds for 
$r^{-1}_a$ which is a monotonic decreasing function of the $a$ variable, 
and takes all values between $\infty $ and $0$. Thus, we are just claiming 
that ${\varphi}_2$ is not square integrable near infinity nor near zero 
(which was proved at the end of Sect. 2). Therefore, for any chosen 
increasing and divergent sequence $\left\{b_n\right\}$ there always exists 
a corresponding decreasing sequence $\left\{a_n\right\}$ such that 
$a_n\to 0$. The monotonic behavior of the functions $r^{-1}_a$ and $r^{-1}_b$, 
jointly with their common lower bound, tell us that for every $b_k\in 
\left\{b_n\right\}$ there exists one value $a_k$ such that a monotonic 
decreasing sequence $\left\{a_n\right\}$ is defined, providing the 
fulfilment of the first equation in Eqs. (3.47).

The second and the third equation in Eqs. (3.47) impose a strong 
restriction on the locus of the centres ${\tilde{m}'}_{a_n}$ and 
${\tilde{m}}_{b_n}$. Notice that for fixed value of $K,\ {\xi }_1$ and 
${\xi }_2$, the cyclic group is uniquely defined and thus, the locus of 
points above which ${\tilde{m}'}_{a_n}$ and ${\tilde{m}}_{b_n}$ are required 
to lie (from the definition claimed), must be a straight line. 
In fact, let us consider
$$
{\mu }_n={({\xi_2-K^n\xi_1})\over ({K^n-1})},
$$ 
into Eqs. (3.47) which, for a fixed hyperbolic cyclic group with $K=A>1$ 
(and fixed point chosen), describe a set of points of the complex plane 
obtained by letting the index $n$ vary. All these points have the same phase:
$$
{\rm{arg} (\mu_{n+1}-\mu_n) }={\rm const.},
$$ 
thus they have to lay on a straight line as first mentioned. Then, by setting
$$
\mu \left(t\right)={{{\xi}_2-t{\xi}_1}\over {t-1}}, \; t\in ]1,\infty[,
$$ 
one can see that this function is a continuous curve on the complex plane 
which has one end at the point 
$$
\mu \left(\infty \right)=-{\xi}_1,
$$ 
hence $\mu(t)$ describes a segment on the complex plane and all points 
${\mu}_n$ lie on such a segment. The same arguments hold for the points
$$
{\eta}_n={{{\xi}_1-K^n{\xi}_2}\over {K^n-1}},
$$ 
which have to lie on another segment of the complex plane. The limit-point 
condition at $r=0$ and $r=\infty$, suggests that such points must lie on 
opposite halfplanes, i.e. ${\mu}_n$ has positive imaginary part for every 
$n$ (resp. negative imaginary part for every $n$) and ${\eta}_n$ has negative 
imaginary part for every $n$ (resp. positive imaginary part for every $n$). 
We also observe that, by setting
\begin{equation}
D_n=\left|{\mu}_{n+1}-{\mu}_n\right|=\left|{\xi}_1-{\xi}_2\right|\left[
{{K^n(K-1)}\over {(K^{n+1}-1)(K^n-1)}}\right]=\left|{\eta}_{n+1}
-{\eta}_n\right|={D'}_n, 
\label{(3.48)} 
\end{equation}
the sequence of distances $\left\{D_n={D'}_n\right\}$ is a monotonic decreasing 
sequence for sufficiently large $K$. This last claim can be easily proved by 
simply setting $K^n=t$ and by differentiating the function
$$
D\left(t\right)={{at(K-1)}\over {(Kt-1)(t-1)}}, \;  t>1, \;  a>0.
$$ 
One obtains 
$$
{{dD}\over {dt}}\left(K^n\right)<0 \; \Longleftrightarrow \; 
K^n>{{1}\over {\sqrt{K-1}}},
$$ 
which is always satisfied for large $n$; but if we want that such an inequality 
should hold for every $n\in \Bbb{N}$, then we should search for the 
real roots of the equation
$$
K^3-K^2-1=0.
$$ 
\begin{figure}[h]
\begin{center}
\includegraphics[scale=0.7]{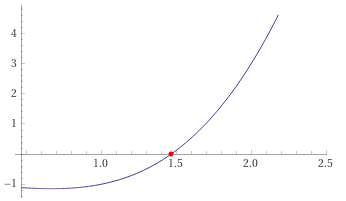}
\caption{Solution of the equation $K^{3}-K^{2}-1=0$}
\end{center}
\end{figure}
Such an equation has only one real root which is 
$$
\tilde K\approx 1.4656,
$$ 
and this is the value which provides a lower bound above which the 
sequence $\left\{D_n\right\}$ is a monotonic decreasing sequence.

The fourth equation in Eqs. (3.47) can be satisfied by any monotonic 
sequence $\left\{b_n\right\}$ (once the hyperbolic cyclic group is 
chosen, i.e. ${\xi}_1,\ {\xi}_2$ and $K$ are fixed) provided that a 
function ${\varphi}_2$ not in ${\cal{L}}^2(c,\infty )$ nor in 
${\cal{L}}^2(0,c)$ could be constructed somehow. We notice that it is 
rather convenient to start with a hyperbolic cyclic group instead of 
solving some limit-point, limit-circle problem on the positive real line. 
This is because the restrictions on the functions ${\varphi}_1$ and 
${\varphi}_2$ given in the second and the third equations of Eqs. (3.47) 
are so strict that it is hopeless trying to fulfil them once ${\varphi}_1$ 
and ${\varphi}_2$ are already known as independent solutions of a 
limit-point problem. We suggest that the function ${\varphi}_2$ should be 
obtained by starting from the fourth equation of Eqs. (3.47) once that 
${\xi}_1,{\xi}_2$ and $K=A\in {\Bbb{R}}^+-\{1\}$ have been fixed. 
The problem here is to construct a pair of functions ${\varphi}_1$ and 
${\varphi}_2$ which are not square integrable near infinity nor near zero, 
and which satisfy Eqs. (3.47) jointly with the system of conditions at 
$c\in {\delta }_n=[a_n,b_n]$
$$
{\varphi}_1\left(c,l\right)=1, \; {\varphi}_2\left(c,l\right)=0,
$$ 
$$
p(c)) {\varphi '}_1 \left(c,l\right)=0, \;  p(c) 
{\varphi '}_2 \left(c,l\right)=1.
$$ 
In this way, ${\varphi}_1$ and ${\varphi}_2$ can be viewed as a system of 
independent solutions for the equation
$$
Lx=lx,
$$ 
and the form so chosen for them, should force the coefficients of the 
operator $L$ to have a particular functional expression, and this is the 
main goal which one can hope to accomplish in a more advanced theory. 
Actually, in the context of the limit-point, limit-circle theory for 
second-order singular self adjoint-problems, a lot of efforts have 
been produced in this direction and a fruitful theory has been developed in 
Refs. \cite{Gelfand1951,Brown2009}. 
Here it is shown how to construct the differential equation of 
a second-order singular self-adjoint problem by starting from the knowledge of 
its spectral function (see Sect. 2). In the literature, functions in Eq. 
(2.51) are also known as {\it Weyl-Titchmarsh functions} and they have been 
widely studied in the context of limit-point, limit-circle theory. We write them 
here below for problems with singular behaviour at both ends of the domain 
of definition for the operator $L$:
$$
M_{11}\left(l\right)={{1}\over {{m'}_{\infty}\left(l\right)
-m_{\infty}\left(l\right)}},
$$ 
$$
M_{12}\left(l\right)=M_{21}\left(l\right)={{1}\over {2}}
{{{m'}_{\infty}\left(l\right)+m_{\infty}\left(l\right)}
\over {{m'}_{\infty }\left(l\right)-m_{\infty}\left(l\right)}},
$$ 
$$
M_{22}\left(l\right)={{{m'}_{\infty}\left(l\right)m_{\infty}
\left(l\right)}\over {{m'}_{\infty}\left(l\right)-m_{\infty}\left(l\right)}}. 
$$
One can thus see that these functions only depend upon the limit points 
$m_{\infty}\left(l\right)$ and ${m'}_{\infty}\left(l\right)$ on the complex 
plane. In Ref. \cite{Gelfand1951} 
it is shown how to construct the function $q(r)$ occurring in 
the operator $L$ in Eq. (3.41), by starting only from the knowledge of 
Weyl-Titchmarch functions. This makes us hope that further developments might 
also be achieved in the context in which Eqs. (3.47) are meaningful.
\vskip 0.3cm
\noindent
{\it Existence of monotonic sequences satisfying the second and the third equations}
\vskip 0.3cm
We now revert to the second and third equations in Eqs. (3.47). Let us refer only 
to the third equation for simplicity of reasoning
$$
{\mu}_n={{({\xi}_2-K^n{\xi}_1)}\over {\left(K^n-1\right)}}\triangleq 
{{\left[{\varphi}_1 {\varphi}_2 \right]\left(b_n\right)}\over 
{\left[{\varphi}_2 {\varphi}_2 \right]\left(b_n\right)}}=-{\tilde{m}}_{b_n}.
$$ 
We know that this equation defines a countable sequence of points which 
lie on a segment of the complex plane. Let us consider the case in which 
$K=A>1$: one of the ends of such a segment is the point $-{\xi}_1$ while the 
distances between successive points are $D_n$ given in Eq. (3.48) where 
$D_n\to 0$ for $n\to \infty $ (if $K>\tilde{K}\approx 1.4656$, then 
$\left\{D_n\right\}$ is a monotonic decreasing sequence approaching zero). 
Therefore, we must require that the function 
\begin{equation}
-{\tilde{m}}_r={{\left[{\varphi}_1 {\varphi}_2\right](r)}
\over {\left[{\varphi}_2 {\varphi}_2 \right](r)}} 
\label{(3.49)} 
\end{equation}
should intersect such a segment for $r>c\in {\delta}_n$ in correspondence 
of the values $r=b_n$ for $n\in \Bbb{N}$. The function in Eq. (3.49) 
is a continuous parametric curve of the $r$ variable on the complex plane 
which follows at once from the continuity of the functions 
${\varphi}_1,{\varphi}_2,{\varphi '}_1$and ${\varphi '}_2$. 
Hence, we are dealing with a continuous curve of the complex plane 
which intersects a given segment at most in a countable sequence of points 
$\left\{{\mu }_n\right\}$, while the distance between consecutive points 
of intersection decreases as $n\to \infty$. Of course, although this 
segment lies on the straight line which passes through the point 
$-{\xi }_1$ and which form an angle 
$$
\vartheta ={{\rm arg} \left({\mu}_{n+1}-\mu_n \right)}={\rm const.}
$$ 
with the positive direction of the real line (we will denote such a 
straight line with $\ell $), such a curve can always be viewed as the 
transformed curve of another curve which in turn intersects the real 
axes precisely at points ${\check{\mu}}_n$ for which 
\begin{equation}
\left|{\check{\mu}}_{n+1}-{\check{\mu}}_n\right|
=\left|{\mu}_{n+1}-{\mu}_n\right|=D_n. 
\label{(3.50)} 
\end{equation}
and it is obtained by an isometry which brings the straight line 
$\ell$ into the real axes. Thus, it will be convenient to study the 
curve which intersects the real axis instead of Eq. (3.49).

Examples of curves which intersect the real axis in a countable sequence 
of points for which distances decrease as the points approach a finite 
limit, can be easily provided. Take for example the following curve 
of the complex plane: 
$$
{\widetilde \gamma}(r)=u(r)+iv(r), 
$$
\begin{equation}
\left \{ \begin{array}{rr}  
u(r)=1-e^{-r} \\ 
& v(r)={\sin \left[\left(1-e^{-r}\right)
\left(\prod^5_{n=1}{r-1-}{{1}\over {{(\sigma)}^n}}\right)\right]\ }
\end{array} 
\right . 
, \;  \sigma >1,\; r\in [0,\infty [ 
\label{(3.51)} 
\end{equation}
whose imaginary part is plotted in Figs. 11 and 12 for $\sigma=1.1$ 
(small and large values of $r$, respectively)
\begin{figure}[h]
\begin{center}
\includegraphics[scale=0.7]{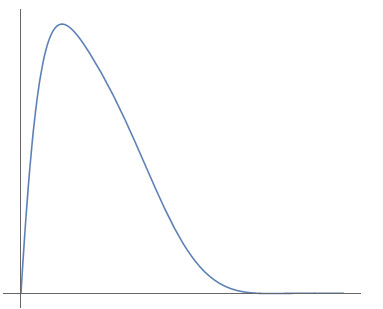}
\caption{Plot of the function $v(r)$ for small values of the independent variable}
\end{center}
\end{figure}
\begin{figure}[h]
\begin{center}
\includegraphics[scale=0.8]{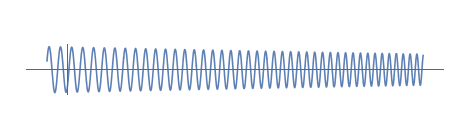}
\caption{Plot of the function $v(r)$ for large values of the independent variable}
\end{center}
\end{figure}
This function has three useful properties: in the limit $r\to \infty $, the 
curve reaches the point $\check{z}=1$; the distance between two subsequent 
points of intersection with the positive real line, decreases when such 
points approach $\check{z}=1$; among all points of intersection with the 
real line, five of them, i.e.
$$
{\check{\mu}}_n=1+{{1}\over {{(\sigma)}^n}},
$$ 
can be arbitrarily chosen by letting vary the number $\sigma >1$. 

Despite the properties mentioned earlier, the function (3.51) is not 
of the type we need because we cannot choose a number $\sigma$ such that 
Eq. (3.50) is satisfied for every $n$, thus we should look for another 
procedure which makes it possible to construct the function (3.49) rather 
than simply guess its functional form. Notice that the points of intersection 
with the real line for the curve (3.51), correspond to the zeros of the 
function $v(r)$ defined therein. Hence, when we act on Eq. (3.49) with an 
isometry which brings $\ell$ into the real line, we obtain a curve of the form
\begin{equation}
\gamma(r)=x(r)+iy(r), 
\label{(3.52)} 
\end{equation}
and we should impose that the function $y$ has an infinite number of zeros 
among which it is possible to find a countable sequence 
$\left\{b_n\right\}$ such that 
\begin{equation}
x\left(b_n\right)={\check{\mu}}_n, 
\label{(3.53)} 
\end{equation}
and the sequence $\left\{{\check{\mu}}_n\right\}$ is necessarily written as below:
\begin{equation}
{\check{\mu}}_n={\check{\mu}}_1+\sum^{n-1}_{k=1}{D_k}, 
\label{(3.54)} 
\end{equation}
where $D_k$ is given in Eq. (3.48). Therefore, once a hyperbolic cyclic 
group is chosen and a function $y(r)$ is so constructed that the 
set of its zeros contains at least the sequence $\left\{b_n\right\}$, 
then the third equation in Eqs. (3.47) can be easily fulfilled. 
Notice that we can choose at will the value of  ${\check{\mu}}_1\in \Bbb{R}$ in 
Eq. (3.54) but there is also another constraint which arises form the 
third equation in Eqs. (3.47). We must have
$$
{\mathop{\rm{lim}}_{n\to \infty } \left|{\mu }_n-{\mu }_1\right|\ }
=\left|{\xi}_1+{{{\xi}_2 -K{\xi}_1}\over {K-1}}\right|
={{\left|{\xi}_1-{\xi}_2 \right|}\over {K-1}}, \; K=A>1,
$$ 
and hence we must also have
\begin{equation}
{\mathop{\rm {lim}}_{n\to \infty } {\check{\mu }}_n}
={\check{\mu}}_1+{{\left|{\xi}_1-{\xi}_2 \right|}\over {K-1}}, 
\label{(3.55)} 
\end{equation}
which is finite for every chosen cyclic hyperbolic group. However, Eq. 
(3.53) defines the sequence $\{b_n\}$ and it is not difficult to find a 
function $x$ such that this sequence can be made monotonic and divergent. 
Of course, this sequence is defined by
\begin{equation}
b_n=x^{-1}\left({\check{\mu}}_n\right). 
\label{(3.56)} 
\end{equation}
Thus, the function 
\begin{equation}
x \; : \; r \in [c,\infty [\longrightarrow \left[{\check{\mu}}_1,
{\check{\mu}}_1+{{\left|{\xi}_1-{\xi}_2\right|}\over {K-1}}\right],
\; c \in {\delta}_n, \; n\in \Bbb{N}, 
\label{(3.57)} 
\end{equation}
is required to be strictly monotonic and thus invertible in its domain. 
This is the only requirement upon the function $x$. Once $x$ is so 
constructed, then there must exist a divergent monotonic sequence 
$\left\{b_n\right\}$. As an example of constructive process for the 
$x$ function, consider the $u$ function into Eq. (3.51) defined on 
$r\in [0,\infty [$ and its values are taken in the inteval $[0,1]$.

If we can easily obtain some functional form for Eq. (3.57), then it 
is a rather difficult task to guess a suitable form of $y$ by 
simply knowing which are its zeros defined in Eq. (3.56). We can 
nevertheless achieve its construction with the help of the theory of canonical 
products in complex analysis \cite{Ahlfors}. 
As we will soon see, given the divergent 
series (3.56), there always exists a representation of an entire function with 
zeros at $\left\{b_n\right\}$ and no other zeros.

Take for example the following infinite product of complex numbers
\begin{equation}
{P=\mathop{\rm {lim}}_{n\to \infty } P_n=\ } {\mathop{\rm {lim}}_{n\to \infty } 
\prod^n_{k=1}{p_k}\ }=\prod^{\infty }_{k=1}{p_k}. 
\label{(3.58)} 
\end{equation}
If such a product is convergent, then $p_k$ must tend to $1$. This is clear from
$$
p_n={{P_n}\over {P_{n-1}}}.
$$ 
In view of this fact, it is convenient to write Eq. (3.58) as
\begin{equation}
P={\mathop{\rm lim}}_{n\to \infty} P_n=\prod^{\infty}_{k=1}
{\left(1+{\omega}_k\right)}, 
\label{(3.59)} 
\end{equation}
where ${\omega}_k\to 0$ is a necessary condition for its convergence. 
Take the infinite sum 
\begin{equation}
S=\sum^{\infty}_{k=1}{ \log (1+{\omega}_k)}, 
\label{(3.60)} 
\end{equation}
suppose that $S$ is finite and denote its partial sum by $S_n$. From 
the fact that ${\omega}_n$ are complex numbers, we must choose the principal 
branch of the logarithm in each term of Eq. (3.60). We obviously have
$$
P_n=e^{S_n},
$$ 
and in the case in which $S_n\to S$, we also have $P_n\to P=e^S\ne 0$. 
Thus, the convergence of the series (3.60) is a sufficient condition 
for the convergence of the infinite product (3.60). It can be proved that 
such a condition is also necessary. We can state the following theorem:
\begin{thm}
The infinite product $\prod^{\infty}_{k=1}
{\left(1+{\omega}_k\right)}$ with $\left(1+{\omega}_n\right)
\ne 0$ converges simultaneously with the series $\sum^{\infty}_{k=1}
{{\log (1+{\omega}_k)}}$ whose terms represent the values of 
the principal branch of the logarithm.
\end{thm}

This theorem proves that the problem of convergence for an infinite 
product can be always reduced to the more familiar question concerning 
the convergence of a series. It can also be noticed that the series 
(3.60) converges absolutely and simultaneously with the simpler series 
$\sum^{\infty}_{k=0}{\left|{\omega}_k\right|}$. This can be deduced by the limit
$$
{\mathop{\rm lim}}_{z\to 0} {\log(1+z)\over z}=1,
$$ 
jointly with the double inequality
$$
\left(1-\epsilon \right)\left|{\omega}_n\right|
<\left|{\rm{log} (1+{\omega}_n)}\right|
<\left(1+\epsilon \right)\left|{\omega}_n\right|,
$$ 
which holds for $\epsilon >0$ and large $n$.
\begin{thm}
A necessary and sufficient condition for the absolute 
convergence of the product $\prod^{\infty}_{k=1}{\left(1+{\omega}_k\right)}$
is the convergence of the series $\sum^{\infty}_{k=1}{\left|{\omega}_k\right|}$.
\end{thm}

Nevertheless, some examples can be found which show that the convergence of 
the series $\sum^{\infty}_{k=1}{{\omega}_k}$ is neither sufficient nor 
necessary for the convergence of the infinite product (3.59).
We can now revert to the problem of the convergence for infinite products 
whose factors are functions of a variable. This will be extremely useful 
for us because we might, in this way, obtain some representation of the 
function $y(r)$ in Eq. (3.52) by using a generalization of the fundamental 
theorem of algebra as we will see.

Take an entire function $g(z)$ on the complex plane. Then $e^{g(z)}\ne 0$ 
is an entire function as well. Conversely, if $f(z)\ne 0$ is an entire function, 
we can show that it can be always represented as $e^{g(z)}$. 
We point out that the logarithmic derivative
$$
{{d}\over {dz}} \left(\log f(z)\right)={{f'(z)}\over {f(z)}},
$$ 
is analytic in the whole plane and thus it is the derivative of an 
entire function $g(z)$. From this fact, by direct computation, we can 
state that $f(z)e^{-g(z)}$ has everywhere vanishing derivative, thus 
$f(z)$ should be a constant multiple of $e^{g(z)}$. The constant factor 
can be absorbed in the definition of the function $g$. This method leads 
us to a powerful representation of entire functions which may have zeros 
on the complex plane. Assume that $f(z)$ has a zero at $z=0$ of 
multiplicity $s$ and a finite sequence of zeros $b_1,\ b_2,\dots ,b_N$ 
(multiple zeros being repeated). From the above discussion, we can write 
such a function \cite{Ahlfors,Greco} as
\begin{equation}
f(z)=z^se^{g(z)}\prod^N_{k=1}{\left(1-{{z}\over {b_k}}\right)}. 
\label{(3.61)} 
\end{equation}
If there exist infinitely many zeros, we can try to obtain a generalization of 
Eq. (3.61) by introducing an infinite product
\begin{equation}
f(z)=z^se^{g(z)}\prod^{\infty}_{k=1}{\left(1-{{z}\over 
{b_k}}\right)}. 
\label{(3.62)} 
\end{equation}
This last representation is valid if and only if the infinite product 
converges uniformly on every compact set of the complex plane. In fact, 
if this is the case, the infinite product occurring in Eq. (3.62) represents 
an entire function with zeros at the same points of $f(z)$ and same 
multiplicity as $f(z)$, and their quotient can be expressed as
$$
z^s e^{g(z)}={{f(z)}\over {\prod^{\infty}_{k=1}{\left(1-{{z}\over {b_k}}\right)}}}.
$$ 
The product in Eq. (3.62), converges absolutely if and only if 
$\sum^{\infty}_{k=1}{{{1}\over {\left|b_k\right|}}}$ is convergent and in 
this case the convergence is also uniform in every compact disc with 
$\left|z\right|<R$ for some $R$. It is only under this special condition that we 
can obtain a representation of the form (3.62). But a method is available which 
makes it possible to introduce some convergence-producing factors for treating 
the general case. One can prove the existence of polynomials $p_k(z)$ such that, 
for any chosen sequence of complex numbers $\left\{b_k\right\}$ 
and $b_n\to \infty$, the function
\begin{equation}
\prod^{\infty}_{k=1}{\left(1-{{z}\over {b_k}}\right)}e^{p_k(z)}, 
\label{(3.63)} 
\end{equation}
converges to an entire function and the product converges together with the series
\begin{equation}
\sum^{\infty}_{k=1}{r_k(z)}=\sum^{\infty}_{k=1} 
\left[\log \left(1-{z \over b_{k}}\right)
+p_{k}(z)\right],
\label{(3.64)} 
\end{equation}
where the branch of the logarithm shall be chosen so that the imaginary 
part of the leading term $r_k(z)$ lies in $[-\pi ,\pi ]$. For a given $R$ we 
can consider the only terms with $\left|b_k\right|>R$. In the region 
$\left|z\right|<R$, the principal branch of 
$\log \left(1-{{z}\over {b_k}}\right)$ 
can be expanded in a Taylor series
$$
{\log \left(1-{{z}\over {b_k}}\right)\ }=-{{z}\over {b_k}}
-{{1}\over {2}}{\left({{z}\over {b_k}}\right)}^2-{{1}\over {3}}
{\left({{z}\over {b_k}}\right)}^3-\dots .
$$ 
We reverse the signs and choose $p_k(z)$ as partial sums
$$
p_k\left(z\right)=-{{z}\over {b_k}}-{{1}\over {2}}{\left({{z}\over 
{b_k}}\right)}^2-\dots {{1}\over {s_k}}{\left({{z}\over {b_k}}\right)}^{s_k}.
$$ 
In this way the leading term of the series (3.64) has the representation
$$
r_k\left(z\right)=-{{1}\over ({s_k+1})}{\left({{z}\over {b_k}}\right)}^{s_k+1}
-{{1}\over ({s_k+2})}{\left({{z}\over {b_k}}\right)}^{s_k+2}-\dots 
$$ 
and we easily obtain the estimate
\begin{equation}
\left|r_k(z)\right|\le {{1}\over ({s_k+1})}{\left({{R}\over 
{\left|b_k\right|}}\right)}^{s_k+1}{\left(1-{{R}\over 
{\left|b_k\right|}}\right)}^{-1}. 
\label{(3.65)} 
\end{equation}
From the previous estimate, it follows $r_k(z)\to 0$ and by 
supposing that the series 
\begin{equation}
\sum^{\infty}_{k=1}{{{1}\over ({s_k+1})}}{\left({{R}\over 
{\left|b_k\right|}}\right)}^{s_k+1}, 
\label{(3.66)} 
\end{equation}
converges, it follows that $\sum^{\infty}_{k=1}{r_k(z)}$ is absolutely 
and uniformly convergent for $\left|z\right|\le R$, and thus the product 
(3.63) represents an analytic function is such a disk. It remains to prove 
that the series (3.66) is convergent, but this is trivial because it has 
a majorant geometric series with convergence ratio less then $1$.
\begin{thm}
There exists an entire function with arbitrarily 
prescribed zeros $b_n$ provided that, in the case of infinitely 
many zeros, $b_n\to \infty$. Every entire function with these and 
no other zeros can be written in the form
\begin{equation}
f(z)=z^se^{g(z)}\prod^{\infty }_{k=1}{\left(1-{{z}\over 
{b_k}}\right)}e^{\left[{{z}\over {b_k}}+{{1}\over {2}}{\left({{z}\over 
{b_k}}\right)}^2+\dots +{{1}\over {s_k}}{\left({{z}\over {b_k}}
\right)}^{s_k}\right]}, 
\label{(3.67)} 
\end{equation}
where the product is taken over all $b_n\ne 0$, the $s_k$   
are certain integers, and $g(z)$ is an entire function.
\end{thm}
This theorem is due to Weierstrass and it answers a problem which in the 
literature is known as the Weierstrass problem \cite{Greco} for the 
representation of entire functions starting from the knowledge of 
their zeros. Functions of type (3.67) are also called entire transcendental functions.
The representation (3.67) can be made considerably more interesting 
if we can choose all $s_k$ equal to each other. In the previous proof 
it has been shown that the product
$$
\prod^{\infty}_{k=1}{\left(1-{{z}\over {b_k}}\right)}
e^{\left[{{z}\over {b_k}}+{{1}\over {2}}{\left({{z}\over 
{b_k}}\right)}^2+\dots +{{1}\over {h}}{\left({{z}\over 
{b_k}}\right)}^{h}\right]},
$$
which is commonly called {\it canonical product} if $h$ is the smallest 
integer which makes covergent the following series for all $R$: 
$$
\sum^{\infty}_{k=1}{{{1}\over ({h+1})}}{\left({{R}\over 
{\left|b_k\right|}}\right)}^{h+1}<\infty,
$$
which may occur if 
$$
\sum_{k=1}^{\infty}{1\over |b_k|^{h+1}}<\infty.
$$
The integer $h$ is called the {\it genus} of the sequence $\{b_k\}$. Whenever 
possible it is rather convenient to use the canonical product into 
the representation (3.67) (which is uniquely determined). If in this 
representation $g(z)$ reduces to a polynomial, the function $f(z)$ is 
said to be of finite genus, and its genus is equal to the degree of 
this polynomial or equal to the genus of the canonical product, 
whichever is the larger. For istance, a function of genus zero is of the form
$$
Cz^m\prod^{\infty}_{k=1}{\left(1-{{z}\over {b_k}}\right)},
$$
with 
$$
\sum_{k=1}^{\infty}{1\over |b_k|}<\infty.
$$
A function with genus one has two representations. It can be of the form
$$
Cz^me^{\alpha z}\prod^{\infty }_{k=1}{\left(1-{{z}\over 
{b_k}}\right)}e^{z\over b_{k}},
$$
with 
$$
\sum_{k=1}^{\infty}{1\over |b_k|} = \infty,\;  
\sum_{k=1}^{\infty}{1\over |b_k|^2}<\infty.
$$
or of the form
$$
Cz^me^{\alpha z}\prod^{\infty }_{k=1}{\left(1-{{z}\over {b_k}}\right)},
$$
with 
$$
\sum_{k=1}^{\infty}{1\over |b_k|}<\infty, \; \alpha\neq 0.
$$
As we can see, this Theorem 3.3 fits our expectations: we were interested 
in constructing the function $y$ occurring in Eq. (3.52) by starting 
from the knowledge of its zeros $b_n$, which are given in Eq. (3.56). 
We have already mentioned that such a sequence of zeros, can be always 
chosen as a strictly monotonic divergent sequence on the positive real 
line by simply requiring the monotonicity of the function $x$ occurring in 
Eq. (3.52) and defined in Eq. (3.57). Recall that the strict monotonicity 
of $\left\{b_n\right\}$ ensures that the circle $C_{b_{n+1}}$is contained 
in the interior of $C_{b_n}$ for each $n$ and thus we must require so for 
the accomplishment of the identification between isometric circles of a 
cyclic hyperbolic group and that of a limit-point theory. By using Eq. 
(3.56) jointly with Eq. (3.54), our $y(r)$ function 
admits the following representation:
\begin{equation}
y\left(r\right)=e^{g(r)}\prod^{\infty}_{k=1}{\left(1-{{r}\over 
{x^{-1}({\check{\mu}}_k)}}\right)}e^{\left[{{r}\over {x^{-1}
({\check{\mu}}_k)}}+{{1}\over {2}}{\left({{r}\over {x^{-1}
({\check{\mu}}_k)}}\right)}^2+\dots +{{1}\over {s_k}}
{\left({{r}\over {x^{-1}({\check{\mu}}_k)}}\right)}^{s_k}\right]}, 
\label{(3.68)} 
\end{equation}
in which we have supposed that the value of ${\check{\mu}}_1$ into 
Eq. (3.54) is so chosen that none of the ${\check{\mu}}_n$ is equal to 
zero (this can be easily obtained by simply setting ${\check{\mu}}_1>0$ 
and invoking the freedom we have upon such a variable). 

A few more remarks are now in order. The first concerns the oscillations 
of the function $y(r)$ which must tend to zero in the limit $r\to \infty $.
This is evident from Eqs. (3.52), (3.53) and (3.54) by taking the limit
$$
\lim_{n \to \infty} \gamma(b_n)
={\check{\mu}}_1+{{\left|{\xi}_1-{\xi}_2 \right|}\over {K-1}}\in \Bbb{R},
$$ 
thus 
$$
\lim_{r \to \infty} y(r)=0.
$$ 
This last condition, for example, can be fulfilled by requiring 
$g(r)=-r$ in Eq. (3.68).

The second remark concerns the possibility of using the representation 
given in Eq. (3.62), which is easier to handle, instead of that given 
in Eq. (3.68). Equation (3.62) can be used if and only if the series
$$
\sum^{\infty}_{k=1}{{{1}\over {\left|b_k\right|}}}
=\sum^{\infty}_{k=1}{{{1}\over {\left|x^{-1}\left(
{\check{\mu}}_k\right)\right|}}},
$$ 
is convergent. For example, if we choose 
$$
x(r)=\left({\check{\mu}}_1+{{\left|{\xi}_1-{\xi}_2\right|}
\over {K-1}}\right)\left(1-e^{-r}\right), \; {\check{\mu}}_1>0,
$$ 
(and this function satisfies Eqs. (3.54), (3.55), (3.57) and it is 
also monotonic in its domain, thus it provides the divergent series 
given in Eq. (3.56)) then, by setting $\nu =\left({\check{\mu}}_1
+{{\left|{\xi}_1-{\xi}_2 \right|}\over {K-1}}\right)$, there 
remains defined the sequence of values
\begin{equation}
b_n= \log \left( {\nu \over \nu-\beta_{n}}\right),
\label{(3.69)} 
\end{equation}
where the ${\check{\mu}}_n$ form a bounded sequence from Eqs. 
(3.54) and (3.55). In this case, we can adopt the representation 
(3.62), if and only if the series
\begin{equation}
\sum^{\infty}_{k=1}{{{1}\over {\left| \log 
\left({{\nu}\over {\nu -{\check{\mu}}_k}}\right)
\right|}}}, 
\label{(3.70)} 
\end{equation}
converges. But this does not converge from the fact that there always 
exists some integer $N$ such that for each $n>N$ the following inequality holds:
\begin{equation}
{\left| \log \left({{\nu}\over {\nu -{\check{\mu}}_n}}\right)
\right|}^{-1}>\ {{1}\over {n}}, \;  n>N. 
\label{(3.71)} 
\end{equation}
Equation (3.71) follows at once by noticing that ${\check{\mu}}_k\to \nu$, 
thus the terms in the series (3.70) are bounded from below by the 
terms of a harmonic series. In this case we cannot use the representation 
(3.62) but we hope that a suitable choice for the monotonic function 
$x$ can be always made in such a way that the resulting series of the 
type (3.70) can be made convergent. In every case, the representation 
(3.68) can be always used.

The last remark concerns the possibility of treating in the same way 
the second equation in Eqs. (3.47). The point here is that we cannot 
apply directly Theorem 3.3, because in such a circumstance there should 
remain defined a monotonic decreasing sequence of points $\left\{a_n\right\}$ 
instead of a monotonic increasing sequence, which is one of the 
hypotheses of Theorem 3.3. Nevertheless, we may expect that all the 
constructive procedure adopted in the last few pages for the fulfilment 
of the third equation in Eqs. (3.47), should also be applicable for 
the fulfilment of the second equation among Eqs. (3.47). This goal 
should be achieved by simply repeating all the reasoning here 
developed, but now for the function 
\begin{equation}
-{\tilde{m}'}_{\rho }={{\left[{\varphi}_1{\varphi}_2\right]
\left({{1}\over {\rho }}\right)}\over {\left[{\varphi}_2
{\varphi}_2 \right]\left({{1}\over {\rho }}\right)}}, \;  
r={{1}\over {\rho }}, \;  r\in ]0,c], \;  c\in {\delta}_n 
\label{(3.72)} 
\end{equation}
where the definition made for the $r$ variable, enables us to replace 
the limit $r\to 0$ (which is the case of the second equation in Eqs. 
(3.47)) with the limit $\rho \to \infty$. Therefore, Eq. (3.7) defines 
a parametric continuous curve on the complex plane in the $\rho$ variable, 
while the second equation in Eqs. (3.47) forces such a curve to intersect 
a particular segment of the complex plane which lies on a straight line 
$\ell '$, which passes through the point $-{\xi }_2$ and forms an angle 
$$
\vartheta' = \rm{arg}\left(\eta_{n+1}-\eta_{n}\right),
$$ 
with the positive direction of the real axis. The result of the 
analysis should possibly end with the construction of a curve 
$$
\gamma '(\rho)=x'(\rho)+iy'(\rho),
$$ 
(here the primed functions do not refer to their derivatives) where 
$x'(\rho)$ should be taken as a monotonic invertible function to 
guarantee the existence of a decreasing sequence $\left\{a_n\right\}$ 
such that $a_n\to 0$. The function $y'$ should also be obtained from 
Eq. (3.68) (or Eq. (3.62) when the infinite product can be made 
convergent) by substituting the $r$ variable with the $\rho$ variable. 

\subsection {Further Considerations}

In the previous subsection we have built a method which might lead 
to the fulfilment of the system of equations (3.47). The aim of the 
previous pages, is to set up a technique which can prove the 
existence of solutions for the variables  
\begin{equation}
{\tilde{m}'}_{a_n},\ \ \ {\tilde{m}}_{b_n}, \; a_n, \;  
b_n, \; \forall n\in \Bbb{N}, 
\label{(3.73)} 
\end{equation}
where $a_n$ and $b_n$ satisfy the strict monotonicity of Eqs. (3.44). 
We have shown, by a constructive process, that such solutions always 
exist and $a_n$ and $b_n$ can always be chosen to fulfil Eqs. (3.44). 
This has been achieved by the exclusive use of the third and the 
second equations in Eqs. (3.47). The first and the fourth equations 
in Eqs. (3.47) can be satisfied in a very easy way once the solutions 
$a_n$ and $b_n$ are known. These latter equations bring us to another 
constructive process which, this time, involves the function 
${\varphi}_2$ for values $r>c$ and $r<c$ (taking care to guarantee its 
differentiability at the point $c\in {\delta}_n$). Thus, after the 
functional form of ${\varphi}_2$ is taken, we should try to obtain 
the functional form of ${\varphi}_1$ for values $r<c$ and $r>c$ by 
retaining its differentiability at the point $c\in {\delta}_n$ and by 
making use of the functions ${\tilde{m}}_r$ and ${\tilde{m}'}_r$. 
In doing this, we should require that 
$$
{\varphi }_1\left(c,l\right)=1, \; 
{\varphi }_2\left(c,l\right)=0,
$$ 
\begin{equation}
p\left(c\right){\varphi '}_1 \left(c,l\right)=0, \;  p
\left(c\right){\varphi '}_2 \left(c,l\right)=1, 
\label{(3.74)} 
\end{equation}
where this time $p(c)$ should be considered as an unknown positive 
real number which represents the value of the coefficient 
$p(r)>0$ of the equation 
\begin{equation}
Lx(r)=-\left(px(r)\right)'+qx(r)
=lx(r), \;  l\in \Bbb{C}, \;  r>0, 
\label{(3.75)} 
\end{equation}
at the point $c$. Equations (3.74) are extremely important because 
they are necessary conditions to let the functions ${\varphi}_1$ and 
${\varphi}_2$ be independent solutions of Eq. (3.75). At this stage, 
the problem is to find the coefficients $p(r)>0$ and $q(r)$ 
in such a way that ${\varphi}_1$ and ${\varphi}_2$, which satisfy 
conditions (3.74), could be independent solutions of the second order 
singular equation given in Eq. (3.75).
As one can see, this is the inverse problem of finding solutions 
of a given differential equation. Such a problem has been studied 
in Refs. \cite{Gelfand1951,Brown2009}.
We note that our method depends on the value of the imaginary 
part of the complex number $l$ occurring in (3.75). There are good 
possibilities to make the fixed points ${\xi}_1$ and ${\xi}_2$ 
dependent on the number $l$, thus one should be able to obtain the 
spectral matrix from Eq. (2.50). This makes us hope that the problem of 
finding the differential equation by starting from the knowledge of 
an independent set of functions ${\varphi}_1$ and ${\varphi}_2$, 
in the context of the present section, might be solvable 
or at least partially solvable in future.

There are two main advantages which arise from the fulfilment of 
the system (3.47) and which should be merely taken in consideration 
from a purely physical point of view. The first one is that such a 
system of equations guarantees the essential self-adjointness of a 
second-order, singular operator at both ends of the positive real line 
of type (3.75). This cannot be only a mathematical property of operators, 
although the retained self-adjointness arising from a Kleinian group 
deserves by itself a careful consideration. The point here is that the 
limit-point theory has been extensively used in the context of the quantum 
theory for operators of type (2.6) (for example in 
Refs. \cite{ReedSimon1975,Simon2015} and it would
be very surprising if some of them could give rise to a pair of independent 
solutions ${\varphi}_1$ and ${\varphi}_2$ and a pair of limit points 
$m_{\infty}={\xi}_1$ and ${m'}_{\infty}={\xi}_2$ which fulfil Eqs. 
(3.47) for some value of the parameter $K$, because of the strict constraints 
inherent to the system (3.47). Anyway, we suggest the possibility of 
retaining some sort of generality in the decomposition of the functional 
space upon which quantum Hamiltonians should be defined. For example, 
we might require that
\begin{equation}
{\cal{L}}^2\left({\Bbb{R}}^n,d^nx\right)={\cal{L}}^2\left({\Bbb{R}}^+,
\zeta(r)dr\right)\otimes {\cal{L}}^2
\left(\cal{T},\cal{O}\right), 
\label{(3.76)} 
\end{equation}
where $\cal{T}$ is a smooth topological manifold homeomorphic to the 
$S^{(n-1)}$ sphere, $d\cal{O}$ is the measure on $\cal{T}$, while 
$\zeta(r)$ is a function of $r\in {\Bbb{R}}^+$ such that 
$d^nx=\zeta (r)drd\cal{O}$. If this decomposition can be achieved 
from a functional analytic point of view, then we could define a set 
of coordinates above the $\cal{T}$ manifold and we could express the 
Laplacian $P=-\Delta$ in terms of these coordinates by starting from 
its expression in orthonormal coordinates
$$
P=-\rm{\Delta }=-{\rm div \; grad}=-\sum^n_{k=1}{{{{\partial }^2}
\over {\partial x^2_k}}}.
$$ 
This procedure should lead to a decomposition of the $\Delta$ 
operator which in turn might be written as
\begin{equation}
\Delta ={\Delta}_r+{\Delta}_{\cal{T}}, 
\label{(3.77)} 
\end{equation}
in which ${\rm{\Delta}}_r$ depends only on the $r$ coordinate 
(and this is the case if the validity of Eq. (3.76) can be ensured) 
while ${\rm{\Delta}}_{\cal{T}}$ is dependent on the other coordinates 
previously defined on $\cal{T}$. Note that the operator 
${\rm{\Delta}}_{\cal{T}}$ reduces to the operator
$$
{\Delta}_S=-{{{\hat{L}}^2}\over {{\hslash }^2 r^2}},
$$ 
when $\cal{T}$ is chosen to be the $S^{(n-1)}$ sphere (${\hat{L}}^2$ 
is the squared angular momentum of the particle in $n$ dimensions). 
Therefore, when the decomposition (3.77) is allowed, we can regard 
the operator ${\Delta}_{\cal{T}}$ to be strictly related to some sort 
of generalized angular momentum of a quantum particle which is also a 
conserved quantity, and thus we can study its self-adjointness properties 
as well as its spectrum. If it is self-adjoint, then the spectrum could 
also be obtained leading to a second-order differential equation 
(as it happens for Eq. (2.3)) in the radial variable, which can be 
treated with the machinery of the limit-point, limit-circle theory. 

All the generality retained so far for the Laplacian operator, leads 
us to the second advantage which arises from the fulfilment of Eq. (3.47). 
In this case, a hyperbolic cyclic group is defined, and its discreteness 
is a necessary condition for the existence of non-constant automorphic 
functions. Let us denote with $T_1, T_2,\dots$ all the elements of 
a generic Kleinian group \cite{Maskit}.
Then we can define an automorphic function 
\cite{Ford} as
\begin{defn}
A function $f$ of the complex 
variable $z$, is said to be automorphic with respect to a group of 
linear transformations $T_1, T_2,\dots$ provided that
\vskip 0.2cm
\noindent 
(1) $f(z)$ is a single-valued analytic function.
\vskip 0.2cm
\noindent 
(2) If $z$ lies in the domain of existence of $f(z)$, 
the same holds for $T_{n}(z)$.
\vskip 0.2cm
\noindent 
(3) $f(T_{n}(z))=f(z)$.
\end{defn}

It turns out that an automorphic function can be non-constant if and only 
if the group of transformations is a properly discrete group, i.e. 
there are no infinitesimal transformations.
The existence for the domain of such functions is ensured by the following theorem:
\begin{thm}
The domain of existence of an automorphic function 
extends into the neighborhood of every limit point of the group.
\end{thm}

The hyperbolic cyclic groups considered in this section are Kleinian 
groups with two limit points, i.e. ${\xi}_1$ and ${\xi}_2$, thus, 
if an automorphic function exists, then its domain of existence must 
extend into the neighborhoods of the limit points of the group. 
Other properties of automorphic functions can be stated as well. 
For example, the limit points of the group are essential singularities 
for the automorphic functions.

In general, given a Kleinian group, nothing ensures the unicity of 
its automorphic function, but one can always establish their existence. 
In this context it can be extremely useful to consider the 
{\it Poincar\'{e} $\theta$ series} which is defined as
\begin{equation}
\theta(z)=\sum^{\infty}_{k=1}{{\left
(c_k+d_k\right)}^{-2m}H(T_{k}(z))}, 
\label{(3.78)} 
\end{equation}
where $c_k$ and $d_k$ are the coefficients of the fractional linear trasformation
\begin{equation}
T_{k}(z)={{a_{k} z+b_k}\over {c_{k} z+d_k}}, \; a_{k} d_k-b_{k} c_k=1, 
\label{(3.79)} 
\end{equation}
$m$ is the number of transformations which the Kleinian group contains 
and $H(z)$ is any rational function of the $z$ variable none of 
whose poles is at a limit point of the group. In this context, the group 
is viewed as being finite, but one can derive a convergent series 
(3.78) when the limit $m\to \infty$ is taken. It can be shown that  
$$
\theta \left(T_j(z)\right)={\left(c_{j} z+d_{j}\right)}^{2m}
\theta \left(z\right),
$$ 
and hence, the quotient between two Poincar\'{e} $\theta$ series 
${\theta}_1(z)$ and ${\theta}_2(z)$ leads to an automorphic function
\begin{equation}
F\left(T_j\left(z\right)\right)={{{\theta }_1(T_j(z))}\over 
{{\theta }_2(T_J(z))}}={{{\left(c_jz+d_j\right)}^{2m}{\theta }_1
\left(z\right)}\over {{\left(c_jz+d_j\right)}^{2m}{\theta }_2
\left(z\right)}}={{{\theta }_1(z)}\over {{\theta }_2(z)}}=F\left(z\right). 
\label{(3.80)} 
\end{equation}
We can thus state the following theorem:
\begin{thm}
If $m\ge 2$ and if the point at infinity 
is an ordinary point of the group, then the $\theta$ series (3.78)
defines a function which is analytic 
except possibly for poles in any connected region not containing 
limit points of the group in its interior.
\end{thm}
 
From the fact that in a hyperbolic cyclic group the point at infinity 
is an ordinary point, this last theorem enables us to state that a 
convergent theta series can be always written down for this type of 
groups and thus, an automorphic function given in Eq. (3.80), 
can be always constructed for them.

When we are in the context of the isometries of an asymptotically 
flat space-time, we expect that a suitable group of isometries should 
give rise to some conserved quantity, i.e. a quantity which does not 
change when we transform some suitable system of coordinates by applying 
the isometries of which such a group is composed. It should be clear 
that conserved quantities are not merely constants if and only if 
the group of isometries is a discrete group. This is the second 
advantage of treating cyclic hyperbolic groups in physical context. 
If we can interpret discrete Kleinian groups 
\cite{Maskit} as discrete groups of 
isometries of an asymptotically flat space-time (and thus we are just 
treating discrete subgroups of the BMS group) then there remains defined 
a non-constant function which is unaffected when the system of coordinates 
is transformed according to such a discrete Kleinian group. We could also 
call such a ``conserved'' function a ``constant of motion'' for a 
particle whose motion is in accordance with the discreteness inherent to the 
Kleinian group considered, on the conformal infinity of an asymptotically flat space-time.  

\section {Concluding Remarks}

In our paper we have studied the nature of fractional linear transformations in
a general relativity context as well as in a quantum theoretical framework. 
Two features deserve special attention: the first is the possibility of separating
the limit-point condition at infinity into loxodromic, hyperbolic, parabolic and
elliptic cases. This is useful in a context in which one wants to look for a
correspondence between essentially self-adjoint spherically symmetric Hamiltonians
of quantum physics and the theory of Bondi-Metzner-Sachs transformations in 
general relativity \cite{AE2018,EA2018}. The analogy therefore arising, suggests 
that further investigations might be performed for a theory in which the role of
fractional linear maps is viewed as a bridge between the quantum theory and
general relativity.

The second aspect to point out is the possibility of interpreting the limit-point
condition at both ends of the positive real line, for a second-order singular
differential operator, which occurs frequently in applied quantum mechanics, as
the limiting procedure arising from a very particular Kleinian group which is the
hyperbolic cyclic group. In this framework, we have found in Sect. 3 that a
consistent system of equations can be derived and studied. Hence we are led to
consider the entire transcendental functions, from which it is possible to
construct a fundamental system of solutions of a second-order differential
equation with singular behavior at both ends of the positive real line, which 
in turn satisfy the limit-point conditions. Further developments in this
direction might also be obtained by constructing a fundamental system of solutions
and then deriving the differential equation whose solutions are the independent
system first obtained. This guarantees two important facts at the same time: 
the essential self-adjointness of a second-order differential operator and the
existence of a conserved quantity which is an automorphic function for the
cyclic group chosen. By accomplishing this process, we hope that some sort of 
interpretation, in terms of fiscrete symmetries of space-time, might also be
established. Moreover, it remains to be seen whether all basic properties of
(global) general relativity have a quantum counterpart, if the correspondence
that we have suggested is found to be viable.

\section*{acknowledgments}
G. E. is grateful to the Dipartimento di Fisica ``Ettore Pancini''
for hospitality and support.


\begin{thebibliography}{99}

\bibitem{Enriques}
F. Enriques, {\it Lezioni di Geometria Proiettiva} (Zanichelli, Bologna, 1920).

\bibitem{Spampinato}
N. Spampinato, {\it Lezioni di Geometria Superiore, Vols. 1-9}
(Raffaele Pironti, Napoli, 1948-1950).

\bibitem{FubiniCech}
S. Fubini and E. Cech, {\it Geometria Proiettiva Differenziale, Vols. 1,2}
(Zanichelli, Bologna, 1926).

\bibitem{AE2018}
F. Alessio and G. Esposito, On the structure and applications of the
Bondi-Metzner-Sachs group, {\it Int. J. Geom. Methods Mod. Phys.}
{\bf 15} (2018) 1830002.

\bibitem{EA2018}
G. Esposito and F. Alessio, From parabolic to loxodromic BMS transformations,
{\it Gen. Rel. Grav.} {\bf 50} (2018) 141.

\bibitem{L1}
P.J. McCarthy and E. Melas, On irreducible representations of the ultrahyperbolic
BMS group, {\it Nucl. Phys. B} {\bf 653} (2003) 369. 

\bibitem{L2}
A. Campoleoni, H.A. Gonzalez, B. Oblak and M. Riegler, Rotating higher spin
partition functions and extended BMS symmetries,
{\it JHEP} {\bf 04} (2016) 034.

\bibitem{L3}
S. W. Hawking, M. Perry and A. Strominger, Superrotation charge and
supertranslation hair on black holes, {\it JHEP} {\bf 05} (2017) 161.

\bibitem{L4} 
G. Barnich, L. Donnay, J. Matulich and R. Troncoso, Super-${\rm BMS}_{3}$
invariant boundary theory from three-dimensional flat supergravity,
{\it JHEP} {\bf 01} (2017) 029.

\bibitem{L5}
E. Melas, On the representation theory of the Bondi-Metzner-Sachs group
and its variants in three space-time dimensions, 
{J. Math. Phys.} {\bf 58} (2017) 071705.

\bibitem{L6}
M. Henneaux and C. Troessaert, BMS group at spatial infinity: the Hamiltonian
(ADM) approach, {\it JHEP} {\bf 03} (2018) 147.

\bibitem{L7}
M. Henneaux and C. Troessaert, Asymptotic symmetries of electromagnetism at 
spatial infinity, {\it JHEP} {\bf 05} (2018) 137.

\bibitem{L8}
M. Henneaux and C. Troessaert, Hamiltonian structure and asymptotic symmetries
of the Einstein-Maxwell system at spatial infinity, 
{\it JHEP} {\bf 07} (2018) 171.

\bibitem{L9}
M. Henneaux and C. Troessaert, Asymptotic structure of a massless scalar field 
and its dual two-form field at spatial infinity, 
{\it JHEP} {\bf 05} (2019) 147.

\bibitem{L10}
M. Henneaux and C. Troessaert, Asymptotic structure of electromagnetism in higher
spacetime dimensions, {\it Phys. Rev. D} {\bf 99} (2019) 125006.

\bibitem{L11}
F. Alessio and M. Arzano, Note on the symplectic structure of asymptotically 
flat gravity and BMS symmetries, {\it Phys. Rev. D} {\bf 100} (2019) 044028.

\bibitem{L12}
S. Bakhoda, F. Mahdich and H. Shojaie, Asymptotic conformal symmetry at
spatial infinity, {\it Phys. Rev. D} {\bf 100} (2019) 124051.

\bibitem{L13}
S. Pasterski, Implications of superrotations, {\it Phys. Rept.}
{\bf 829} (2019) 1.

\bibitem{L14}
M. Henneaux, J. Matulish and J. Neogi, Asymptotic realization of the 
super-BMS algebra at spatial infinity, {\it Phys. Rev. D} {\bf 101}
(2020) 126016.

\bibitem{L15}
O. Fuentealba, M. Henneaux, S. Majumdar, J. Matulich, C. Troessaert,
{\it Class. Quantum Grav.} {\bf 37} (2020) 235011.

\bibitem{L16}
L. Donnay, G. Giribet and F. Rosso, Quantum BMS transformations in 
conformally flat space-times and holography, 
{\it JHEP} {\bf 12} (2020) 102.

\bibitem{L17}
G. Barnich and R. Ruzziconi, Coadjoint representation of the BMS group
on celestial Riemann surfaces, arXiv:2103.11253 [gr-qc].
 
\bibitem{Chatterjee1990}
A. Chatterjee, Large-N expansions in quantum mechanics, {\it Phys. Rep.}
{\bf 186} (1990) 249.

\bibitem{ReedSimon1975}
M. Reed and B. Simon, {\it Methods of Modern Mathematical Physics. II. Fourier
Analysis and Self-Adjointness} (Academic Press, New York, 1975).

\bibitem{deAlfaroRegge1965}
V. de Alfaro and T. Regge, {\it Potential Scattering} (North Holland, 
Amsterdam, 1965).

\bibitem{Simon2015}
B. Simon, {\it A Comprehensive Course in Analysis, Part 4, Operator Theory}
(American Mathematical Society, Providence, 2015).

\bibitem{Weyl1910}
H. Weyl, On ordinary differential equations with singularities and the
associated expansions of arbitrary functions, {\it Math. Annal.}
{\bf 68} (1910) 220.

\bibitem{BE2019}
V. F. Bellino and G. Esposito, Revisited version of Weyl's limit-point
limit-circle criterion for essential self-adjointness,
{\it J. Phys. Comm.} {\bf 3} (2019) 035017.

\bibitem{CoddingtonLevinson1955}
E. A. Coddington and N. Levinson, {\it Theory of Ordinary Differential
Equations} (McGraw-Hill, New York, 1955).

\bibitem{Tit1962}
E. C. Titchmarsh, {\it Eigenfunction Expansions Associated with Second-Order
Differential Equations} (Oxford University Press, Oxford, 1962).

\bibitem{Ford}
L. Ford, {\it Automorphic Functions} (Chelsea Publishing Company,
New York, 1929).

\bibitem{Gelfand1951}
I. M. Gelfand and B. M. Levitan, On the determination of a differential 
equation from its spectral function, 
{\it Am. Math. Transl.} {\bf 1} (1951) 253.

\bibitem{Brown2009}
M. Brown, J. Hinchcliffe, M. Marletta, S. Naboko and I. Wood, 
The abstract Titchmarsh-Weyl M-function for adjoint operator pairs and
its relation to the spectrum, {\it Int. Eqs. Op. Theory}
{\bf 63} (2009) 297.

\bibitem{Ahlfors}
L. V. Ahlfors, {\it Complex Analysis} (McGraw-Hill, New York, 1966).

\bibitem{Greco}
D. Greco, {\it Complementi di Analisi} (Liguori Editore, Napoli, 1983).

\bibitem{Maskit}
B. Maskit, {\it Kleinian Groups} (Springer, Berlin, 1988).

\end{thebibliography}
\end{document}